\documentclass[longauth]{aa}

\usepackage{graphicx}
\usepackage{natbib}
\usepackage{scalerel}
\usepackage{xcolor}
\usepackage{orcidlink}
\usepackage{soul}

\definecolor{purple1}{rgb}{0.5,0,0.87}

\usepackage{txfonts}
\hypersetup{
    colorlinks=true,
    linkcolor=blue,
    filecolor=magenta,      
    urlcolor=blue,
    citecolor=blue,
    pdftitle={Euclid: Early Release Observations – The intracluster light and intracluster globular clusters of the Perseus cluster},
    pdfauthor={Kluge et al.}
}
\makeatletter
\renewcommand*\aa@pageof{, page \thepage{} of \pageref*{LastPage}}
\makeatother

\usepackage[utf8]{inputenc}

\usepackage[switch, modulo]{lineno}

\usepackage{euclid}

\begin{document}

\title{\Euclid: Early Release Observations --  The intracluster light and intracluster globular clusters of the Perseus cluster\thanks{This paper is published on behalf of the Euclid Consortium.}}

\newcommand{\orcid}[1]{} 
\author{M.~Kluge\orcidlink{0000-0002-9618-2552}\inst{\ref{aff1}}
\and N.~A.~Hatch\orcidlink{0000-0001-5600-0534}\thanks{\email{nina.hatch@nottingham.ac.uk}}\inst{\ref{aff2}}
\and M.~Montes\orcidlink{0000-0001-7847-0393}\inst{\ref{aff3},\ref{aff4}}
\and J.~B.~Golden-Marx\orcidlink{0000-0002-6394-045X}\inst{\ref{aff2}}
\and A.~H.~Gonzalez\orcidlink{0000-0002-0933-8601}\inst{\ref{aff5}}
\and J.-C.~Cuillandre\orcidlink{0000-0002-3263-8645}\inst{\ref{aff6}}
\and M.~Bolzonella\orcidlink{0000-0003-3278-4607}\inst{\ref{aff7}}
\and A.~Lan\c{c}on\orcidlink{0000-0002-7214-8296}\inst{\ref{aff8}}
\and R.~Laureijs\inst{\ref{aff9}}
\and T.~Saifollahi\orcidlink{0000-0002-9554-7660}\inst{\ref{aff8},\ref{aff10}}
\and M.~Schirmer\orcidlink{0000-0003-2568-9994}\inst{\ref{aff11}}
\and C.~Stone\orcidlink{0000-0002-9086-6398}\inst{\ref{aff12}}
\and A.~Boselli\orcidlink{0000-0002-9795-6433}\inst{\ref{aff13},\ref{aff14}}
\and M.~Cantiello\orcidlink{0000-0003-2072-384X}\inst{\ref{aff15}}
\and J.~G.~Sorce\orcidlink{0000-0002-2307-2432}\inst{\ref{aff16},\ref{aff17},\ref{aff18}}
\and F.~R.~Marleau\orcidlink{0000-0002-1442-2947}\inst{\ref{aff19}}
\and P.-A.~Duc\orcidlink{0000-0003-3343-6284}\inst{\ref{aff20}}
\and E.~Sola\orcidlink{0000-0002-2814-3578}\inst{\ref{aff21}}
\and M.~Urbano\orcidlink{0000-0001-5640-0650}\inst{\ref{aff8}}
\and S.~L.~Ahad\orcidlink{0000-0001-6336-642X}\inst{\ref{aff22},\ref{aff23}}
\and Y.~M.~Bah\'{e}\orcidlink{0000-0002-3196-5126}\inst{\ref{aff24}}
\and S.~P.~Bamford\orcidlink{0000-0001-7821-7195}\inst{\ref{aff2}}
\and C.~Bellhouse\orcidlink{0000-0002-6179-8007}\inst{\ref{aff2}}
\and F.~Buitrago\orcidlink{0000-0002-2861-9812}\inst{\ref{aff25},\ref{aff26}}
\and P.~Dimauro\orcidlink{0000-0001-7399-2854}\inst{\ref{aff27},\ref{aff28}}
\and F.~Durret\orcidlink{0000-0002-6991-4578}\inst{\ref{aff29}}
\and A.~Ellien\orcidlink{0000-0002-1038-3370}\inst{\ref{aff30}}
\and Y.~Jimenez-Teja\orcidlink{0000-0002-6090-2853}\inst{\ref{aff31},\ref{aff28}}
\and E.~Slezak\orcidlink{0000-0003-4771-7263}\inst{\ref{aff32}}
\and N.~Aghanim\inst{\ref{aff17}}
\and B.~Altieri\orcidlink{0000-0003-3936-0284}\inst{\ref{aff33}}
\and S.~Andreon\orcidlink{0000-0002-2041-8784}\inst{\ref{aff34}}
\and N.~Auricchio\orcidlink{0000-0003-4444-8651}\inst{\ref{aff7}}
\and M.~Baldi\orcidlink{0000-0003-4145-1943}\inst{\ref{aff35},\ref{aff7},\ref{aff36}}
\and A.~Balestra\orcidlink{0000-0002-6967-261X}\inst{\ref{aff37}}
\and S.~Bardelli\orcidlink{0000-0002-8900-0298}\inst{\ref{aff7}}
\and R.~Bender\orcidlink{0000-0001-7179-0626}\inst{\ref{aff1},\ref{aff38}}
\and D.~Bonino\orcidlink{0000-0002-3336-9977}\inst{\ref{aff39}}
\and E.~Branchini\orcidlink{0000-0002-0808-6908}\inst{\ref{aff40},\ref{aff41},\ref{aff34}}
\and M.~Brescia\orcidlink{0000-0001-9506-5680}\inst{\ref{aff42},\ref{aff43},\ref{aff44}}
\and J.~Brinchmann\orcidlink{0000-0003-4359-8797}\inst{\ref{aff45},\ref{aff46}}
\and S.~Camera\orcidlink{0000-0003-3399-3574}\inst{\ref{aff47},\ref{aff48},\ref{aff39}}
\and G.~P.~Candini\orcidlink{0000-0001-9481-8206}\inst{\ref{aff49}}
\and V.~Capobianco\orcidlink{0000-0002-3309-7692}\inst{\ref{aff39}}
\and C.~Carbone\orcidlink{0000-0003-0125-3563}\inst{\ref{aff50}}
\and J.~Carretero\orcidlink{0000-0002-3130-0204}\inst{\ref{aff51},\ref{aff52}}
\and S.~Casas\orcidlink{0000-0002-4751-5138}\inst{\ref{aff53}}
\and M.~Castellano\orcidlink{0000-0001-9875-8263}\inst{\ref{aff27}}
\and S.~Cavuoti\orcidlink{0000-0002-3787-4196}\inst{\ref{aff43},\ref{aff44}}
\and A.~Cimatti\inst{\ref{aff54}}
\and G.~Congedo\orcidlink{0000-0003-2508-0046}\inst{\ref{aff55}}
\and C.~J.~Conselice\orcidlink{0000-0003-1949-7638}\inst{\ref{aff56}}
\and L.~Conversi\orcidlink{0000-0002-6710-8476}\inst{\ref{aff57},\ref{aff33}}
\and Y.~Copin\orcidlink{0000-0002-5317-7518}\inst{\ref{aff58}}
\and F.~Courbin\orcidlink{0000-0003-0758-6510}\inst{\ref{aff24}}
\and H.~M.~Courtois\orcidlink{0000-0003-0509-1776}\inst{\ref{aff59}}
\and M.~Cropper\orcidlink{0000-0003-4571-9468}\inst{\ref{aff49}}
\and A.~Da~Silva\orcidlink{0000-0002-6385-1609}\inst{\ref{aff60},\ref{aff61}}
\and H.~Degaudenzi\orcidlink{0000-0002-5887-6799}\inst{\ref{aff62}}
\and J.~Dinis\orcidlink{0000-0001-5075-1601}\inst{\ref{aff60},\ref{aff61}}
\and C.~A.~J.~Duncan\inst{\ref{aff56},\ref{aff63}}
\and X.~Dupac\inst{\ref{aff33}}
\and S.~Dusini\orcidlink{0000-0002-1128-0664}\inst{\ref{aff64}}
\and M.~Farina\orcidlink{0000-0002-3089-7846}\inst{\ref{aff65}}
\and S.~Farrens\orcidlink{0000-0002-9594-9387}\inst{\ref{aff6}}
\and S.~Ferriol\inst{\ref{aff58}}
\and P.~Fosalba\orcidlink{0000-0002-1510-5214}\inst{\ref{aff66},\ref{aff67}}
\and M.~Frailis\orcidlink{0000-0002-7400-2135}\inst{\ref{aff68}}
\and E.~Franceschi\orcidlink{0000-0002-0585-6591}\inst{\ref{aff7}}
\and M.~Fumana\orcidlink{0000-0001-6787-5950}\inst{\ref{aff50}}
\and S.~Galeotta\orcidlink{0000-0002-3748-5115}\inst{\ref{aff68}}
\and B.~Garilli\orcidlink{0000-0001-7455-8750}\inst{\ref{aff50}}
\and W.~Gillard\orcidlink{0000-0003-4744-9748}\inst{\ref{aff69}}
\and B.~Gillis\orcidlink{0000-0002-4478-1270}\inst{\ref{aff55}}
\and C.~Giocoli\orcidlink{0000-0002-9590-7961}\inst{\ref{aff7},\ref{aff70}}
\and P.~G\'omez-Alvarez\orcidlink{0000-0002-8594-5358}\inst{\ref{aff71},\ref{aff33}}
\and B.~R.~Granett\orcidlink{0000-0003-2694-9284}\inst{\ref{aff34}}
\and A.~Grazian\orcidlink{0000-0002-5688-0663}\inst{\ref{aff37}}
\and F.~Grupp\inst{\ref{aff1},\ref{aff38}}
\and L.~Guzzo\orcidlink{0000-0001-8264-5192}\inst{\ref{aff72},\ref{aff34}}
\and S.~V.~H.~Haugan\orcidlink{0000-0001-9648-7260}\inst{\ref{aff73}}
\and J.~Hoar\inst{\ref{aff33}}
\and H.~Hoekstra\orcidlink{0000-0002-0641-3231}\inst{\ref{aff74}}
\and W.~Holmes\inst{\ref{aff75}}
\and I.~Hook\orcidlink{0000-0002-2960-978X}\inst{\ref{aff76}}
\and F.~Hormuth\inst{\ref{aff77}}
\and A.~Hornstrup\orcidlink{0000-0002-3363-0936}\inst{\ref{aff78},\ref{aff79}}
\and P.~Hudelot\inst{\ref{aff80}}
\and K.~Jahnke\orcidlink{0000-0003-3804-2137}\inst{\ref{aff11}}
\and E.~Keih\"anen\orcidlink{0000-0003-1804-7715}\inst{\ref{aff81}}
\and S.~Kermiche\orcidlink{0000-0002-0302-5735}\inst{\ref{aff69}}
\and A.~Kiessling\orcidlink{0000-0002-2590-1273}\inst{\ref{aff75}}
\and T.~Kitching\orcidlink{0000-0002-4061-4598}\inst{\ref{aff49}}
\and R.~Kohley\inst{\ref{aff33}}
\and B.~Kubik\orcidlink{0009-0006-5823-4880}\inst{\ref{aff58}}
\and M.~K\"ummel\orcidlink{0000-0003-2791-2117}\inst{\ref{aff38}}
\and M.~Kunz\orcidlink{0000-0002-3052-7394}\inst{\ref{aff82}}
\and H.~Kurki-Suonio\orcidlink{0000-0002-4618-3063}\inst{\ref{aff83},\ref{aff84}}
\and O.~Lahav\orcidlink{0000-0002-1134-9035}\inst{\ref{aff85}}
\and S.~Ligori\orcidlink{0000-0003-4172-4606}\inst{\ref{aff39}}
\and P.~B.~Lilje\orcidlink{0000-0003-4324-7794}\inst{\ref{aff73}}
\and V.~Lindholm\orcidlink{0000-0003-2317-5471}\inst{\ref{aff83},\ref{aff84}}
\and I.~Lloro\inst{\ref{aff86}}
\and E.~Maiorano\orcidlink{0000-0003-2593-4355}\inst{\ref{aff7}}
\and O.~Mansutti\orcidlink{0000-0001-5758-4658}\inst{\ref{aff68}}
\and O.~Marggraf\orcidlink{0000-0001-7242-3852}\inst{\ref{aff87}}
\and K.~Markovic\orcidlink{0000-0001-6764-073X}\inst{\ref{aff75}}
\and N.~Martinet\orcidlink{0000-0003-2786-7790}\inst{\ref{aff13}}
\and F.~Marulli\orcidlink{0000-0002-8850-0303}\inst{\ref{aff88},\ref{aff7},\ref{aff36}}
\and R.~Massey\orcidlink{0000-0002-6085-3780}\inst{\ref{aff89}}
\and S.~Maurogordato\inst{\ref{aff32}}
\and H.~J.~McCracken\orcidlink{0000-0002-9489-7765}\inst{\ref{aff80}}
\and E.~Medinaceli\orcidlink{0000-0002-4040-7783}\inst{\ref{aff7}}
\and S.~Mei\orcidlink{0000-0002-2849-559X}\inst{\ref{aff90}}
\and M.~Melchior\inst{\ref{aff91}}
\and Y.~Mellier\inst{\ref{aff29},\ref{aff80}}
\and M.~Meneghetti\orcidlink{0000-0003-1225-7084}\inst{\ref{aff7},\ref{aff36}}
\and E.~Merlin\orcidlink{0000-0001-6870-8900}\inst{\ref{aff27}}
\and G.~Meylan\inst{\ref{aff24}}
\and M.~Moresco\orcidlink{0000-0002-7616-7136}\inst{\ref{aff88},\ref{aff7}}
\and L.~Moscardini\orcidlink{0000-0002-3473-6716}\inst{\ref{aff88},\ref{aff7},\ref{aff36}}
\and E.~Munari\orcidlink{0000-0002-1751-5946}\inst{\ref{aff68},\ref{aff92}}
\and R.~C.~Nichol\inst{\ref{aff93}}
\and S.-M.~Niemi\inst{\ref{aff9}}
\and J.~W.~Nightingale\orcidlink{0000-0002-8987-7401}\inst{\ref{aff94},\ref{aff95}}
\and C.~Padilla\orcidlink{0000-0001-7951-0166}\inst{\ref{aff96}}
\and S.~Paltani\orcidlink{0000-0002-8108-9179}\inst{\ref{aff62}}
\and F.~Pasian\orcidlink{0000-0002-4869-3227}\inst{\ref{aff68}}
\and K.~Pedersen\inst{\ref{aff97}}
\and W.~J.~Percival\orcidlink{0000-0002-0644-5727}\inst{\ref{aff22},\ref{aff23},\ref{aff98}}
\and V.~Pettorino\inst{\ref{aff9}}
\and S.~Pires\orcidlink{0000-0002-0249-2104}\inst{\ref{aff6}}
\and G.~Polenta\orcidlink{0000-0003-4067-9196}\inst{\ref{aff99}}
\and M.~Poncet\inst{\ref{aff100}}
\and L.~A.~Popa\inst{\ref{aff101}}
\and L.~Pozzetti\orcidlink{0000-0001-7085-0412}\inst{\ref{aff7}}
\and G.~D.~Racca\inst{\ref{aff9}}
\and F.~Raison\orcidlink{0000-0002-7819-6918}\inst{\ref{aff1}}
\and R.~Rebolo\inst{\ref{aff3},\ref{aff4}}
\and A.~Renzi\orcidlink{0000-0001-9856-1970}\inst{\ref{aff102},\ref{aff64}}
\and J.~Rhodes\orcidlink{0000-0002-4485-8549}\inst{\ref{aff75}}
\and G.~Riccio\inst{\ref{aff43}}
\and H.-W.~Rix\orcidlink{0000-0003-4996-9069}\inst{\ref{aff11}}
\and E.~Romelli\orcidlink{0000-0003-3069-9222}\inst{\ref{aff68}}
\and M.~Roncarelli\orcidlink{0000-0001-9587-7822}\inst{\ref{aff7}}
\and E.~Rossetti\orcidlink{0000-0003-0238-4047}\inst{\ref{aff35}}
\and R.~Saglia\orcidlink{0000-0003-0378-7032}\inst{\ref{aff38},\ref{aff1}}
\and D.~Sapone\orcidlink{0000-0001-7089-4503}\inst{\ref{aff103}}
\and B.~Sartoris\orcidlink{0000-0003-1337-5269}\inst{\ref{aff38},\ref{aff68}}
\and M.~Sauvage\orcidlink{0000-0002-0809-2574}\inst{\ref{aff6}}
\and R.~Scaramella\orcidlink{0000-0003-2229-193X}\inst{\ref{aff27},\ref{aff104}}
\and P.~Schneider\orcidlink{0000-0001-8561-2679}\inst{\ref{aff87}}
\and T.~Schrabback\orcidlink{0000-0002-6987-7834}\inst{\ref{aff19}}
\and A.~Secroun\orcidlink{0000-0003-0505-3710}\inst{\ref{aff69}}
\and G.~Seidel\orcidlink{0000-0003-2907-353X}\inst{\ref{aff11}}
\and M.~Seiffert\orcidlink{0000-0002-7536-9393}\inst{\ref{aff75}}
\and S.~Serrano\orcidlink{0000-0002-0211-2861}\inst{\ref{aff66},\ref{aff105},\ref{aff106}}
\and C.~Sirignano\orcidlink{0000-0002-0995-7146}\inst{\ref{aff102},\ref{aff64}}
\and G.~Sirri\orcidlink{0000-0003-2626-2853}\inst{\ref{aff36}}
\and J.~Skottfelt\orcidlink{0000-0003-1310-8283}\inst{\ref{aff107}}
\and L.~Stanco\orcidlink{0000-0002-9706-5104}\inst{\ref{aff64}}
\and P.~Tallada-Cresp\'{i}\orcidlink{0000-0002-1336-8328}\inst{\ref{aff51},\ref{aff52}}
\and A.~N.~Taylor\inst{\ref{aff55}}
\and H.~I.~Teplitz\orcidlink{0000-0002-7064-5424}\inst{\ref{aff108}}
\and I.~Tereno\inst{\ref{aff60},\ref{aff26}}
\and R.~Toledo-Moreo\orcidlink{0000-0002-2997-4859}\inst{\ref{aff109}}
\and F.~Torradeflot\orcidlink{0000-0003-1160-1517}\inst{\ref{aff52},\ref{aff51}}
\and I.~Tutusaus\orcidlink{0000-0002-3199-0399}\inst{\ref{aff110}}
\and E.~A.~Valentijn\inst{\ref{aff10}}
\and L.~Valenziano\orcidlink{0000-0002-1170-0104}\inst{\ref{aff7},\ref{aff111}}
\and T.~Vassallo\orcidlink{0000-0001-6512-6358}\inst{\ref{aff38},\ref{aff68}}
\and G.~Verdoes~Kleijn\orcidlink{0000-0001-5803-2580}\inst{\ref{aff10}}
\and A.~Veropalumbo\orcidlink{0000-0003-2387-1194}\inst{\ref{aff34},\ref{aff41},\ref{aff112}}
\and Y.~Wang\orcidlink{0000-0002-4749-2984}\inst{\ref{aff108}}
\and J.~Weller\orcidlink{0000-0002-8282-2010}\inst{\ref{aff38},\ref{aff1}}
\and O.~R.~Williams\orcidlink{0000-0003-0274-1526}\inst{\ref{aff113}}
\and G.~Zamorani\orcidlink{0000-0002-2318-301X}\inst{\ref{aff7}}
\and E.~Zucca\orcidlink{0000-0002-5845-8132}\inst{\ref{aff7}}
\and A.~Biviano\orcidlink{0000-0002-0857-0732}\inst{\ref{aff68},\ref{aff92}}
\and C.~Burigana\orcidlink{0000-0002-3005-5796}\inst{\ref{aff114},\ref{aff111}}
\and G.~De~Lucia\orcidlink{0000-0002-6220-9104}\inst{\ref{aff68}}
\and K.~George\orcidlink{0000-0002-1734-8455}\inst{\ref{aff38}}
\and V.~Scottez\inst{\ref{aff29},\ref{aff115}}
\and P.~Simon\inst{\ref{aff87}}
\and A.~Mora\orcidlink{0000-0002-1922-8529}\inst{\ref{aff116}}
\and J.~Mart\'{i}n-Fleitas\orcidlink{0000-0002-8594-569X}\inst{\ref{aff116}}
\and F.~Ruppin\orcidlink{0000-0002-0955-8954}\inst{\ref{aff58}}
\and D.~Scott\orcidlink{0000-0002-6878-9840}\inst{\ref{aff117}}}
										   
\institute{Max Planck Institute for Extraterrestrial Physics, Giessenbachstr. 1, 85748 Garching, Germany\label{aff1}
\and
School of Physics and Astronomy, University of Nottingham, University Park, Nottingham NG7 2RD, UK\label{aff2}
\and
Instituto de Astrof\'isica de Canarias, Calle V\'ia L\'actea s/n, 38204, San Crist\'obal de La Laguna, Tenerife, Spain\label{aff3}
\and
Departamento de Astrof\'isica, Universidad de La Laguna, 38206, La Laguna, Tenerife, Spain\label{aff4}
\and
Department of Astronomy, University of Florida, Bryant Space Science Center, Gainesville, FL 32611, USA\label{aff5}
\and
Universit\'e Paris-Saclay, Universit\'e Paris Cit\'e, CEA, CNRS, AIM, 91191, Gif-sur-Yvette, France\label{aff6}
\and
INAF-Osservatorio di Astrofisica e Scienza dello Spazio di Bologna, Via Piero Gobetti 93/3, 40129 Bologna, Italy\label{aff7}
\and
Observatoire Astronomique de Strasbourg (ObAS), Universit\'e de Strasbourg - CNRS, UMR 7550, Strasbourg, France\label{aff8}
\and
European Space Agency/ESTEC, Keplerlaan 1, 2201 AZ Noordwijk, The Netherlands\label{aff9}
\and
Kapteyn Astronomical Institute, University of Groningen, PO Box 800, 9700 AV Groningen, The Netherlands\label{aff10}
\and
Max-Planck-Institut f\"ur Astronomie, K\"onigstuhl 17, 69117 Heidelberg, Germany\label{aff11}
\and
Department of Physics, Universit\'{e} de Montr\'{e}al, 2900 Edouard Montpetit Blvd, Montr\'{e}al, Qu\'{e}bec H3T 1J4, Canada\label{aff12}
\and
Aix-Marseille Universit\'e, CNRS, CNES, LAM, Marseille, France\label{aff13}
\and
INAF - Osservatorio Astronomico di Cagliari, Via della Scienza 5, 09047 Selargius (CA), Italy\label{aff14}
\and
INAF - Osservatorio Astronomico d'Abruzzo, Via Maggini, 64100, Teramo, Italy\label{aff15}
\and
Univ. Lille, CNRS, Centrale Lille, UMR 9189 CRIStAL, 59000 Lille, France\label{aff16}
\and
Universit\'e Paris-Saclay, CNRS, Institut d'astrophysique spatiale, 91405, Orsay, France\label{aff17}
\and
Leibniz-Institut f\"{u}r Astrophysik (AIP), An der Sternwarte 16, 14482 Potsdam, Germany\label{aff18}
\and
Universit\"at Innsbruck, Institut f\"ur Astro- und Teilchenphysik, Technikerstr. 25/8, 6020 Innsbruck, Austria\label{aff19}
\and
Universit\'e de Strasbourg, CNRS, Observatoire astronomique de Strasbourg, UMR 7550, 67000 Strasbourg, France\label{aff20}
\and
Institute of Astronomy, University of Cambridge, Madingley Road, Cambridge CB3 0HA, UK\label{aff21}
\and
Waterloo Centre for Astrophysics, University of Waterloo, Waterloo, Ontario N2L 3G1, Canada\label{aff22}
\and
Department of Physics and Astronomy, University of Waterloo, Waterloo, Ontario N2L 3G1, Canada\label{aff23}
\and
Institute of Physics, Laboratory of Astrophysics, Ecole Polytechnique F\'ed\'erale de Lausanne (EPFL), Observatoire de Sauverny, 1290 Versoix, Switzerland\label{aff24}
\and
Departamento de F\'{i}sica Te\'{o}rica, At\'{o}mica y \'{O}ptica, Universidad de Valladolid, 47011 Valladolid, Spain\label{aff25}
\and
Instituto de Astrof\'isica e Ci\^encias do Espa\c{c}o, Faculdade de Ci\^encias, Universidade de Lisboa, Tapada da Ajuda, 1349-018 Lisboa, Portugal\label{aff26}
\and
INAF-Osservatorio Astronomico di Roma, Via Frascati 33, 00078 Monteporzio Catone, Italy\label{aff27}
\and
Observatorio Nacional, Rua General Jose Cristino, 77-Bairro Imperial de Sao Cristovao, Rio de Janeiro, 20921-400, Brazil\label{aff28}
\and
Institut d'Astrophysique de Paris, 98bis Boulevard Arago, 75014, Paris, France\label{aff29}
\and
OCA, P.H.C Boulevard de l'Observatoire CS 34229, 06304 Nice Cedex 4, France\label{aff30}
\and
Instituto de Astrof\'isica de Andaluc\'ia, CSIC, Glorieta de la Astronom\'\i a, 18080, Granada, Spain\label{aff31}
\and
Universit\'e C\^{o}te d'Azur, Observatoire de la C\^{o}te d'Azur, CNRS, Laboratoire Lagrange, Bd de l'Observatoire, CS 34229, 06304 Nice cedex 4, France\label{aff32}
\and
ESAC/ESA, Camino Bajo del Castillo, s/n., Urb. Villafranca del Castillo, 28692 Villanueva de la Ca\~nada, Madrid, Spain\label{aff33}
\and
INAF-Osservatorio Astronomico di Brera, Via Brera 28, 20122 Milano, Italy\label{aff34}
\and
Dipartimento di Fisica e Astronomia, Universit\`a di Bologna, Via Gobetti 93/2, 40129 Bologna, Italy\label{aff35}
\and
INFN-Sezione di Bologna, Viale Berti Pichat 6/2, 40127 Bologna, Italy\label{aff36}
\and
INAF-Osservatorio Astronomico di Padova, Via dell'Osservatorio 5, 35122 Padova, Italy\label{aff37}
\and
Universit\"ats-Sternwarte M\"unchen, Fakult\"at f\"ur Physik, Ludwig-Maximilians-Universit\"at M\"unchen, Scheinerstrasse 1, 81679 M\"unchen, Germany\label{aff38}
\and
INAF-Osservatorio Astrofisico di Torino, Via Osservatorio 20, 10025 Pino Torinese (TO), Italy\label{aff39}
\and
Dipartimento di Fisica, Universit\`a di Genova, Via Dodecaneso 33, 16146, Genova, Italy\label{aff40}
\and
INFN-Sezione di Genova, Via Dodecaneso 33, 16146, Genova, Italy\label{aff41}
\and
Department of Physics "E. Pancini", University Federico II, Via Cinthia 6, 80126, Napoli, Italy\label{aff42}
\and
INAF-Osservatorio Astronomico di Capodimonte, Via Moiariello 16, 80131 Napoli, Italy\label{aff43}
\and
INFN section of Naples, Via Cinthia 6, 80126, Napoli, Italy\label{aff44}
\and
Instituto de Astrof\'isica e Ci\^encias do Espa\c{c}o, Universidade do Porto, CAUP, Rua das Estrelas, PT4150-762 Porto, Portugal\label{aff45}
\and
Faculdade de Ci\^encias da Universidade do Porto, Rua do Campo de Alegre, 4150-007 Porto, Portugal\label{aff46}
\and
Dipartimento di Fisica, Universit\`a degli Studi di Torino, Via P. Giuria 1, 10125 Torino, Italy\label{aff47}
\and
INFN-Sezione di Torino, Via P. Giuria 1, 10125 Torino, Italy\label{aff48}
\and
Mullard Space Science Laboratory, University College London, Holmbury St Mary, Dorking, Surrey RH5 6NT, UK\label{aff49}
\and
INAF-IASF Milano, Via Alfonso Corti 12, 20133 Milano, Italy\label{aff50}
\and
Centro de Investigaciones Energ\'eticas, Medioambientales y Tecnol\'ogicas (CIEMAT), Avenida Complutense 40, 28040 Madrid, Spain\label{aff51}
\and
Port d'Informaci\'{o} Cient\'{i}fica, Campus UAB, C. Albareda s/n, 08193 Bellaterra (Barcelona), Spain\label{aff52}
\and
Institute for Theoretical Particle Physics and Cosmology (TTK), RWTH Aachen University, 52056 Aachen, Germany\label{aff53}
\and
Dipartimento di Fisica e Astronomia "Augusto Righi" - Alma Mater Studiorum Universit\`a di Bologna, Viale Berti Pichat 6/2, 40127 Bologna, Italy\label{aff54}
\and
Institute for Astronomy, University of Edinburgh, Royal Observatory, Blackford Hill, Edinburgh EH9 3HJ, UK\label{aff55}
\and
Jodrell Bank Centre for Astrophysics, Department of Physics and Astronomy, University of Manchester, Oxford Road, Manchester M13 9PL, UK\label{aff56}
\and
European Space Agency/ESRIN, Largo Galileo Galilei 1, 00044 Frascati, Roma, Italy\label{aff57}
\and
Universit\'e Claude Bernard Lyon 1, CNRS/IN2P3, IP2I Lyon, UMR 5822, Villeurbanne, F-69100, France\label{aff58}
\and
UCB Lyon 1, CNRS/IN2P3, IUF, IP2I Lyon, 4 rue Enrico Fermi, 69622 Villeurbanne, France\label{aff59}
\and
Departamento de F\'isica, Faculdade de Ci\^encias, Universidade de Lisboa, Edif\'icio C8, Campo Grande, PT1749-016 Lisboa, Portugal\label{aff60}
\and
Instituto de Astrof\'isica e Ci\^encias do Espa\c{c}o, Faculdade de Ci\^encias, Universidade de Lisboa, Campo Grande, 1749-016 Lisboa, Portugal\label{aff61}
\and
Department of Astronomy, University of Geneva, ch. d'Ecogia 16, 1290 Versoix, Switzerland\label{aff62}
\and
Department of Physics, Oxford University, Keble Road, Oxford OX1 3RH, UK\label{aff63}
\and
INFN-Padova, Via Marzolo 8, 35131 Padova, Italy\label{aff64}
\and
INAF-Istituto di Astrofisica e Planetologia Spaziali, via del Fosso del Cavaliere, 100, 00100 Roma, Italy\label{aff65}
\and
Institut d'Estudis Espacials de Catalunya (IEEC),  Edifici RDIT, Campus UPC, 08860 Castelldefels, Barcelona, Spain\label{aff66}
\and
Institut de Ciencies de l'Espai (IEEC-CSIC), Campus UAB, Carrer de Can Magrans, s/n Cerdanyola del Vall\'es, 08193 Barcelona, Spain\label{aff67}
\and
INAF-Osservatorio Astronomico di Trieste, Via G. B. Tiepolo 11, 34143 Trieste, Italy\label{aff68}
\and
Aix-Marseille Universit\'e, CNRS/IN2P3, CPPM, Marseille, France\label{aff69}
\and
Istituto Nazionale di Fisica Nucleare, Sezione di Bologna, Via Irnerio 46, 40126 Bologna, Italy\label{aff70}
\and
FRACTAL S.L.N.E., calle Tulip\'an 2, Portal 13 1A, 28231, Las Rozas de Madrid, Spain\label{aff71}
\and
Dipartimento di Fisica "Aldo Pontremoli", Universit\`a degli Studi di Milano, Via Celoria 16, 20133 Milano, Italy\label{aff72}
\and
Institute of Theoretical Astrophysics, University of Oslo, P.O. Box 1029 Blindern, 0315 Oslo, Norway\label{aff73}
\and
Leiden Observatory, Leiden University, Einsteinweg 55, 2333 CC Leiden, The Netherlands\label{aff74}
\and
Jet Propulsion Laboratory, California Institute of Technology, 4800 Oak Grove Drive, Pasadena, CA, 91109, USA\label{aff75}
\and
Department of Physics, Lancaster University, Lancaster, LA1 4YB, UK\label{aff76}
\and
Felix Hormuth Engineering, Goethestr. 17, 69181 Leimen, Germany\label{aff77}
\and
Technical University of Denmark, Elektrovej 327, 2800 Kgs. Lyngby, Denmark\label{aff78}
\and
Cosmic Dawn Center (DAWN), Denmark\label{aff79}
\and
Institut d'Astrophysique de Paris, UMR 7095, CNRS, and Sorbonne Universit\'e, 98 bis boulevard Arago, 75014 Paris, France\label{aff80}
\and
Department of Physics and Helsinki Institute of Physics, Gustaf H\"allstr\"omin katu 2, 00014 University of Helsinki, Finland\label{aff81}
\and
Universit\'e de Gen\`eve, D\'epartement de Physique Th\'eorique and Centre for Astroparticle Physics, 24 quai Ernest-Ansermet, CH-1211 Gen\`eve 4, Switzerland\label{aff82}
\and
Department of Physics, P.O. Box 64, 00014 University of Helsinki, Finland\label{aff83}
\and
Helsinki Institute of Physics, Gustaf H{\"a}llstr{\"o}min katu 2, University of Helsinki, Helsinki, Finland\label{aff84}
\and
Department of Physics and Astronomy, University College London, Gower Street, London WC1E 6BT, UK\label{aff85}
\and
NOVA optical infrared instrumentation group at ASTRON, Oude Hoogeveensedijk 4, 7991PD, Dwingeloo, The Netherlands\label{aff86}
\and
Universit\"at Bonn, Argelander-Institut f\"ur Astronomie, Auf dem H\"ugel 71, 53121 Bonn, Germany\label{aff87}
\and
Dipartimento di Fisica e Astronomia "Augusto Righi" - Alma Mater Studiorum Universit\`a di Bologna, via Piero Gobetti 93/2, 40129 Bologna, Italy\label{aff88}
\and
Department of Physics, Centre for Extragalactic Astronomy, Durham University, South Road, DH1 3LE, UK\label{aff89}
\and
Universit\'e Paris Cit\'e, CNRS, Astroparticule et Cosmologie, 75013 Paris, France\label{aff90}
\and
University of Applied Sciences and Arts of Northwestern Switzerland, School of Engineering, 5210 Windisch, Switzerland\label{aff91}
\and
IFPU, Institute for Fundamental Physics of the Universe, via Beirut 2, 34151 Trieste, Italy\label{aff92}
\and
School of Mathematics and Physics, University of Surrey, Guildford, Surrey, GU2 7XH, UK\label{aff93}
\and
School of Mathematics, Statistics and Physics, Newcastle University, Herschel Building, Newcastle-upon-Tyne, NE1 7RU, UK\label{aff94}
\and
Department of Physics, Institute for Computational Cosmology, Durham University, South Road, DH1 3LE, UK\label{aff95}
\and
Institut de F\'{i}sica d'Altes Energies (IFAE), The Barcelona Institute of Science and Technology, Campus UAB, 08193 Bellaterra (Barcelona), Spain\label{aff96}
\and
Department of Physics and Astronomy, University of Aarhus, Ny Munkegade 120, DK-8000 Aarhus C, Denmark\label{aff97}
\and
Perimeter Institute for Theoretical Physics, Waterloo, Ontario N2L 2Y5, Canada\label{aff98}
\and
Space Science Data Center, Italian Space Agency, via del Politecnico snc, 00133 Roma, Italy\label{aff99}
\and
Centre National d'Etudes Spatiales -- Centre spatial de Toulouse, 18 avenue Edouard Belin, 31401 Toulouse Cedex 9, France\label{aff100}
\and
Institute of Space Science, Str. Atomistilor, nr. 409 M\u{a}gurele, Ilfov, 077125, Romania\label{aff101}
\and
Dipartimento di Fisica e Astronomia "G. Galilei", Universit\`a di Padova, Via Marzolo 8, 35131 Padova, Italy\label{aff102}
\and
Departamento de F\'isica, FCFM, Universidad de Chile, Blanco Encalada 2008, Santiago, Chile\label{aff103}
\and
INFN-Sezione di Roma, Piazzale Aldo Moro, 2 - c/o Dipartimento di Fisica, Edificio G. Marconi, 00185 Roma, Italy\label{aff104}
\and
Satlantis, University Science Park, Sede Bld 48940, Leioa-Bilbao, Spain\label{aff105}
\and
Institute of Space Sciences (ICE, CSIC), Campus UAB, Carrer de Can Magrans, s/n, 08193 Barcelona, Spain\label{aff106}
\and
Centre for Electronic Imaging, Open University, Walton Hall, Milton Keynes, MK7~6AA, UK\label{aff107}
\and
Infrared Processing and Analysis Center, California Institute of Technology, Pasadena, CA 91125, USA\label{aff108}
\and
Universidad Polit\'ecnica de Cartagena, Departamento de Electr\'onica y Tecnolog\'ia de Computadoras,  Plaza del Hospital 1, 30202 Cartagena, Spain\label{aff109}
\and
Institut de Recherche en Astrophysique et Plan\'etologie (IRAP), Universit\'e de Toulouse, CNRS, UPS, CNES, 14 Av. Edouard Belin, 31400 Toulouse, France\label{aff110}
\and
INFN-Bologna, Via Irnerio 46, 40126 Bologna, Italy\label{aff111}
\and
Dipartimento di Fisica, Universit\`a degli studi di Genova, and INFN-Sezione di Genova, via Dodecaneso 33, 16146, Genova, Italy\label{aff112}
\and
Centre for Information Technology, University of Groningen, P.O. Box 11044, 9700 CA Groningen, The Netherlands\label{aff113}
\and
INAF, Istituto di Radioastronomia, Via Piero Gobetti 101, 40129 Bologna, Italy\label{aff114}
\and
Junia, EPA department, 41 Bd Vauban, 59800 Lille, France\label{aff115}
\and
Aurora Technology for European Space Agency (ESA), Camino bajo del Castillo, s/n, Urbanizacion Villafranca del Castillo, Villanueva de la Ca\~nada, 28692 Madrid, Spain\label{aff116}
\and
Department of Physics and Astronomy, University of British Columbia, Vancouver, BC V6T 1Z1, Canada\label{aff117}}

\abstract{

We study the intracluster light (ICL) and intracluster globular clusters (ICGCs) in the nearby Perseus cluster of galaxies using \Euclid's Early Release Observations. By modelling the isophotal and iso-density contours, we mapped the distributions and properties of the ICL and ICGCs out to radii of $200$--$600$\,kpc (up to $\sim\frac{1}{3}$ of the virial radius, depending on the parameter) from the brightest cluster galaxy (BCG). 
We find that the central 500\,kpc of the Perseus cluster hosts $70\,000\pm2800$ globular clusters, and $1.7\times10^{12}$\,L$_{\odot}$ of diffuse light from the BCG+ICL in the near-infrared \HE. This accounts for $38\pm6$\% of the cluster's total stellar luminosity within this radius.
The ICL and ICGCs share a coherent spatial distribution which suggests that they have a common origin or that a common potential governs their distribution. Their contours on the largest scales ($>200$\,kpc) are not centred on the BCG's core, but are instead offset westwards by 60\,kpc towards several luminous cluster galaxies. This offset is opposite to the displacement observed in the gaseous intracluster medium. 
The radial surface brightness profile of the BCG+ICL is best described by a double S\'ersic model, with $68\pm4$\% of the \HE\ light contained in the extended, outer component. 
The transition between these components occurs at $\approx$60\,kpc, beyond which the isophotes become increasingly elliptical and off-centred. Furthermore, the radial ICGC number density profile closely follows the profile of the BCG+ICL only beyond this 60\,kpc radius, where we find an average of 60--80 globular clusters per $10^9$\,M$_\odot$ of diffuse stellar mass. The BCG+ICL colour becomes increasingly blue with radius, consistent with the stellar populations in the ICL having subsolar metallicities [Fe/H] $\sim-0.6$ to $-1.0$.
The colour of the ICL, and the specific frequency and luminosity function of the ICGCs suggest that the ICL+ICGCs were tidally stripped from the outskirts of massive satellites with masses of a few $\times10^{10}$\,M$_\odot$, with an increasing contribution from dwarf galaxies at large radii.

}

%
%
%
%
   \titlerunning{\Euclid: ERO -- The intracluster stars of Perseus}
   \authorrunning{M.~Kluge et al.}
   \maketitle
%
%
%
%
\begin{figure*}[t]
    \centering
    \includegraphics[width=\linewidth]{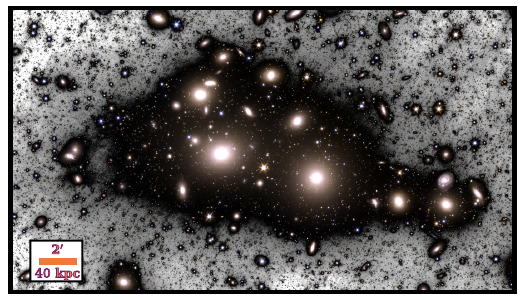}
    \caption{Region of 26\arcmin$\times$15\arcmin\ around the centre of the Perseus cluster of galaxies. The figure is a composite of an RGB image using the \IE, \JE, and \HE\ bands, respectively, and a combined $\IE+\JE+\HE$ image used for the inverted black and white background. North is up, east is left. The brightest two galaxies in this image are the BCG, NGC\,1275, to the left, and NGC\,1272 to the right.}  \label{fig:rgb}
\end{figure*}

\section{\label{sc:Intro}Introduction}

Over the last 20 years, observations have shown that the intracluster light (ICL) is a ubiquitous feature in galaxy clusters \citep[e.g.][]{Feldmeier2004, Kluge2020, golden-marx2023}. As the byproduct of interactions between galaxies in clusters \citep[e.g.][]{Gregg1998, Mihos2005}, the ICL is a fossil record of all the dynamical interactions the system has experienced and offers a holistic view of the cluster's history (see \citealt{Contini2021,Arnaboldi2022,Montes2022} for recent reviews). Thus, the origin and assembly history of the ICL is central to understanding the global evolution of the cluster galaxy population. 

In addition, the ICL has been shown to be a tool for inferring the radius of the cluster and even its dark matter distribution \citep{MT19, Alonso-Asensio2020, Deason2021, Gonzalez2021, Yoo2022, ContrerasSantos2024}. Despite being such a useful tool for understanding our Universe, our knowledge of this component is limited because the ICL is faint \citep[$\mu_V > 26.5$ mag\,arcsec$^{-2}$,][]{Rudick2006} and extended (hundreds of kpc), so we need deep and wide-field observations to study it.

The accretion events that form this diffuse light also bring large numbers of globular clusters (GCs) into the intracluster space, which we refer to as intracluster globular clusters (ICGCs) \citep[e.g.][]{Durrell2014, AlamoMartinez2017, Lee2022}. These luminous tracers are an additional clue to infer the past history of the cluster. Moreover, \citet{Reina-Campos2023} show that the GCs can also be used to trace the dark matter distribution in halos.

The \Euclid \citep{Laureijs11,EuclidSkyOverview} space mission will observe nearly one-third of the sky in four photometric bands: one in the visible (\IE) using the VIS instrument \citep{EuclidSkyVIS} and three in the near-infrared (NIR; \YE, \JE, \HE) using the NISP instrument \citep{EuclidSkyNISP}. The faint detection limit, the design of the wide-field telescope, and the tight control on the scattered light makes \Euclid\ ideal to study the low surface-brightness (LSB) Universe \citep{Scaramella-EP1,Borlaff-EP16}, in particular the ICL. 

In this work, we analyse the ICL and ICGCs of the Perseus cluster of galaxies (Abell 426). The images analysed here are part of the Early Release Observations (ERO) programme, a collection of observations dedicated to showcasing \Euclid's capabilities. Perseus is located close to the Galactic plane and therefore is awash with Galactic cirri, and suffers from high and spatially varying extinction. It is therefore a highly complex case study that we use to demonstrate the potential of \Euclid\ for LSB research, even in difficult conditions.

The Perseus galaxy cluster is one of the most spectacular nearby astronomical objects. It is a low-redshift ($z=0.0179$), massive, rich Bautz-Morgan class II-III galaxy cluster with a velocity dispersion of $1040^{+34}_{-43}$\,km\,s$^{-1}$ \citep{Aguerri2020} and is the brightest X-ray cluster in the sky in terms of flux \citep{Edge1992}. X-ray observations report that the virial radius of this cluster ($r_{200,{\rm c}}$) is 1.79\,Mpc \citep[82\arcmin\ on the sky,][]{Simionescu2011}, which encloses a total mass ($M_{200,{\rm c}}$) of $(6.65\pm0.5) \times10^{14}$\,M$_{\odot}$, making it one of the most massive nearby clusters. The mass and virial radius derived through optical spectroscopy are a factor of 1.8 and 1.2 larger than this, respectively \citep{Aguerri2020}. Although this cluster has been studied in many previous observations, \Euclid's field of view (0.57 deg$^2$), superb spatial resolution and low background means these observations are the first in-depth, high-resolution look at the Perseus cluster as a whole from the optical to the near-infrared, allowing for a detailed study of the intracluster stellar population. 

In the central regions of the cluster, the stellar population belonging to the ICL overlaps with the stellar population belonging to the brightest cluster galaxy (BCG). These two components can be separated kinematically \citep[][]{Dolag2010,Longobardi2013,Longobardi2015,Remus2017,Hartke2022}, but it is not possible to separate the BCG from the ICL using the \Euclid  photometry alone. Nevertheless, it is possible to define regions in space in which the majority of the stellar population belong to the ICL rather than the BCG. For simplicity, in this work we refer to the diffuse light beyond $100$\,kpc of the BCG as ICL. At the same time, the GCs beyond this radius are identified as ICGCs. This definition is supported by observations that find a break radius in the BCG+ICL light profile at $60$--80\,kpc \citep{Zibetti2005,gonzalez05, Iodice2016, zhang19, Montes2021}, and the stellar tracers across this boundary display different kinematics and metallicities \citep[e.g.][]{Longobardi2013, Longobardi2018, Hartke2022, Hartke2023}. By defining the ICL and ICGCs as existing beyond this radius, we make sure that the influence of the light and GCs from the BCG is minimal.

The luminosity distance of the Perseus cluster is $(72\pm3)$\,Mpc, or alternatively, $m-M=34.3\pm0.1$\,mag \citep{Tully2023}, which corresponds to an angular diameter distance of $0.338$\,kpc\,arcsec$^{-1}$. We assume a standard flat $\Lambda$\,CDM cosmology with $\Omega_{\rm m}=0.319$ and $H_0=67$\,km\,s$^{‐1}$Mpc$^{‐1}$ \citep{Planck_2018}, and all magnitudes are given using the AB magnitude system.

\section{Observations}
We used the ERO images \citep{EROcite} of the Perseus cluster taken by ESA's \Euclid\ satellite \citep{EuclidSkyOverview}. The observations were centred on the position RA~$= 3^{\rm h}\,18^{\rm m}\,40$, Dec~$ = 41{\degr}\,39{\arcmin}$\,00, rotated clockwise by 30\degr\ relative to north, and have a field of view of $\sim0.7$\,deg$^{2}$. The images were obtained in a dithered observation sequence that is similar to the Reference Observation Sequence (ROS) that will be used to observe the \Euclid\ Wide Survey \citep[EWS;][]{Scaramella-EP1}. In this sequence, an image in \IE\ is taken simultaneously with slitless grism spectra in the NIR, followed by NIR images taken in \JE\, then \HE, and finally \YE. The telescope is then dithered and the sequence is repeated three further times. Four ROS were combined to make the final Perseus images, whereas the EWS will only be at the depth of one ROS.

The resulting images of the Perseus cluster have a maximum exposure time of 9056\,s in \IE, 1395.2\,s in \YE, \JE, and \HE. The \IE\ images have a pixel scale of 0.1\,arcsec\,pix$^{-1}$, and a spatial resolution (FWHM) of \ang{;;0.16}. The NIR images have a pixel scale of 0.3\,arcsec\,pix$^{-1}$, and a spatial resolution of $\sim\ang{;;0.49}$. Further details on these images are provided in \citet{EROPerseusOverview}. The images were processed according to the method described in the accompanying article \citep{EROData}, which has been optimised to preserve the LSB emission in each image.
Figure \ref{fig:rgb} shows a \YE, \JE, \HE\ false-colour image centred on the two brightest galaxies in the Perseus cluster: the BCG, NGC\,1275, and its companion NGC\,1272. It illustrates that the ICL signal is preserved out to many arcminutes.

\begin{figure*}
    \centering
    \includegraphics[width=1.0\linewidth]{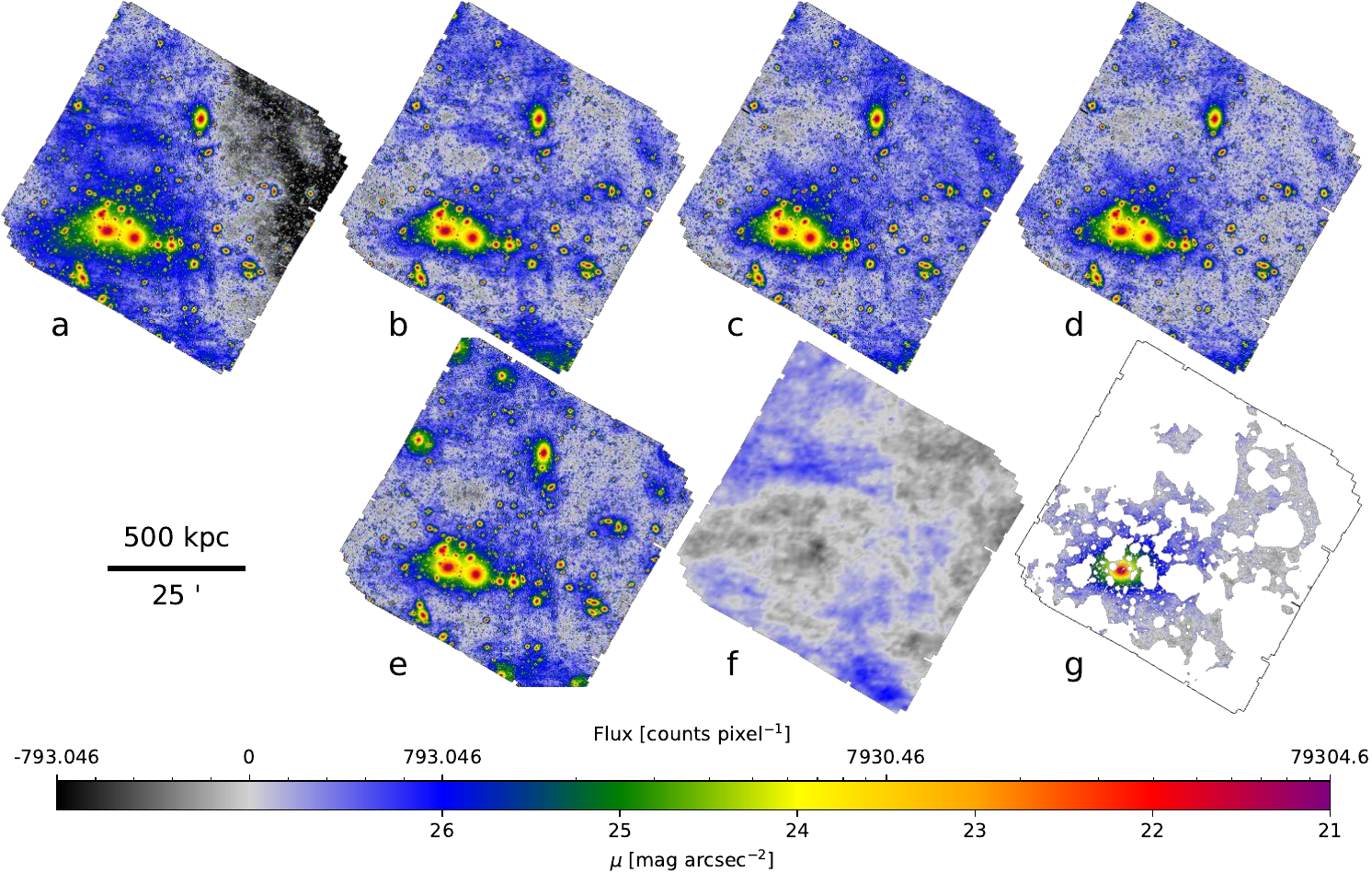}
    \caption{Steps taken to flatten the background in the \IE image. The panels show: (a) the original \IE\ image with an east-west gradient visible in the background, (b) the \IE\ image that has been flattened to match the background properties of the CFHT $i'$-band image (which is shown in panel e), (c) is the image after the subtraction of a scaled 12\micron\ WISE image (shown in panel f) to remove Galactic cirri and the PSF subtraction has been performed on 75 of the brightest stars in the field of view. Panel (d) shows the final processed \IE\ image after a 2D first-order polynomial gradient was subtracted from the background. Panel (g) shows only the area of the image that is left unmasked, after the light profile of NGC\,1272 has been subtracted. The scaling of the colour map transitions smoothly from logarithmic to linear steps at low fluxes in order to highlight the patterns in the background. All panels show the full field of view of the \IE\ image.}
    \label{fig:vis_steps}
\end{figure*}

\section{Data processing} \label{sec:dataprocessing}
The ERO were detrended following the steps outlined in \citet{EROData}, including: overscan correction and bias structure correction, flat fielding on median and large scales using the zodiacal light as a flat illumination source to produce images that appear background-flat, flat fielding on small scales using calibration flats, and detector to detector image scaling.
These processes ensure the continuity of the extended emission but limit the accuracy of the photometry to a few percent. 

The LSB emission in the NISP images was strongly impacted by persistence (see Fig. 11 in \citealt{EROData}), and accurate measurements of the ICL cannot be made without removing this strong signal. Since masking the persistence would lead to large holes in the stacked images, the persistence was modelled and subtracted in the individual exposures before the exposures were stacked. Details of this modelling are provided in \citet{EROData}. Following these processing steps, the $1\sigma$ depths of the LSB-optimised images are $\mu(\IE)=30.5\,\mathrm{mag}\,\mathrm{arcsec}^{-2}$ and $\mu(\YE,\JE,\HE)=28.7,\, 28.9,\, 28.9\,\mathrm{mag}\,\mathrm{arcsec}^{-2}$, measured over a $\ang{;;10}\times\ang{;;10}$ area, and expressed as \textrm{asinh} AB magnitudes.

The Perseus ERO was affected by low-level scattered light in VIS on the level of 3\%--4\% of the zodiacal light \citep{EuclidSkyOverview} which would impact the ICL measurements in \IE. Therefore, in addition to the data processing described in \citet{EROData}, we applied the following processing steps that are essential for accurate diffuse LSB photometry: 
\begin{itemize}
    \item{The modelling and removal of a large-scale gradient in the \IE\ image.}
    \item{The modelling and removal of Galactic cirri from the \IE\ image.}
    \item The modelling and subtraction of the large-scale point spread function (PSF) for the brightest stars in the field of view, which was applied to the \IE\ and NIR images.
\end{itemize}
These steps are visualised in Fig.\,\ref{fig:vis_steps} and explained in detail below.

\begin{figure*}
    \centering
    \includegraphics[width=\linewidth]{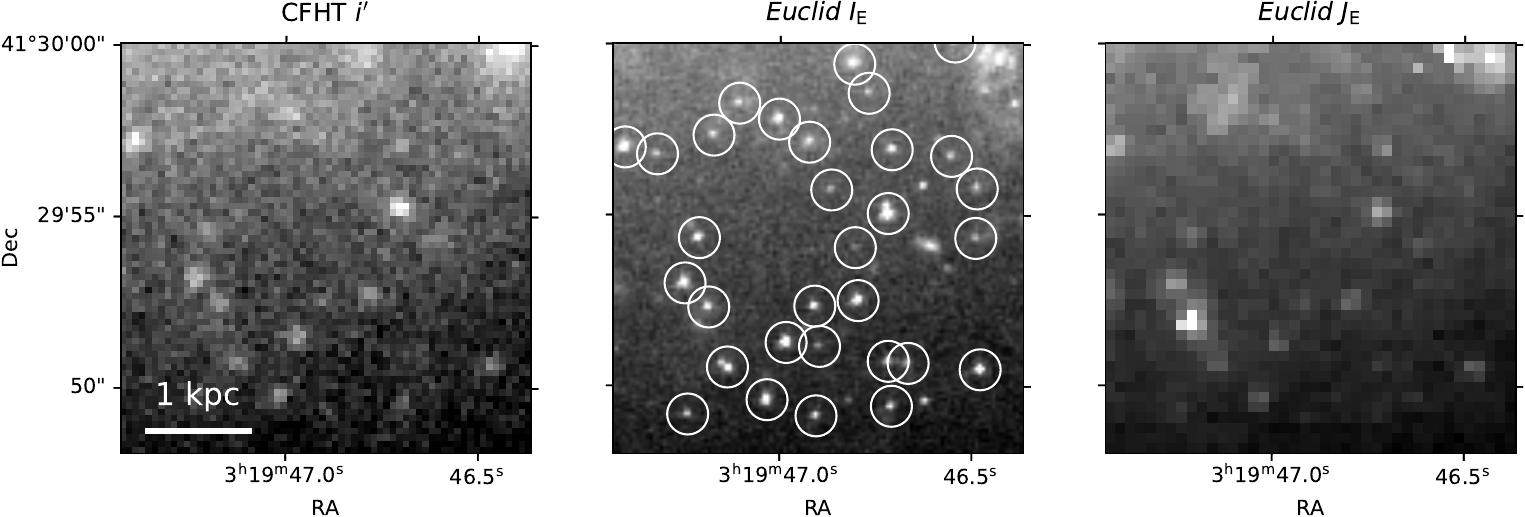}\\
    \caption{Zoomed-in view of the Perseus ERO focusing on a region that lies 17\,kpc southwest of NGC\,1275. The ground-based seeing-limited CFHT $i'$-band image (left panel) is comparable in wavelength to the \Euclid\ \IE-band image (middle panel). However, the superior spatial resolution in \IE\ reveals numerous GCs, which are encircled in white. \Euclid's resolution decreases with wavelength because of diffraction and undersampling of the PSF, such that most of the GCs cannot be identified in the \JE\ band (right panel). Black and white correspond to a surface brightness of $\mu=22$ and $21$\,mag\,arcsec$^{-2}$, respectively, in the left and middle panels, and $\mu(\JE)=23$ and $22$\,mag\,arcsec$^{-2}$ in the right panel.
    }
    \label{fig:cfht_euclid_zoom}
\end{figure*}

The initial \IE\ image had a large-scale east-west gradient with background inhomogeneities of the order of 26\,mag\,arcsec$^{-2}$ (see Fig.\,\ref{fig:vis_steps}a), which corresponds to $3\%-4\%$ of the background flux. Stray light on that level had been noticed in the VIS detector \citep{EuclidSkyOverview}. This strong gradient is neither apparent in the NIR images nor in a deep $i'$-band image of the Perseus cluster (see Fig.\,\ref{fig:vis_steps}e), that was observed with the Canada-France-Hawaii Telescope\footnote{Details of the CFHT images of Perseus, and how they were optimised for LSB emission, can be found in \citet{EROPerseusOverview}. The \IE\ image has advantages over the CFHT $i'$-band image for detecting GCs (see Fig.\,\ref{fig:cfht_euclid_zoom}) and the ICL due to its compact PSF and minimal PSF wings, significantly reducing PSF broadening effects on the surface brightness and colour profiles \citep{Duc2015,Trujillo2016,Kluge2020} and allowing us to explore the ICL in the critical region between NGC\,1275 and NGC\,1272 near a bright contaminating star (see Fig.\,\ref{fig:rgb}).} (CFHT). 
Regardless of its origin (whether astronomical or stray light in the telescope), this light gradient is a contaminant to the ICL, so we removed it.

Unfortunately, the gradient is more complex than can be described by a 2D first-order polynomial because it falls off steeper in the west. Fitting higher-order functions to the background can result in overfitting and removing part of the ICL. To avoid this, we matched the \IE\ background to that of the CFHT image.
This image is well suited for this task because its background is much flatter and it is sufficiently deep: the same structures in the ICL and the cirri are visible in both Figs.\,\ref{fig:vis_steps}a and \ref{fig:vis_steps}e (e.g. diffuse emission with $\mu(\IE)\sim$26\,mag\,arcsec$^{-2}$ in the north, west, and south image corners) and they are fainter than the artificial gradient observed in Fig.\,\ref{fig:vis_steps}a. 

First, the CFHT image was subtracted from the \Euclid\ \IE\ image. Subsequently, a 2D fourth-order polynomial was fit to the residuals. This polynomial was then subtracted from the original \IE\ image, producing the final image shown in Fig.\,\ref{fig:vis_steps}b.\footnote{The amplitude of the linear gradient is comparable to the perturbations described by the higher orders. We cannot conclude whether the high-order terms are significant because instrument-based inhomogeneities in the CFHT background cannot be excluded.
Different PSF shapes have no impact on the matched background pattern. They lead to strong small-scale outliers that were discarded by sigma clipping. Furthermore, prominent PSF ghosts from the CFHT image were masked in the residual image.} 
This step will not be required in future \Euclid\ data releases since the low-level scattered light in VIS will be modelled and removed in the pipeline processing of the data.

Galactic cirri are prominent at optical wavelengths but less so at NIR wavelengths \citep[e.g.][]{Roman2020}. Therefore we modelled and removed Galactic cirri from the \IE\ image, but not the NIR images. 
To do so, we extracted the region of the Perseus ERO from the dust emission maps by \citet[][Fig.\,\ref{fig:vis_steps}f]{Meisner2014}. This map was generated from the WISE 12\,\micron\ imaging data and is free of compact sources and other contaminating artifacts. The angular resolution is limited by the highest HEALPIX resolution of $\rm{NSIDE}=8192$, corresponding to $\sim$26\arcsec\ per pixel.

We normalised this map to match the average background properties of the \IE\ image (Fig.\,\ref{fig:vis_steps}b). We find that the brightest patches of Galactic cirri in this region have a surface brightness of $\mu(\IE)=26\,\mathrm{mag}\,\mathrm{arcsec}^{-2}$ and a size of $\sim$10\arcmin. To create the image shown in Fig.\,\ref{fig:vis_steps}c, we then subtracted this low-resolution cirri image from the \IE\ image. 

Unfortunately, the resulting image (Fig.\,\ref{fig:vis_steps}c) displayed a slight flux gradient that is likely an artifact produced from incorrect cirri subtraction and the need to perform the CFHT-image flattening before the cirri subtraction. We therefore fitted a 2D first-order polynomial gradient\footnote{We did not fit a higher-order function as it can lead to oversubtraction of the ICL.} to the masked image and subtracted it from the final \IE\ image. This gradient is opposite to the first gradient but the amplitude is four times smaller. We used the resulting image, shown in Fig.\,\ref{fig:vis_steps}d, for the remaining analysis.

At the time of analysis, large-scale PSF models were not available, so we constructed empirical PSFs for each image (\IE, \YE, \JE, and \HE). We first created postage stamps of $\sim$100 bright stars with magnitudes in the range $11.5 < \IE < 13.6$.  Our goal was not to accurately subtract the PSF core, which is saturated for these bright stars, but rather to accurately subtract the outer PSF profile where the star's diffuse light can be confused with ICL \citep[e.g.][]{Montes2021}. For \IE\ we used postage stamps of 1400 pixels (140\arcsec) on a side, while for the NIR bands, the postage stamps were 500 pixels (150\arcsec) on a side. We masked bright sources in each postage stamp (excluding the diffraction spikes), then median combined the stamps to obtain a PSF.    

The PSF model was then subtracted from the 75 brightest stars in the Perseus field, excluding those stars located near the edge of the field or those located close to another bright source (which prevented the outer profile of the star from being accurately measured). The same stars were subtracted in each of the \IE, \YE, \JE, and \HE\ images.  We note that some of these stars were used in the construction of the PSF. To ensure the best subtraction of the outer profile, we normalised the PSF profiles using a carefully selected aperture around each star that minimises the chi-squared statistic for the difference between the normalised PSF and the outer light profile of the individual star.

The extended PSF wings from the bright central source in NGC\,1275 can possibly contaminate the ICL. To estimate the impact, we matched the surface brightness of the extended PSF model \citep{EROData} to the non-saturated inner region $1.35"<a<1.5"$ of NGC\,1275. The contribution of the PSF wings to the total light is $<3\%$ ($<0.3\%$) at $a>10"$ ($a>100"$) and, thus, negligible in the surface brightness and colour profiles over the relevant radial range. We also ignored the impact on the surface brightness profile of NGC\,1272 because it does not host a bright central point source.

\section{\label{sc:method} Detecting the intracluster stellar tracers}

We used two tracers of intracluster stars in this analysis: the ICL and ICGCs. The ICL presents as diffuse LSB emission, while GCs at the distance of the Perseus cluster appear as faint point sources embedded in the ICL and throughout the halos of bright cluster galaxies. The method to measure these two sources diverged at this point in the analysis, and we describe both methods in turn. 

\subsection{Isolating the intracluster light} \label{sec:det_icl}

Precise surface photometry of the ICL requires complete and homogeneous masks where only the BCG and ICL remain unmasked, so we aggressively masked the remaining high surface-brightness regions of the images, including all GC candidates. We applied the method described in \cite{Kluge2020} and \cite{Kluge2020_dissertation}. In brief:
\begin{itemize}
    \item Images were ``flattened'' by subtracting a spline-based background to remove the diffuse light or other LSB features. The spline step size varied between 50 and 100 pixels (independent of the pixel scale) for different masks depending on the source sizes. This step was necessary to avoid masking the BCG+ICL.
    \item Flattened images were then smoothed to optimally identify objects on various spatial scales while avoiding the masking of noise peaks misidentified as signals. The standard deviation of the Gaussian smoothing kernel varied between 3 and 15 pixels for different masks, while larger spatial scales were handled by median-binning by $20\,{\mathrm {pixels}}\times 20\,{\mathrm {pixels}}$ and re-applying the masking procedure.
    \item Masks were derived on the smoothed and flattened images by applying signal-to-noise scaled surface brightness thresholds that were optimised for each spatial scale. 
    \item A manual re-masking was applied to the star-forming regions of NGC\,1275 (\citealt{Conselice2001}; see Fig \ref{fig:masked_zoomin}) as well as the outskirts of very extended sources and background inhomogeneities on arcminute scales (due to e.g. Galactic cirri residuals or isolated patches of ICL detached from the BCG). 
    \item This procedure was repeated in the inner regions of NGC\,1275 and NGC\,1272 after the galaxies' models had been subtracted from the images (see Sect.\,\ref{sec:isomodelling}) to ensure that the stellar halos of galaxies near these bright galaxies were also masked.
    \item Masks were generated for each of the four images separately, and the union of these masks was used to create a single mask that was applied to all images for the ICL analysis.
\end{itemize}
The resulting limiting surface brightness for 2D structures can be estimated by inspecting Fig.\,\ref{fig:vis_steps}g. The large-scale variations in the remaining unmasked regions are of the order of 27\,mag\,arcsec$^{-2}$.

\subsection{Detecting intracluster globular clusters}\label{sec:GC_detection}

Most GCs are embedded in the BCG and the surrounding ICL, so we constructed a new source catalogue that includes sources embedded within the high surface-brightness regions. We first removed as much of the diffuse light as possible by applying a ring median filter, with $R_{\rm in} = 2$\,pixels and $R_{\rm out} = 4$\,pixels, to the original \IE\ image (Fig.\,\ref{fig:vis_steps}a) to derive a filtered image of the \IE\ image. This filtered image was subtracted from the original \IE\ image, and the resulting residual image was used for source detection. 

Objects were detected using the {\tt astropy.photutils} python package \citep{Bradley2023} by selecting sources with 3 connected pixels with a flux threshold of at least 3 times the root-mean-square (RMS) statistic of the background. To ensure even detection across the entire field of view we masked regions that were observed for less than 30\% of the total exposure time. 

Photometry and shape measurements of the sources were made on the original image (Fig.\,\ref{fig:vis_steps}d). Background-subtracted photometry was measured in circular apertures with radii of 2 and 5 pixels, with a median background measured in an annulus with $R_{\rm in} = 7$\,pixels and $R_{\rm out} = 13$\,pixels. We corrected for Galactic extinction using the {\it Planck} thermal dustmap \citep{Planck_dust2014}, \citet{Gordon2023} extinction law, and assuming an SED of a 5700\,K blackbody. 

GC candidates were selected as sources with an aperture-corrected magnitude in the range $22.7<\IE <26.3$, an FWHM less than 4\,pixels, elongation (defined as the ratio of the semi-major and semi-minor axis) less than 2, and a concentration index (the difference between the uncorrected aperture magnitude measured at 2 and 5\,pixels, \citealt{Peng2011}) between 0 and 1\,mag. The faint magnitude limit corresponds to the approximate turn-over magnitude for GCs at the distance of the Perseus cluster. The bright limit is equivalent to 3$\sigma$ brighter than the turn-over magnitude assuming the GC luminosity function (GCLF) has an approximately Gaussian spread of $\sigma\sim1.2$\,mag. Figure \ref{fig:cfht_euclid_zoom} shows an exemplary image highlighting the numerous GC candidates that are visible in the high spatial resolution \IE\ band (middle panel), whereas most cannot be detected in the \JE\ band (right panel) and the seeing-limited CFHT image (left panel).

We note that the above criteria also select Milky Way stars and small background galaxies. We removed some of these contaminating sources using an {\tt astropy.photutils} segmentation map of sources that are larger than 4\,arcsec$^{2}$, and that are twice the RMS of the background. We removed any GC candidate that overlaps with these extended sources detected in \IE. This segmentation map also masked the brightest features of the emission-line nebula of NGC\,1275, and the high-velocity system of NGC\,1275 that overlaps the northwestern part of the BCG. There are several bright star clusters embedded in these systems \citep{Canning2010}, but the GCs would only be co-spatial with the line-emitting filaments for a relatively short time ($10^8$\,yr), so we are not likely missing a large fraction of the GCs in this system.

This work focuses on the GCs in the BCG and the intracluster region so we removed GCs associated with the other cluster galaxies. Using the \IE\ image as the detection image (Fig.\,\ref{fig:vis_steps}d), we created a background RMS map by obtaining a median filtered map on the scale of a square box of 1000\,pixels on a side. Large objects were selected as sources that have 50\,000 connected pixels that are 3.5 times the RMS of the background. Candidate GCs within the large objects were removed from the GC candidate catalogue, except for candidate GCs within the BCG and the nearby bright galaxy NGC\,1272. The GCs within NGC\,1272 were modelled and removed statistically (see Sect.\,\ref{sec:isomodelling}), rather than masked, as the galaxy is very close to the cluster core. 

The remaining candidate GCs still include contamination from faint stars and point-like background galaxies, which are assumed to have a uniform distribution over the field. These contaminants were removed via statistical background subtraction in the results shown below. The number density of these contaminants is described in Appendix\,\ref{sec:background_constant}.

\subsection{Modelling of ICL surface brightness profiles and ICGC density profile} \label{sec:isomodelling}

Our approach to modelling the ICL is based on fitting ellipses to lines of constant surface flux (isophotes). The biggest advantage of this approach over 2D parametric image modelling is that we obtain radial profiles of the surface brightness, ellipticity, position angle, and centring. However, a disadvantage arises when galaxies overlap because the isophotes cannot be approximated by ellipses; iterative modelling is necessary in such cases.
In addition to the BCG, the Perseus cluster core contains another bright galaxy, NGC\,1272, that is only 1\,magnitude fainter (in \IE) and lies only 5\arcmin\ (105\,kpc) from NGC\,1275. The close proximity of a bright cluster galaxy complicates the modelling of the BCG+ICL light profile and BCG+ICGC density profile. We therefore modelled and subtracted the second-ranked cluster galaxy (in both diffuse light and GC surface density) before measuring the BCG+ICL light profile and the BCG+ICGC density profile.

We began by modelling NGC\,1275 and subtracting this model from the image. We then modelled NGC\,1272 and subtracted it from the original image before remodelling NGC\,1275. To minimise the contamination due to overlapping isophotes, the surface brightness profile of NGC\,1272 was extrapolated beyond a semi-major axis radius of $a\approx200\arcsec$ using the best-fit S\'ersic profile ($n\approx1.0$).

\begin{figure*}
    \centering
    \includegraphics[width=0.49\linewidth]{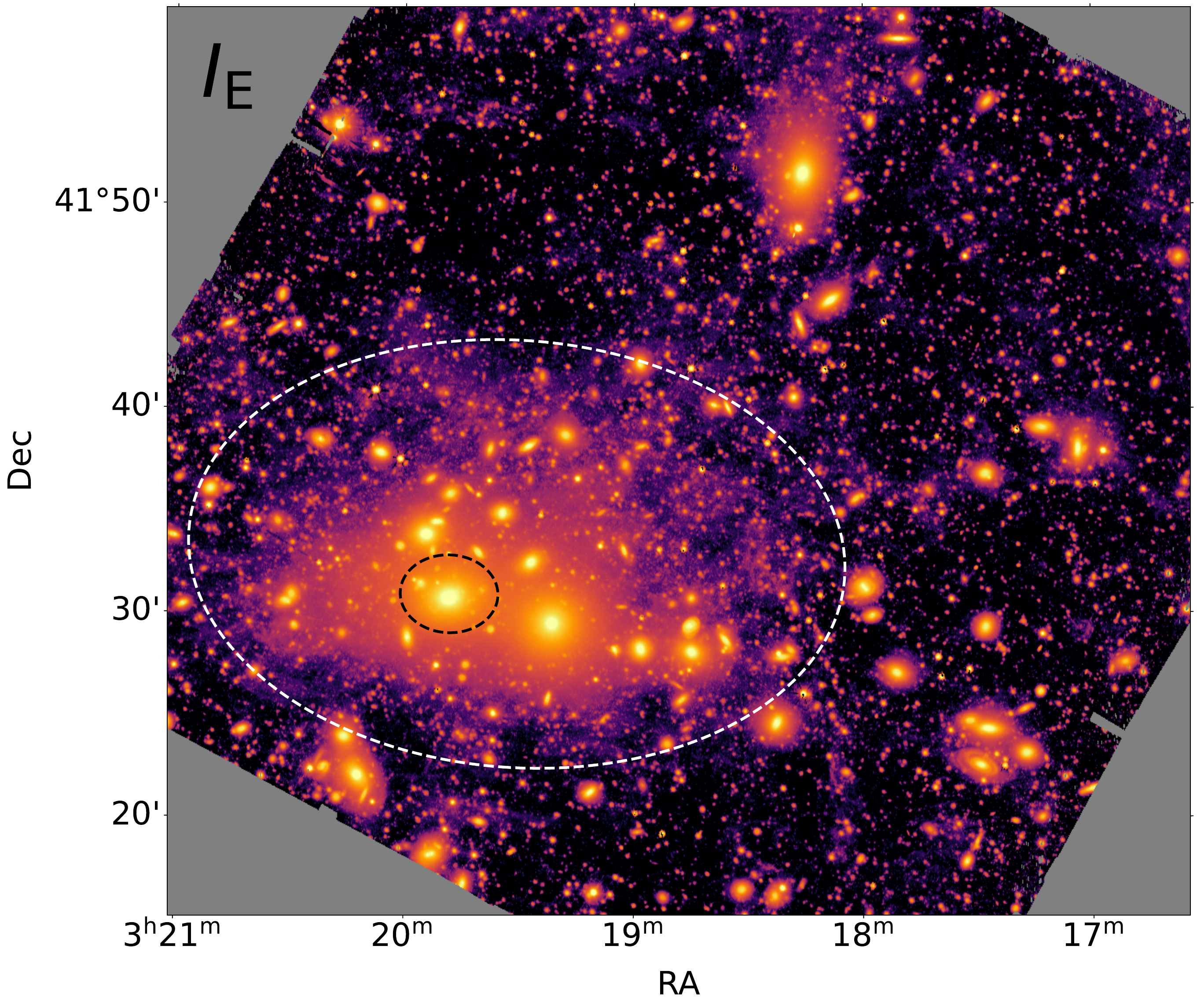}
     \includegraphics[width=0.49\linewidth]{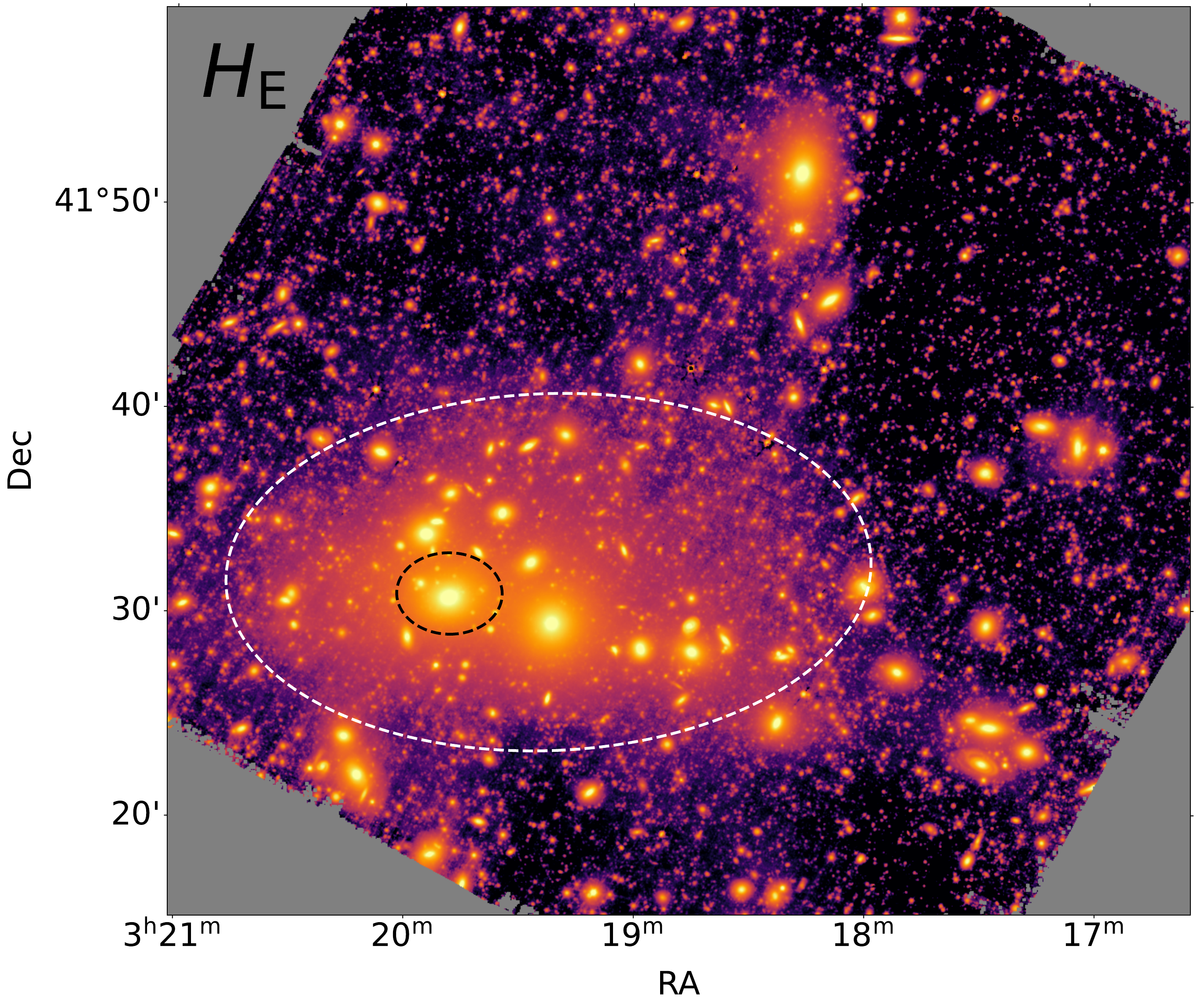}
      \caption{Processed \IE\ image (left) and \HE\ image (right). The log-normal colour scheme is used to emphasise the ICL. We masked small, bright sources and interpolated over the masked pixels using a Gaussian kernel with a standard deviation of $\sigma=30$ and 6\,pixels for the \IE\ and \HE\ images, respectively. To guide the eye, we show the isophotal contours with a semi-major axis of $a=50$\,kpc (black) and 320\,kpc (white).}
    \label{fig:IE_map}
\end{figure*}

\begin{figure}
    \centering
    \includegraphics[width=1\linewidth]{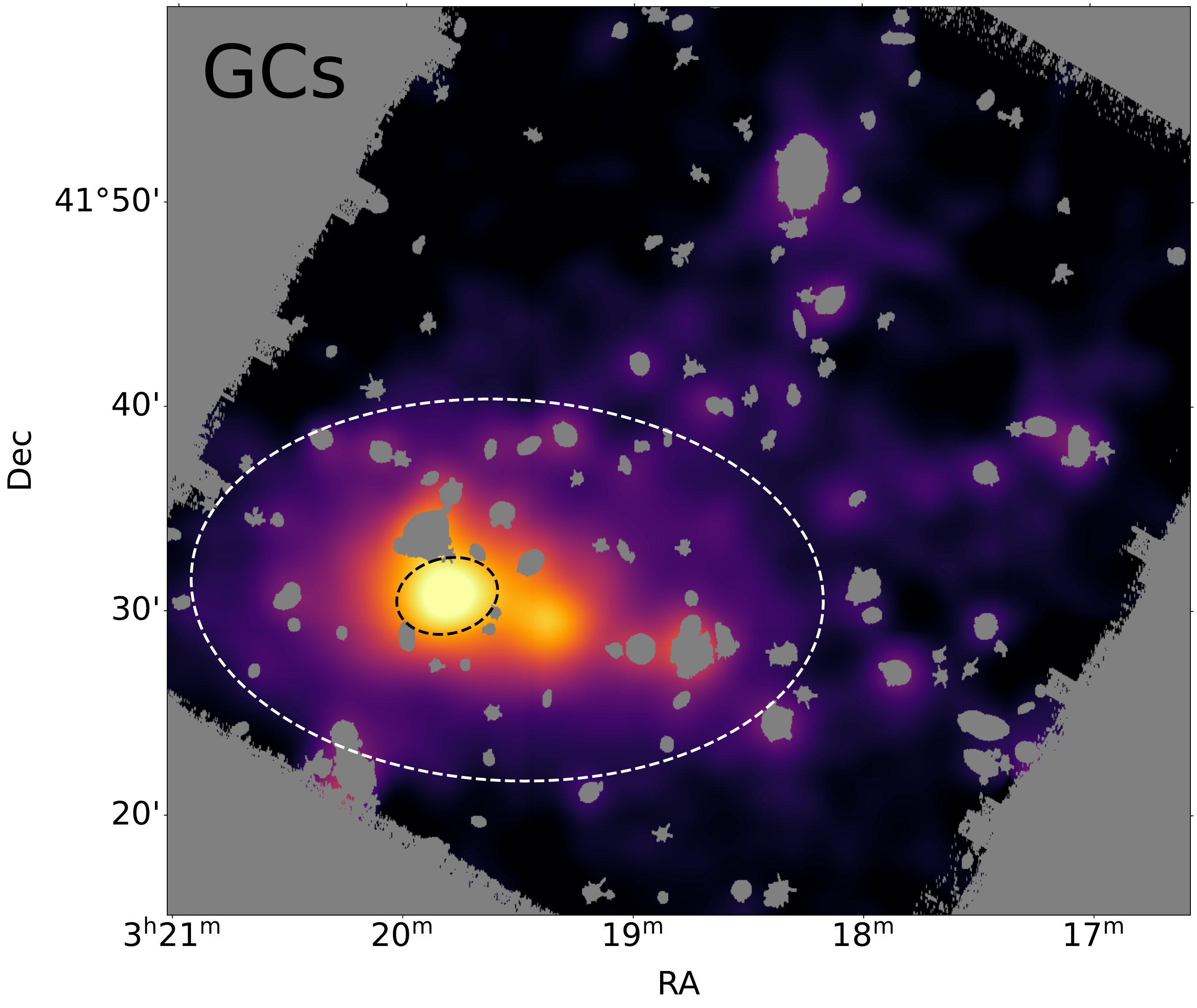}
      \caption{Map of the number density distribution of GC candidates. It was produced by modelling each GC candidate as a single pixel with unity flux and then smoothed with a Gaussian kernel of $\sigma=50$\arcsec. The colour scheme therefore represents the GC number density.  We masked bright stars, diffraction spike residuals, the emission line nebula, and the high-velocity system of NGC\,1275, as well as large cluster galaxies (except for the NGC\,1275 and its nearby companion in the west, NGC\,1272). Iso-density contours with a semi-major axis of $a=50$\,kpc (black) and 320\,kpc (white) are shown for the GC candidates.}
    \label{fig:GC_map}
\end{figure}

We fitted ellipses to the isophotes using the {\tt ellipse} task from the python package {\tt photutils} \citep{Bradley2023}. Each ellipse has four free parameters: the central coordinates, $x_0$ and $y_0$, the ellipticity, $\epsilon=1-b/a$ (where $a$ is the semi-major axis radius and $b$ is the semi-minor axis radius), and the position angle, PA (counting anticlockwise starting from the horizontal line). The semi-major axis radius, $a$, was chosen to increase in logarithmic steps by 10\% for each isophote. As the ICL becomes fainter at larger radii, some ellipse parameters could not be measured robustly. This happened when the parameters changed significantly and randomly for sequential isophotes. At this point, we fixed the parameters to the last robust isophote for which their values were successfully measured. 

The flux along the isophotes where $a<15$\,pixels was measured using {\tt ellipse}. However, for $a>15$\,pixels, we took the median value of all unmasked pixels in an elliptical annulus around that isophote, which produces equivalent results but is computationally more efficient. The width of the annulus is equivalent to the step size at the radius, which was selected to extract the maximum signal-to-noise while avoiding overlaps between different annuli. We generated model BCG+ICL images by interpolating the 1D isophotal shape profiles onto a fine grid with 1000 steps equidistant in $a^{1/4}$. The corresponding annuli were filled with the surface flux at the given radius.

A similar method was used to determine a simplified model of the BCG+ICGC distribution by fitting ellipses to lines of constant GC surface density (iso-density contours). 
GCs are spatially distinct point sources but computing the shape of their global distribution requires a smooth map. Therefore, we computed the surface density of GC candidates in bins of 40\arcsec\ on a side. We then used this map of GC surface density to fit iso-density contours to the GC distribution using the same method as described above. 

To increase the maximum radius to which the GC iso-density contours can be fitted robustly, we smoothed the image using a Gaussian kernel with $\sigma=56$\arcsec\ and repeated the procedure. Both profiles, one representative of the inner region and one of the outer region of the ICGC distribution, were merged at $a=450\arcsec$. We checked that the smoothed profiles converge sufficiently to the unsmoothed profiles at $a=450\arcsec$.

Once the position of the iso-density contours was defined, we calculated the mean density of GCs in an elliptical annulus around the contours using the unsmoothed and unbinned GC candidate catalogues, and taking into account the masked area of each annulus. The width of each annulus is equivalent to the step size at that radius. To construct the profile around NGC\,1275, we masked both NGC\,1272 and  NGC\,1265 (the brightest galaxy in the north of the field of view) with ellipses with $a=100$\,kpc to minimise the contamination of GCs from the halos of these galaxies.

Ideally, the background constant of the BCG+ICL surface brightness and GC surface density would be measured beyond the virial radius of the cluster, well beyond the region where ICL may exist. However, due to the large extent of the Perseus ICL on the sky, which goes beyond the observed field of view, we determined the residual background flux or GC surface density as the residual signal after subtracting extrapolated S\'ersic profiles. Details are given in Appendix \ref{sec:background_constant}.
For the GCs this background value was measured from candidate sources beyond the elliptical aperture with a semi-major axis of 666\,kpc, which resulted in a background density of 52.2 sources per arcmin$^2$. Further details on the background measurement and uncertainties are provided in Appendix \ref{sec:background_constant}. 
The full images, masks, residuals, and models for the BCG+ICL in all four filter bands, and the GC surface density measured in \IE, are shown in Appendix \ref{sec:masks_resids}.

\section{\label{sc:} Results}

\subsection{Distribution of the BCG+ICL and GCs}\label{sec:distribution}
Figure \ref{fig:IE_map} presents the \IE\ and \HE\ images with a colour bar chosen to accentuate the ICL. To further highlight the ICL -- rather than be distracted by small, high surface-brightness sources -- we masked small, bright sources and interpolated over these pixels using a Gaussian kernel of $\sigma=30$\,pixels and 6\,pixels for the \IE\ and \HE\ images, respectively. 

The GC number density map, shown in Fig.\,\ref{fig:GC_map}, was produced by modelling each GC candidate as a single \IE\ pixel with unity flux. We then smoothed this map with a Gaussian kernel with $\sigma=50$\arcsec\ and took care to properly interpolate over the masked regions of the image. The colour scheme therefore represents the GC number density.  NGC\,1272 was not masked in this image because we fitted a model to its GC distribution and subtracted it before measuring the iso-density contours. The GCs of NGC\,1272 are distributed in a compact circular configuration, extending no more than 100\,kpc from the galaxy's core. 

\begin{figure}[t]
    \centering
    \includegraphics[width=\linewidth]{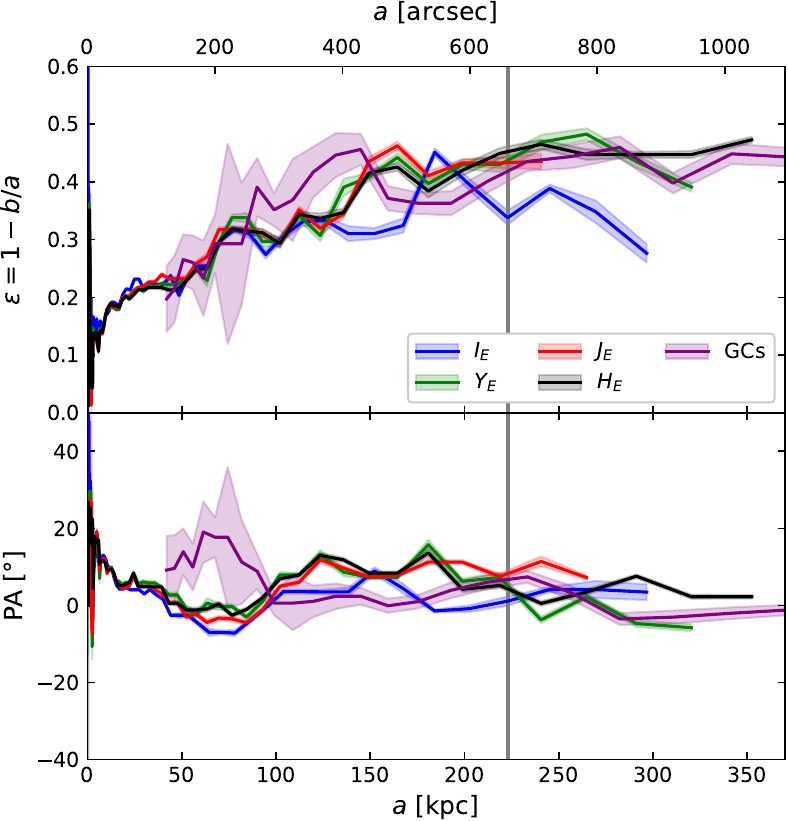}
    \caption{Ellipticities and position angles (PA) of the ellipses fit to isophotes of the BCG+ICL in the \IE, \YE, \JE, and \HE images and iso-density contours of the GC map. The GC distribution was binned to a pixel scale of 40\arcsec, therefore only data beyond 120\arcsec\ radius are shown. The vertical line indicates the radius $a=11\arcmin$ beyond which remaining cirrus could alter the profiles of the ICL (see Appendix \ref{sec:background_constant}).}
    \label{fig:ellp_PA}
\end{figure}

\begin{figure}[t]
    \centering
    \includegraphics[width=\linewidth]{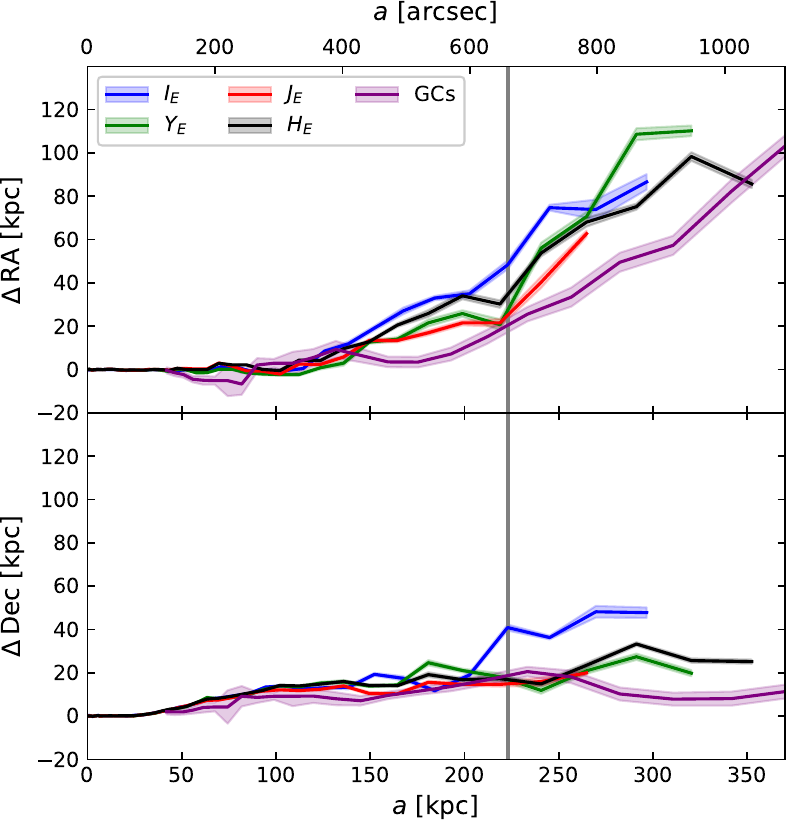}
    \caption{Offset of the RA (top) and Dec (bottom) centroid of the isophotal or isodensity elliptical contours in all four images and the intracluster GCs map. The $\Delta\,{\rm RA}=0$ and $\Delta\,{\rm Dec}=0$ points are defined as the position of the nucleus of the BCG in the \HE\ image. Positive $\Delta\,{\rm RA}$ are counted westward. The vertical line indicates the radius $a=11\arcmin$ beyond which remaining cirrus could alter the profiles of the ICL (see Appendix \ref{sec:background_constant}).}
    \label{fig:offsets}
\end{figure}

Two of the fitted contours are shown on Figs.\,\ref{fig:IE_map} and \ref{fig:GC_map} to illustrate the best-fit ellipses on both small and large scales. The shape of the isophotes in the \IE, \YE, \JE, and \HE images are well matched quantitatively. Figures \ref{fig:ellp_PA} and \ref{fig:offsets} show that the ellipticities, position angles, and centroids of the fitted ellipses to all four images of the BCG+ICL and the iso-density contours of the GC map are similar. The ellipticity increases from $\epsilon=0$ to $\epsilon=0.4$ over the inner 100\,kpc in both light and GC tracers. The position angles of the ellipses are between approximately $-10$\degr\ and $+20$\degr\ and are similar for the two tracers beyond 100\,kpc, whilst within 100\,kpc there is a difference between the two tracers by up to 26\degr. Beyond 100\,kpc, the distribution of both the ICL and the ICGCs deviate significantly from a circular profile with an ellipticity of $\epsilon\sim0.4$.

There is also a close correspondence in the position of the ellipse centroids for the BCG+ICL measured at all four wavelengths.  The small differences in centroids can be explained by the remaining background inhomogeneities in the individual images. The spatial distributions of the lighter blue features at $\mu(\IE)=27$\,mag\,arcsec$^{-2}$ in Figs. \ref{fig:vis_steps}b and \ref{fig:vis_steps}c suggest that inaccurate cirri subtraction may be responsible for the ICL appearing to shift north in the \IE\ image as well as its slightly rounder shape at $a>250$\,kpc.
We marked the corresponding radius of $a=11\arcmin$\ in Figs.\,\ref{fig:ellp_PA} and \ref{fig:offsets} by a vertical line.
Overall, the similar contour shapes in the \IE\ and \HE\ images validate the reliability of the ellipsoidal fits to the BCG+ICL.

One of the most prominent features in Figs.\,\ref{fig:IE_map} and \ref{fig:GC_map} is that ICL and ICGC contours on the largest scales ($a\sim320$\,kpc) are not centred on the BCG, but rather are offset westwards of the BCG core, \text{$\Delta\,{\rm RA}\sim60$\,kpc (3\% of $r_{200,{\rm c}}$)}. Figure \ref{fig:offsets} shows that this offset is seen at all wavelengths and in the ICGC distribution -- although the western offset is about a factor of two smaller in the ICGCs than the diffuse light. We cannot attribute this offset to incomplete or overzealous cirri subtraction given that this offset is also observed in the distribution of ICGCs, which is not influenced by the presence of cirri.
Furthermore, this offset is unlikely to be caused by contamination of light or GCs from NGC\,1272 because our iterative fitting of the two largest galaxies in the core, NGC\,1275 and NGC\,1272 (see Sect.\,\ref{sec:isomodelling}), means that we account for and remove the contribution of NGC\,1272 to the light and the GCs in the intracluster region.

In summary, we have shown in Figs.\,\ref{fig:IE_map}--\ref{fig:offsets} that the ICL and ICGCs share a coherent spatial distribution. The elliptical contours used to model their distributions have similar ellipticities and position angles for the entire radial range for which they can be measured ($a=350$\,kpc). Such a close correspondence suggests that these two tracers of the intracluster stellar population have a common origin and/or they are well mixed and their distribution is governed by a common potential.

\subsection{Radial profile of the intracluster light and intracluster globular clusters}\label{sec:1dprofiles}

\begin{table}[]
  \caption{Directly integrated structural parameters from the BCG+ICL surface brightness profiles in different filter bands.}
    \centering    
    \renewcommand{\arraystretch}{1.4} 
    \begin{tabular}{|l|llll|}
    \hline\hline
    Filter & $a_{\rm e}$ & $\mu_{\rm e}$ & $L_{<500\,{\rm kpc}}$ &$f_{\rm BCG+ICL}$ \\
    ~~ & [kpc] & [mag\,arcsec$^{-2}$] & [$10^{12}$\,L$_\odot$ ]&~~~~~\%  \\
    \hline
    \IE & $92^{+198}_{-62}$ & $21.93^{+3.64}_{-3.43}$ & $0.96^{+1.20}_{-0.42}$ & $44^{+26}_{-22}$\\
    \YE & $143^{+14}_{-12}$ & $24.92^{+0.13}_{-0.14}$ & $1.30^{+0.09}_{-0.06}$ & $42^{+8}_{-7}$\\
    \JE & $116^{+13}_{-10}$ & $24.50^{+0.14}_{-0.14}$ & $1.29^{+0.07}_{-0.06}$ & $33^{+7}_{-6}$ \\
    \HE & $109^{+5}_{-5}$ & $24.31^{+0.05}_{-0.05}$ & $1.67^{+0.04}_{-0.04}$ & $38^{+6}_{-6}$  \\
    \hline\hline
    \end{tabular}
    \tablefoot{Derived parameters are the half-light radius along the semi-major axis, $a_{\rm e}$, the surface brightness at the half-light radius, $\mu_{\rm e}$, the total extinction-corrected luminosity within the BCG+ICL in units of solar luminosities, and the fraction of cluster light in the BCG+ICL component, $f_{\rm BCG+ICL}$. All parameters were computed by considering only the light within the $a=500$\,kpc ellipse.}
    \label{tab:struc_params}
\end{table}

\begin{figure}[t]
    \centering
    \includegraphics[width=\linewidth]{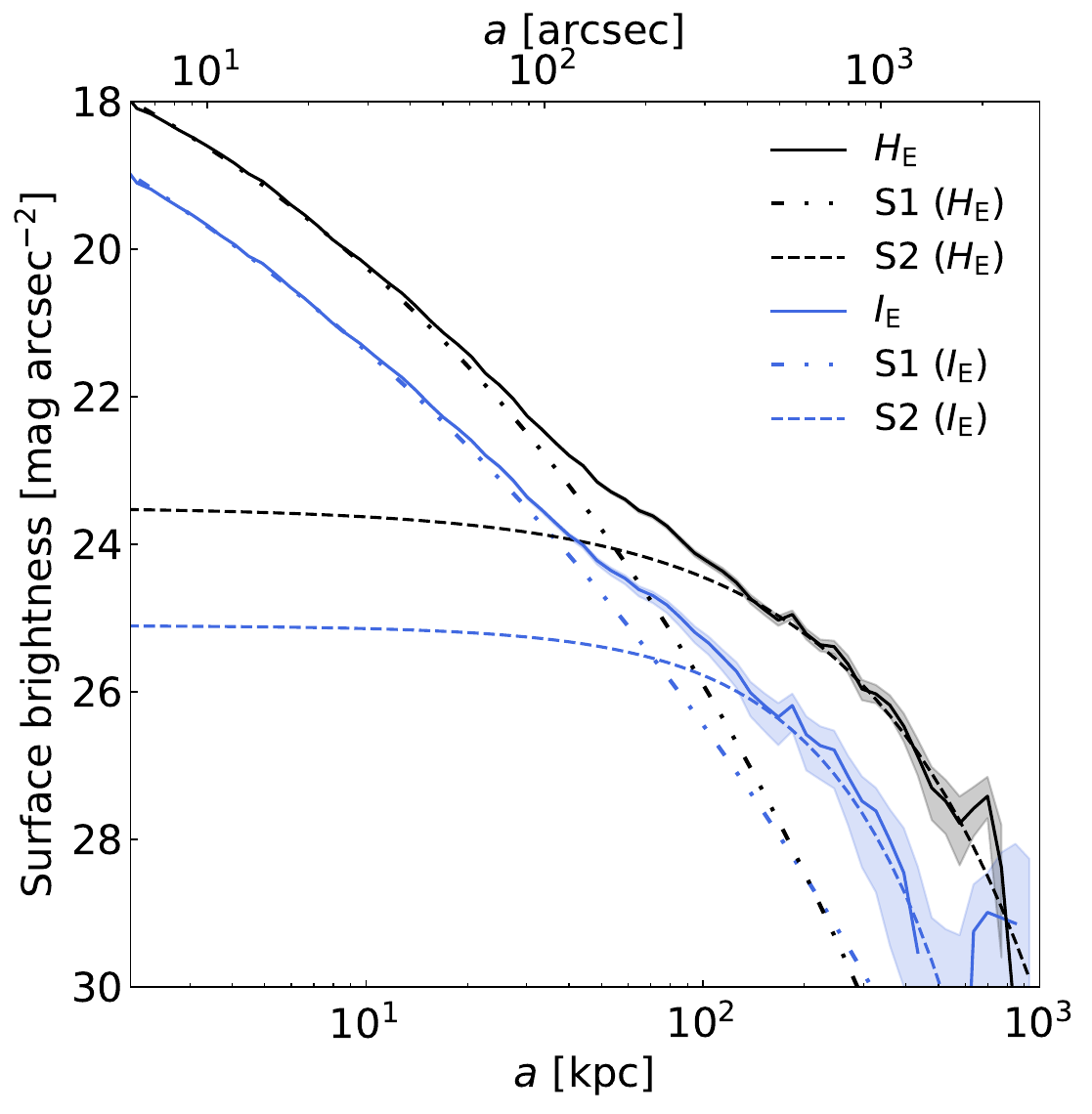}
    \caption{Radial surface brightness profiles of the BCG+ICL in both \IE\ and \HE. The profiles require two S\'ersic components to be adequately fit: a compact component (S1) and an extended component (S2). The transition between the profile being dominated by the compact BCG to being dominated by the extended ICL component occurs at $a=71\pm13$\,kpc in \IE and $a=55\pm6$\,kpc in \HE.
    The radius is not BCG-centric but is given by the semi-major axis radius $a$ of the best-fit ellipse to each isophote with free central coordinates (see Fig. \ref{fig:offsets}).} 
    \label{fig:1D_dist}
\end{figure}

\begin{figure}[t]
    \centering
    \includegraphics[width=\linewidth]{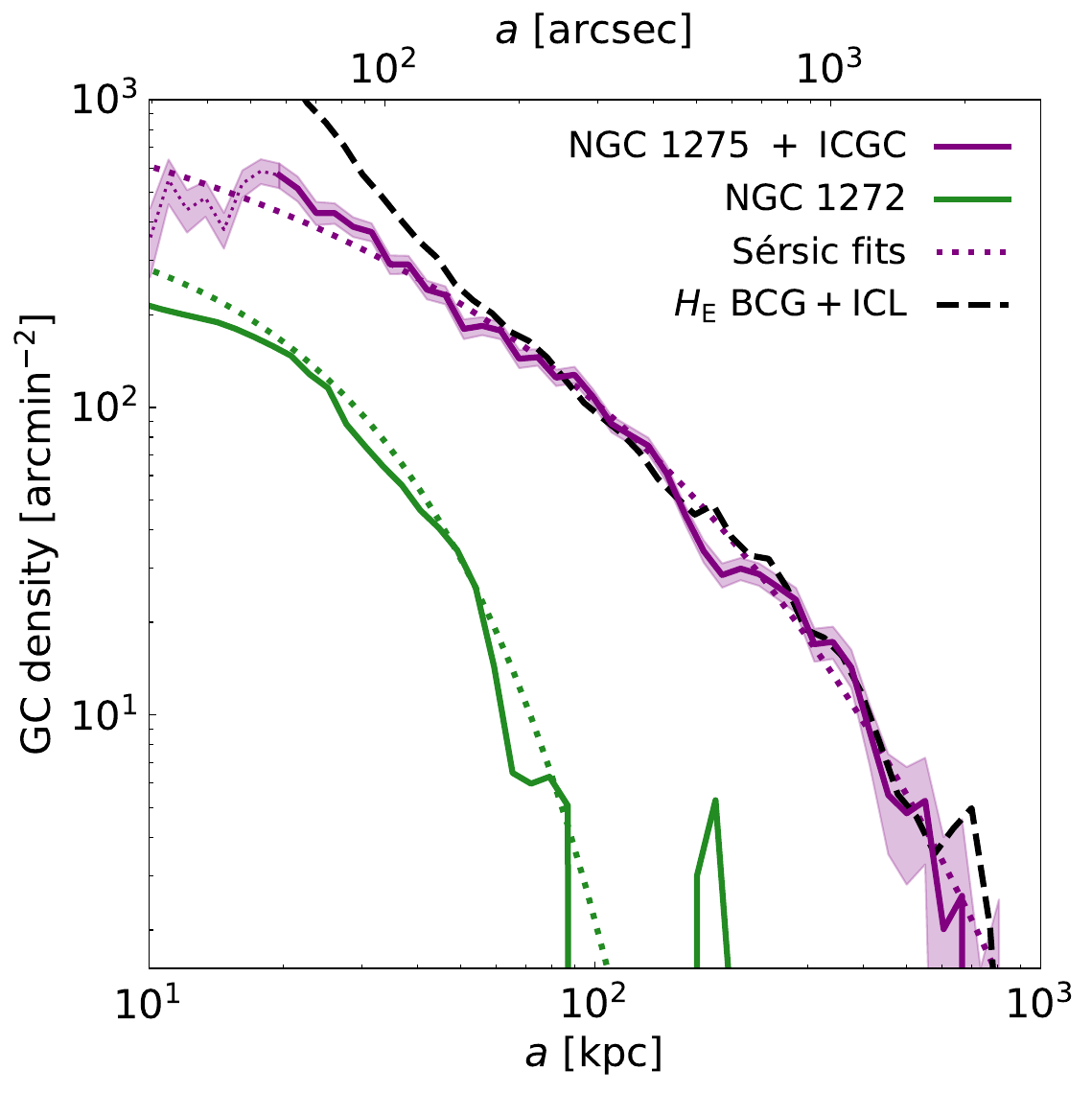}
    \caption{Radial profile of the GC number density surrounding the BCG NGC\,1275 (solid purple line) and the nearby galaxy NGC\,1272 (solid green line). To construct the profile of NGC\,1275, we masked both NGC\,1272 and the NGC\,1265 (the brightest galaxy in the north of the field of view) with ellipses with $a=100$\,kpc. The profile of GCs in both galaxies are well fit by S\'ersic profiles (dotted coloured lines), but the GCs associated with NGC\,1275 are far more extended, with a measurable excess above the background up to 600\,kpc. The radial profile of GCs surrounding NGC\,1275 is well matched to the \HE- band surface brightness profile of the BCG+ICL (black dashed line, scaled arbitrarily) at radii greater than $\sim60$\,kpc from the BCG nucleus. A significant fraction of the GCs may be missed in the highly crowded region within 20\,kpc of the BCG's core.}
    \label{fig:GC_1d}
\end{figure}

\begin{table*}[]
      \caption{Best-fit parameters of the S\'ersic fits to the surface brightness and surface density profiles.}
        \renewcommand{\arraystretch}{1.4}
    \centering
    \begin{tabular}{|l|lllll|}
    \hline\hline
    Diffuse light & $a_{\rm e}$ [kpc] & $\mu_{\rm e}$ [mag\,arcsec$^{-2}$] & $n$ & $L_{\rm tot}$ [$10^{12}$\,L$_\odot$ ] & $L_{< 500\,{\rm kpc}}$ [$10^{12}$\,L$_\odot$ ] \\
    \hline
    S1 (\IE) & $26 \pm 5$ & $23.2 \pm 0.3$ & $5.0 \pm 0.5$ & $0.29 \pm 0.03$ & $0.29 \pm 0.03$ \\
    S2 (\IE) & $186 \pm 23$ & $26.5 \pm 0.2$ & $0.8 \pm 0.3$ & $0.27 \pm 0.05$ & $0.26 \pm 0.04$ \\
    S1 (\HE) & $18 \pm 2$ & $21.4 \pm 0.1$ & $3.5 \pm 0.2$ & $0.62 \pm 0.03$ & $0.62 \pm 0.03$ \\
    S2 (\HE) & $266 \pm 38$ & $25.7 \pm 0.2$ & $1.2 \pm 0.3$ & $1.28 \pm 0.21$ & $1.03 \pm 0.07$ \\
     \hline\hline
     Globular clusters &  & $N_{\rm e}$ [arcmin$^{-2}$] & & $N_{\rm tot}$ & $N_{< 500\,{\rm kpc}}$ \\
    \hline
    NGC\,1275 + ICGCs & $242 \pm 7$ & $27 \pm 2$ & $2.2 \pm 0.1$ & $42\,400 \pm 900$ & $31\,800 \pm 300$ \\
    NGC\,1272 & $31\pm2$ & $90\pm9$ & $1.0\pm0.3$ & $2100\pm170$ & $2100\pm170$ \\
    \hline\hline
    \end{tabular}
    \tablefoot{Best-fit parameters of the double-S\'ersic fit to the surface brightness profile of the BCG+ICL in \IE\ and \HE, and the GC surface density profiles around NGC\,1275+ICGC and the nearby massive elliptical NGC\,1272. For each component, $a_{\rm e}$ and $\mu_{\rm e}$ list the half-light radius along the semi-major axis and the surface brightness at the half-light radius, respectively. Total luminosities, $L_{\rm tot}$, and luminosities within 500\,kpc of the BCG, $L_{<500\,{\rm kpc}}$, are corrected for extinction. $N_{\rm tot}$ ($N_{<500\,{\rm kpc}}$) is the total number of selected GCs with $\IE<26.3$ (and within 500\,kpc), after background subtraction. The uncertainties are either the statistical uncertainties of the best-fit S\'ersic parameters or, for $L$ and $N$, propagated uncertainties of the best-fit S\'ersic parameters using 1000 Monte Carlo realisations.}
    \label{tab:sersic}
\end{table*}

We constructed radial surface brightness profiles of the BCG+ICL using the isophotal contours defined using the \IE\ image and measured the median flux within the annuli in both \IE\ and \HE\ images. These surface brightness profiles are shown in Fig.\,\ref{fig:1D_dist}. The ICL was detected above the noise to a distance of $\sim$600\,kpc from the BCG in \HE, but only $\sim$400\,kpc in \IE. We show in Appendix \ref{sec:background_constant} that the $1\sigma$ limiting surface brightness for the 1D profiles is $\mu(\IE)=28.9$\,mag\,arcsec$^{-2}$ and $\mu(\HE)=28.7$\,mag\,arcsec$^{-2}$. These values should not be considered the typical surface brightness limit of ICL detectable in the EWS. On one hand, the exposure time of these ERO is four times that of the EWS, suggesting they should reach fainter depths than the EWS. On the other hand, the EWS avoids bright Galactic cirri and the continuous coverage of the survey area means the background can be more accurately modelled. The detectability of ICL in the EWS will be explored in an upcoming publication (Bellhouse et al. in prep).

Structural parameters obtained by direct integration of the surface brightness and ellipticity profiles within $a<500$\,kpc are listed in Table\,\ref{tab:struc_params}. Uncertainties were estimated by adding and subtracting the 1$\sigma$ background uncertainties (see Appendix \ref{sec:background_constant}) from the surface flux profiles and integrating the profiles again.
The luminosity\footnote{To transform magnitudes to Solar luminosities, we used the absolute magnitudes of the Sun in the \Euclid\ filters given in Appendix \ref{app:magabs}.} of the BCG+ICL increases with wavelength from $1.0\times10^{12}$\,L$_\odot$ in \IE\ to $1.7\times10^{12}$\,L$_\odot$ in \HE. The high concentration (small $a_{\rm e}$) in \IE\ towards the centre of the BCG is likely due to the diffuse blue light from the recent star formation in this galaxy. However, the uncertainties in $a_{\rm e}$ are large in \IE because of the remaining contamination by cirrus (see App. \ref{sec:correlated_bgscatter}).
On the other hand, $a_{\rm e}$ decreases with wavelength in the NIR filters, indicating that the ICL becomes bluer with radius.

We also calculated the fraction of the total cluster light that is emitted from the BCG+ICL. This is defined as $f_{\rm BCG+ICL}=L_{\rm BCG+ICL}/L_{\rm cluster}$, where the luminosity of the cluster includes the satellite galaxies (identified by \citealt{EROPerseusOverview} and \citealt{EROPerseusDGs}) within the $a=500$\,kpc isophote ellipse defined in the \IE\ image. The uncertainty in these fractions comes from adding, in quadrature, the 1\,$\sigma$ errors on the galaxy luminosity functions and the total ICL luminosities given in Table~1. We find that across the four \Euclid\ wavebands $f_{\rm BCG+ICL}$ is $\sim$40\%. The values at each wavelength are given in Table \ref{tab:struc_params}, and are consistent with both simulations and observations of other clusters \citep[e.g.][]{Zibetti2005, gonzalez2007, zhang19, kluge21, Sampaio-Santos2021, Brough2024}.

The BCG+ICL profile in both \IE\ and \HE\ requires two S\'ersic components to obtain a good fit over most of the radial range: a compact component (labelled S1), and an extended profile (labelled S2). The parameters of both components are listed in Table \ref{tab:sersic}. The transition between light dominated by the S1 component to light dominated by the S2 component occurs at $a=71\pm13$\,kpc in \IE\ and $a=55\pm6$\,kpc in \HE. 

Due to recent star formation occurring in the BCG, the compact S1 component comprises a larger fraction of the BCG+ICL light within 500\,kpc at \IE\ ($53\pm12$\%) compared to \HE\ ($37\pm8$\%). When we extrapolate the BCG+ICL surface brightness profile out to $a=3.4$\,Mpc the amount of light in the S2 component does not increase at \IE-wavelengths which tells us that most of the \IE\ light is limited to within 500\,kpc. On the other hand, the S2(\HE) component increases by 24\%, implying that the NIR ICL contributes more at larger radii. When extrapolated to the whole cluster, the extended component (S2) comprises $67\pm7$\% of the total BCG+ICL \HE light, but we warn that this assumes the outer S\'ersic component remains a good fit even though we did not have a measure of the ICL profile beyond 600\,kpc. 

The GC surface density contours of NGC\,1272 and NGC\,1275 were fit iteratively, allowing us to separate the GCs of these two giant elliptical galaxies. These galaxies only differ by 1\,mag in luminosity, yet the radial distributions of their GCs, plotted in Fig.\,\ref{fig:GC_1d}, are remarkably different. Both profiles reveal a flattening towards their cores, consistent with \citet{Penny2012} and \cite{Harris2017}, and likely due to incompleteness. However, the central 10--20\,kpc region of NGC\,1275 contains over twice the GC density of NGC\,1272.

The most obvious difference between the profiles of NGC\,1275 and NGC\,1272 is the extremely wide distribution of GCs around NGC\,1275, extending out to 600\,kpc. On the other hand, NGC\,1272 presents a relatively compact distribution that only extends out to 100\,kpc. NGC\,1272, which lies 105\,kpc west of NGC\,1275, contributes negligibly to the GCs within the inner 40\,kpc of NGC\,1275; however, the ICGCs surrounding NGC\,1275 contaminate the GCs in the core of NGC\,1272 by up to 25\%. 

Both the GCs in NGC\,1272 and NGC\,1275+ICGC can be modelled by a single S\'ersic profile, whose parameters are listed in Table\,\ref{tab:sersic}. NGC\,1272 follows an exponential profile, with half the GCs enclosed within 31\,kpc of the nucleus, while NGC\,1275+ICGCs follow a profile with $n=2.2\pm0.1$, and half of the GCs lie at a distance greater than $242\pm7$\,kpc, which agrees with the half-light radius for the S2 component in \HE\ (Table\,\ref{tab:sersic}). The wide profile of the NGC\,1275+ICGCs is remarkably similar to that of the BCG+ICL at radial distances larger than 60\,kpc. The \HE\ BCG+ICL surface brightness profile, shown in Fig.\,\ref{fig:1D_dist}, has been arbitrarily normalised and overlaid in Fig.\,\ref{fig:GC_1d} to illustrate the similarity in the radial profiles over the radial range $60<a/{\rm kpc}<600$. 

By directly integrating the structural parameters of the density profiles we find $31\,800\pm 300$ GCs in the NGC\,1275+ICGC within 500\,kpc. This is 15 times the number within NGC\,1272. Our GC candidates were limited to sources with $\IE<26.3$, so to extrapolate this measurement to the total number of GCs in the cluster, we must first determine the GC luminosity function (GCLF).

\subsection{Luminosity function of the globular clusters}\label{sec:GCLF_res}

\begin{figure}
    \centering
    \includegraphics[width=1\linewidth]{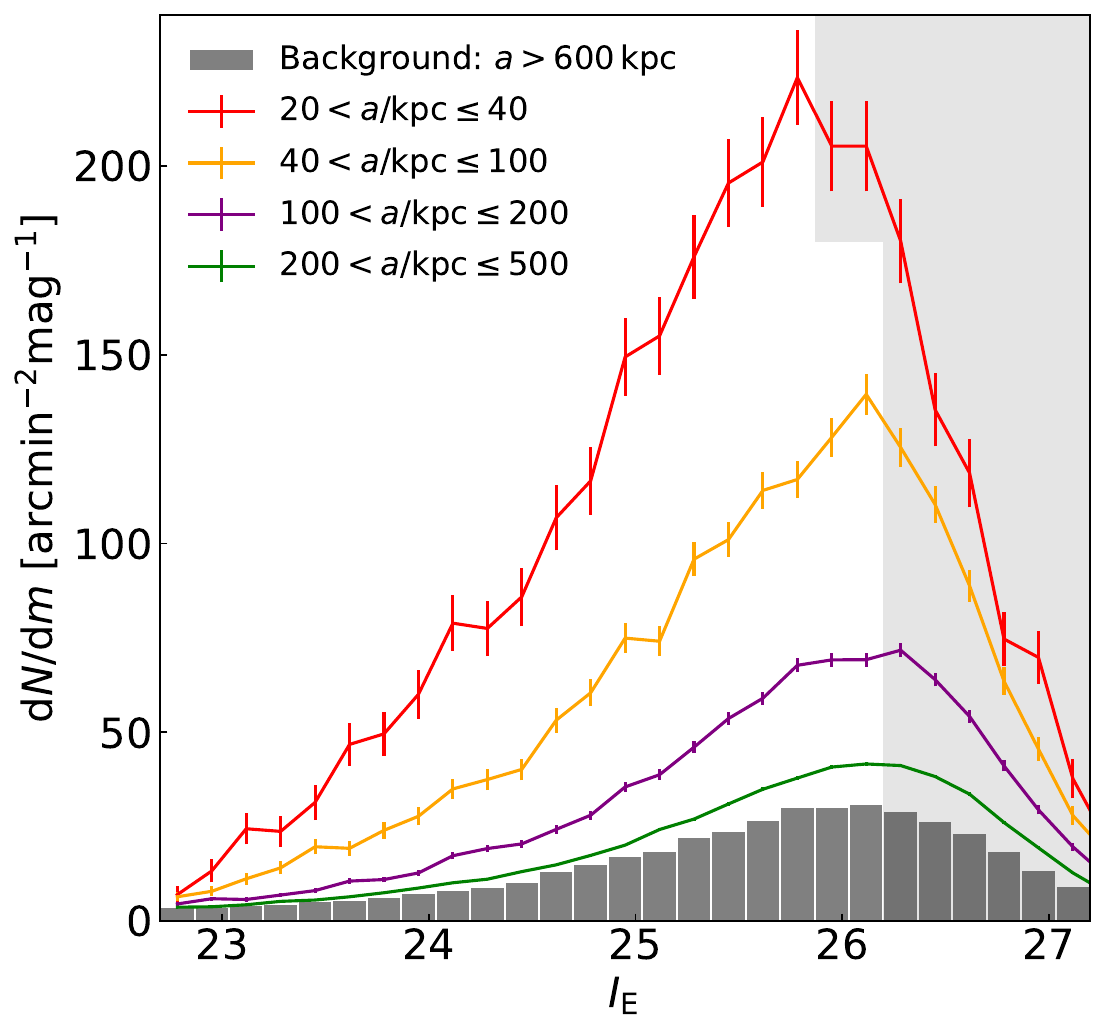}
    \caption{Luminosity function of sources that match the compactness criteria laid out in Sect.\,\ref{sec:GC_detection} in various elliptical annuli surrounding the BCG (red, yellow, purple, and green lines). These sources include contamination from background sources as well as GC-candidates. The luminosity function of background sources (grey histagram) has been calculated from sources in the area beyond $a>600$\,kpc which also match the compactness criteria laid out in Sect.\,\ref{sec:GC_detection}. The greyed-out area marks the portions of the luminosity functions where the source selection is less than 95\% complete.} 
    \label{fig:GCLF_non-bkg}
\end{figure}

\begin{figure*}
    \centering
    \includegraphics[width=0.49\linewidth]{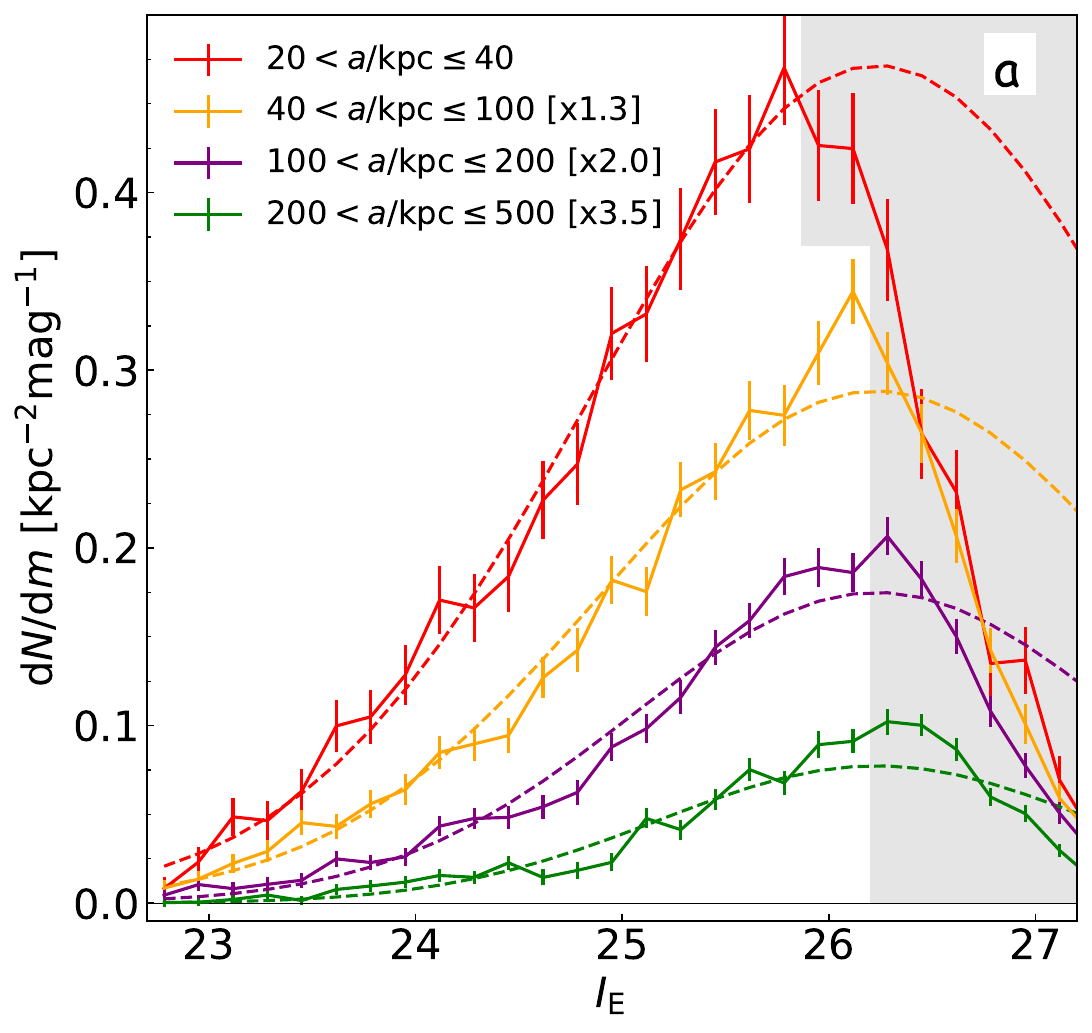}
  \includegraphics[width=0.49\linewidth]{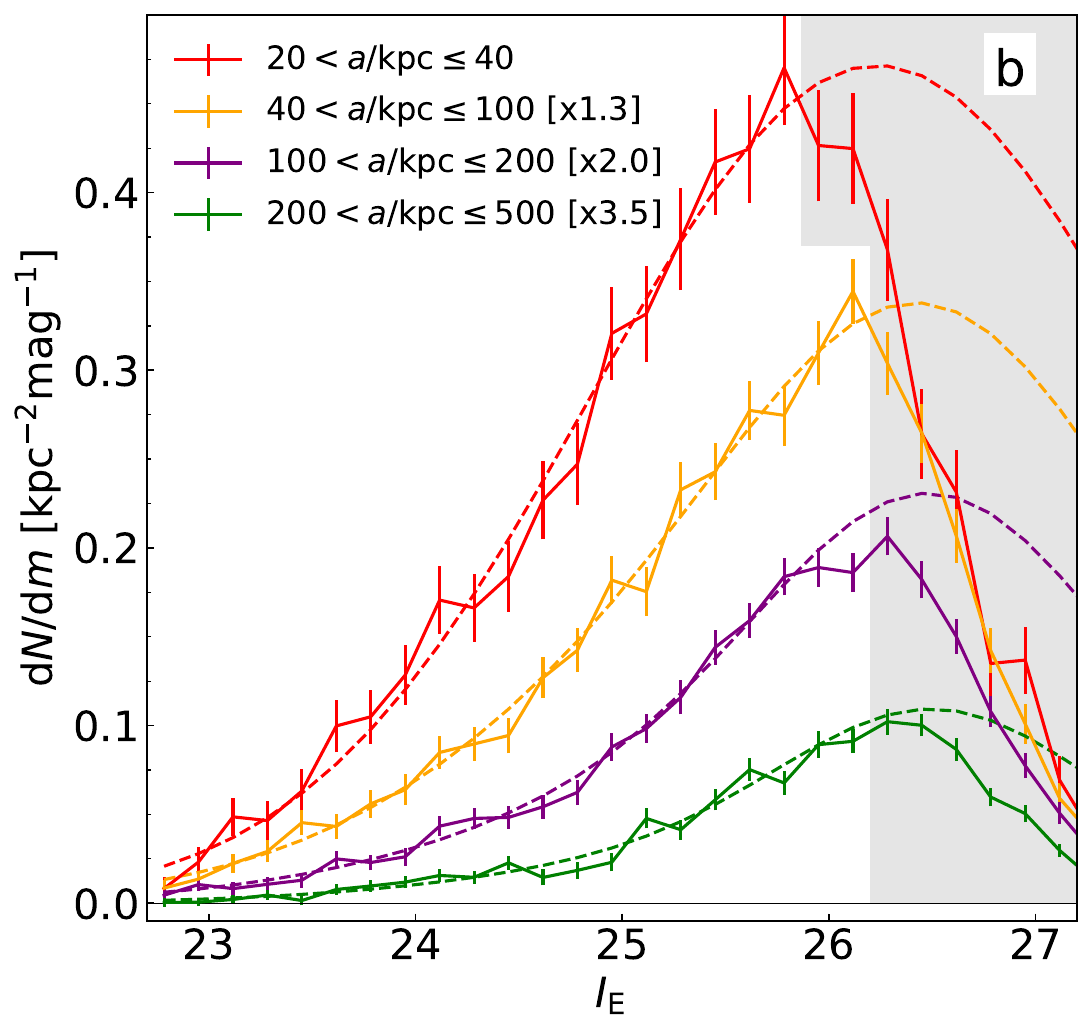}
    \caption{Luminosity function of GC candidates in various elliptical annuli surrounding the BCG. The luminosity function of background sources has been scaled to the area within each annulus and subtracted from each luminosity function. The data have been scaled for clarity according to the numbers shown in the legend. The greyed-out area marks the region where the GC candidate selection is less than 95\% complete. Regions of high surface brightness such as the centre of the BCG are incomplete to a brighter point source magnitude. In panel a (left), the dashed lines display the single Gaussian that is the best-fit to the luminosity function in each annulus, where the fit is limited to only magnitudes that are $>95$\% complete (i.e. non-greyed-out regions). The best-fit Gaussian parameters are listed in Table \ref{tab:GCLF}. In panel b (right), the yellow, purple and green dashed lines show the 2-component Gaussian distributions that are the best-fit to the luminosity function at $\IE<26.2$ (non-greyed-out regions) for the annuli with $a>40$\,kpc. The best-fit 2-component Gaussian parameters are listed in Table \ref{tab:GCLF2}.} 
    \label{fig:GCLF1}
\end{figure*}

We calculated the GCLF within NGC\,1275 and in three large annuli at large radial distances from the BCG. The detection of GCs is a strong function of local surface brightness, with high surface-brightness regions, such as the centre of the BCG, resulting in incompleteness to bright GC magnitudes. Therefore, we limited our analysis of the GCLF of NGC\,1275 in the elliptical annulus at $20 < a/ {\rm kpc}\leq40$, that is, beyond the bright region within 20\,kpc of the nucleus but still within the BCG region where the light from the compact component (S1) dominates over the light from the extended (S2) component (see Fig.\,\ref{fig:1D_dist}). We then measured the GCLFs within three regions: an annulus of 40 to 100\,kpc which covers the outer halo of the BCG, and two annuli covering ICL at 100 to 200\,kpc and 200 to 500\,kpc from the BCG.

We extended the GC-candidate sample to include all sources with $22.7<\IE<28.7$ that matched the compactness criteria laid out in Sect.\,\ref{sec:GC_detection}. The luminosity distributions of these sources in each of the four annuli of interest are shown in Fig.\,\ref{fig:GCLF_non-bkg}. These sources include contamination from background sources as well as GC-candidates, so we measured the luminosity function of the background sources and subtracted it from the luminosity function of the GC candidates within the four annuli of interest.

From the radial profile shown in Fig.\,\ref{fig:GC_1d}, we detected an excess of GCs out to a semi-major axis of $a=600$\,kpc. We therefore used the region that lies beyond the ellipse with $a=600$\,kpc to select background sources. This background region covers 609\,arcmin$^2$, and encompasses 31\,911 objects selected by our compactness criteria laid out in Sect.\,\ref{sec:GC_detection}. The luminosity function of these background sources is shown as the grey histogram in Fig.\,\ref{fig:GCLF_non-bkg}. We estimated that fewer than $4\%$ of the objects selected in this background region are ICGCs since Fig.\,\ref{fig:GC_1d} shows that the surface number density of GCs beyond this semi-major axis is less than 2\,arcmin$^{-2}$. 

We estimated the uncertainty of the luminosity function of background sources within each of our four annuli separately. For each annulus, we bootstrapped the sources from the background region 100 times but limited the sample size in each bootstrap to the expected number of contaminating sources in the annulus (determined as the annulus area times the surface density of the background sources). The luminosity distribution and uncertainty of the background sources within each annulus are defined as the mean and standard deviation of these 100 realisations. These luminosity functions of the background sources were then subtracted from the luminosity function within each annulus.

The resulting background-subtracted GCLFs measured at $20 < a/ {\rm kpc}\leq40$, 40 to 100\,kpc, 100 to 200\,kpc, and 200 to 500\,kpc are shown in  Fig.\,\ref{fig:GCLF1}. The uncertainty is defined as the standard deviation of 100 realisations of the background in each luminosity bin added in quadrature to the Poisson uncertainty of the number of GC candidates in each luminosity bin of the annulus. However, we note that our background region is significantly smaller than the 200--500\,kpc annulus, which makes the uncertainties on the background in this area unreliable. To account for this, we multiplied our background uncertainty for the 200--500\,kpc annulus by 1.56, which is the ratio of the area of the 200--500\,kpc annulus to the background region. 

Before exploring the ICGC luminosity function, we first check whether the GCLF we measured within NGC\,1275 (i.e. $20 < a/ {\rm kpc}\leq40$) matches the expected shape found in previous studies. GCs within giant elliptical galaxies have a luminosity function that follows a Gaussian form \citep[e.g.][]{Jordan07}. At the distance of Perseus, this Gaussian is expected to peak at a magnitude of $\IE\sim26.3$ with a $\sigma\sim1.3$\,mag, which is consistent with measurements of NGC\,1275 by \citet{Harris2017}. 

We performed a least-squares fit of a Gaussian to this distribution, allowing the mean, amplitude, and width of the Gaussian to be free parameters. The detection of point-like sources within this annulus is complete to $\IE=25.8$ so we limited the fit to $22.7<\IE<25.8$. We derived a turn-over magnitude of $\IE =26.24\pm 0.12$ and $\sigma =1.39\pm 0.05$\,mag, which is consistent with empirical expectations \citep{Jordan07,Villegas2010,Harris14} and previous measurements of the GCLF within this galaxy \citep{Harris2017}. The goodness-of-fit of the Gaussian model is given by the reduced $\chi^2$ metric. In a fit with 15 degrees of freedom, the reduced $\chi^2$ of $1.01$ means the data are consistent with the Gaussian model. 

\begin{table}[]
 \caption{Results of the single Gaussian fits to the GCLF in each annulus.}
    \centering    
    \renewcommand{\arraystretch}{1.4}
    \begin{tabular}{|c|ccc|}
    \hline\hline
    Annulus  [kpc]& $\mu$  [mag]& $\sigma$  [mag] &  reduced $\chi^2$ \\ 
    \hline 
    20--40\, &    $26.24\pm{0.12}$ & $1.39\pm{0.05}$ & 1.01\\
     40--100 & $26.24^*$ & $1.33\pm{0.01}$ &  1.38  \\ 
     100--200   & $26.24^*$ & $1.18\pm{0.02}$  & 2.60 \\ 
     200--500  & $26.24^*$& $1.05\pm{0.03}$ &  2.28\\
    \hline\hline
    \end{tabular}
    \tablefoot{Parameters marked with an asterisk were fixed during the fit.  The goodness-of-fit of the model is given by the reduced $\chi^2$ metric.}
    \label{tab:GCLF}
\end{table}

 \begin{table*}[t]
 \caption{Results of the single and double Gaussian fits to the GCLF.}
    \centering    
    \renewcommand{\arraystretch}{1.4} 
    \begin{tabular}{|c|ccccc|}
    \hline\hline
    Annulus [kpc] & $N_{\rm tot}$ & $N_2/N_{\rm tot}$ & $\sigma_1 $ [mag] ($\mu_1\equiv26.24$)& $ \sigma_2 $  [mag] ($\mu_2\equiv26.54$)   &  reduced $\chi^2$ \\ 
    \hline 
     20--40 &$6100 \pm 400$& -- &$1.39\pm{0.05}$ & -- & 1.01\\
     40--100& $12\,600\pm800~~$& $0.25\pm0.05$ & $1.46\pm{0.04}$& $0.84\pm{0.07}$ &  0.93  \\ 
     100--200 &  $17\,200\pm1400$ & $0.44\pm0.08$  & $1.48\pm{0.07}$ & $0.80\pm{0.05}$  & 1.06 \\ 
     200--500  &  $34\,100\pm2300$ & $0.51\pm0.06$ & $1.42\pm{0.07}$ & $0.72\pm{0.04}$ &  1.54\\
     \hline
     20--500 &$70\,000\pm2800$  & $0.40\pm0.04$ &  & &  \\
    \hline\hline
    \end{tabular}
    \tablefoot{Double Gaussian fits were performed on the GCLF in the three outer annuli, while a single Gaussian fit was applied to the GCLF in the 20--40\,kpc annulus. $N_{\rm Tot}$ refers to the total number of GCs in each annulus extrapolated from $\IE < 26.2$ assuming the GCLF follows the Gaussian (or double Gaussian) model. The mean of each Gaussian was fixed to the listed $\mu$. $N_2/N_{\rm tot}$ lists the fraction of GCs in the narrow Gaussian component in each annulus. $\sigma_1$ and $\sigma_2$ are the standard deviations of the Gaussian distributions fixed to a mean of $\IE=26.24$ and $\IE=26.54$, respectively. The goodness-of-fit of the model is given by the reduced $\chi^2$ metric having 16 degrees of freedom.}
    \label{tab:GCLF2}
\end{table*}

Having established that our GC selection and statistical background subtraction are robust, we then measured the GCLF of the intracluster region, which has never been explored to this depth at large radii from NGC\,1275. We performed a least-squares fit of a Gaussian to the luminosity distributions at 40 to 100\,kpc, 100 to 200\,kpc, and 200 to 500\,kpc. We limited the fitting to the slightly wider range $22.7<\IE<26.2$, since the GCs in the intracluster region are complete to fainter magnitudes than the GCs in the BCG, although this does not significantly affect our results. We find that the fits did not converge when we allowed the mean, amplitude, and width of the Gaussian to be free parameters. Therefore, we fixed the mean of the Gaussian to be the same as that measured in the 20--40\,kpc annulus ($\IE=26.24$) and solved only for the width and amplitude of the Gaussian. 

The best-fit $\sigma$ of the Gaussian distributions range between $1.05<\sigma/{\mathrm{mag}}<1.33$, and although the fits appear reasonable (see Fig.\,\ref{fig:GCLF1}a), the reduced $\chi^2$ in Table\,\ref{tab:GCLF} is high. In each of the fits to the three outer annuli, there are 18 degrees of freedom, so the 95\%, 99\%, and 99.9\% critical values of the $\chi^2$ distribution correspond to a reduced $\chi^2$ of 1.60, 1.93, and 2.35, respectively. Thus the data in the 40--100\,kpc annulus is consistent with the Gaussian model, but the model is inconsistent with the data of the 100--200\,kpc and 200--500\,kpc annuli to $>99$\% confidence. 
In each annulus, there is a systematic deficit of GC candidates at $\IE \sim25$ and an excess at $\IE \sim26.1$ compared with the best-fitting Gaussian model which suggests that a single Gaussian is not a good model for the luminosity function in the three outermost annuli.

Since the data are incompatible with a single Gaussian model, we attempted to fit the three outer annuli with a slightly more complex model. We constructed a new model, motivated by observations of the GCLF, consisting of two Gaussian distributions: one with a turn-over magnitude fixed to the BCG value ($\IE=26.24$), and the other fixed\footnote{The model failed to converge if we left the second turn-over magnitude to be a free parameter.} to 0.3\,mag fainter ($\IE=26.54$), which is the measured turn-over magnitude in dwarf galaxies \citep{Villegas2010, Carlsen2022}. The widths ($\sigma_1$ and $\sigma_2$) and amplitudes of the Gaussian distributions were allowed to vary\footnote{We also attempted to have both Gaussian distributions centred on $\IE=26.24$, and allowed the widths and amplitudes to vary, but this still left an excess of GCs above the model at magnitudes fainter than the completeness limit.}.  
The results of the least-squares fits are shown in Table\,\ref{tab:GCLF2} and Fig.\,\ref{fig:GCLF1}b. The fits to the double Gaussian model have 16 degrees of freedom, so the data are consistent with the model in all annuli (i.e. the reduced $\chi^2$ is within the 95\% critical value of the $\chi^2$ distribution for all fits).

In all annuli, the standard deviations of the Gaussian fixed to $\mu_{\IE}=26.24$ are very similar ($\sigma_1\sim1.44$) and are consistent with the expectation for massive elliptical galaxies. In addition, the standard deviations of the Gaussian fixed to $\mu_{\IE}=26.54$ are also very similar and are consistent with the expectation for dwarf galaxies ($\sigma_2\sim0.8$) in all annuli. We reiterate that the annuli were fit independently and, therefore, the similarity in the standard deviations of the best-fit models in each annulus lends credence to the fidelity of our chosen model. 

The population of GCs within the narrow GCLF comprises $40\pm4$\% of the total GC population in the BCG and intracluster regions. However, there is a strong gradient in the fraction of narrow component to total GCs: $25\pm5$\% in the 40--100\,kpc annulus, increasing to $51\pm6$\% in the 200--500\,kpc annulus. 

The background subtraction becomes increasingly important with increasing radius from the BCG, which is similar to the increase in the fraction of narrow component to total GCs. We therefore examined whether the narrow GCLF component may be due to an erroneous feature in the derived background luminosity function that was becoming more influential at large radii. If this were the case, we would expect the number of GCs in the narrow GCLF component to increase in proportion to the importance of the background subtraction, which is proportional to the ratio of the annulus area to the background region area. This ratio is 0.05, 0.19, and 1.56 for the annuli 40--100\,kpc, 100--200\,kpc, and 200--500\,kpc, respectively. We find, however, that the number of GCs in the narrow component is $[0.05, 0.14, 0.31]\times 47\,770$ in the 40--100\,kpc, 100--200\,kpc, and 200--500\,kpc annuli, respectively\footnote{To obtain these numbers we only counted the number of GCs in the unmasked regions of the annulus, and hence they differ from the numbers quoted in Table\,\ref{tab:GCLF2} for which we interpolated over the masked regions.}. There are almost four times fewer narrow-component GCs in the outermost annulus than expected if this feature was caused by incorrect background subtraction. We therefore surmise that the GCs in the narrow Gaussian component of the luminosity function are not due to an erroneous feature in the background luminosity function.

Having measured the luminosity function of the GCs, we next corrected our estimates of the total number of GCs for incompleteness. In Table \ref{tab:GCLF2}, we list the extrapolated number of GCs expected in each elliptical annulus around the BCG and provide uncertainties from the errors on the 2-component Gaussian fit to the luminosity function. In total, we find the BCG+ICGC hosts $70\,000 \pm 2800$ GCs between $20<a/{\rm kpc}\leq500$. If we conservatively define the intracluster region as the region beyond 100\,kpc of the BCG nucleus, then the number of ICGCs is $51\,300\pm2700$, which is 73\% of this system's total GC population.

\subsection{Radial colour profile of the BCG+ICL}\label{sec:colours}

Radial colour gradients provide valuable constraints on the formation processes of galaxies and, in this case, the BCG and ICL \citep{Zibetti2005, MT14, DeMaio2015, Morishita2017, Mihos2017, DeMaio2018, Contini2019}. 
Different formation mechanisms are imprinted in the stellar populations of galaxies and, thus, in their radial colour profiles.

The radial colour profiles in $\YE-\JE$ and $\JE-\HE$ have been obtained from the surface brightness profiles that were consistently measured using the BCG+ICL elliptical contours defined using the \IE\ image. Although the BCG+ICL radial surface brightness profiles can be measured up to 600\,kpc, the greater uncertainties at large radii mean that colour measurements are unreliable at those radii. Therefore, we limited the colour profiles to the inner 200\,kpc. We did not use \IE\ to measure the colours as cirri can strongly contaminate the signal, even after we attempted to subtract it. Cirri were not subtracted from any of the NIR images, but we expect a minimal level of contamination\footnote{The Galactic cirri that contaminate the \IE\ image are not seen as strongly in the NIR (see Fig.\,\ref{fig:rgb}). This agrees with the modelling of the Galactic cirri in \citet{Zhang2023}, where the scattered light peaks around $\sim7000$\,\AA\ and drops rapidly at redder wavelengths. \Euclid\ will provide key information about the properties of the cirri in the NIR \citep[see also][]{Roman2020}.} and thus a minimal influence on the BCG+ICL colours. 

\begin{figure}
    \centering
    \includegraphics[width=\linewidth]{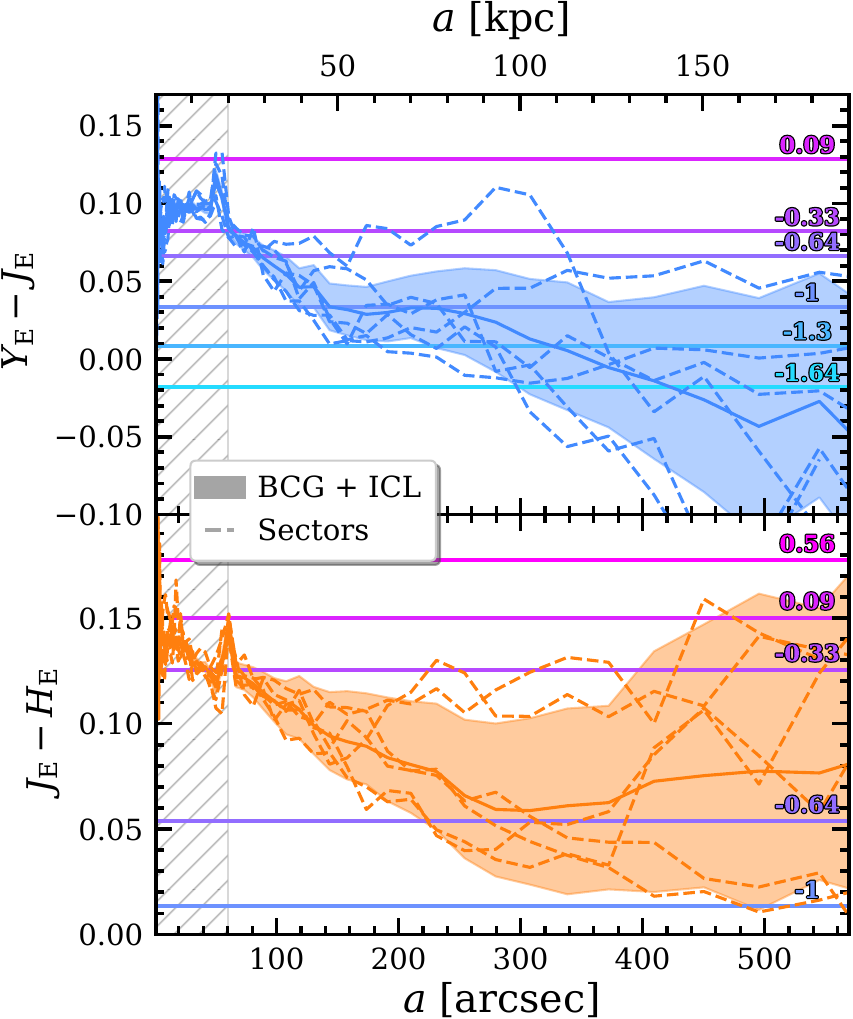}
    \caption{Radial colour profiles (solid non-horizontal lines) of the BCG+ICL. The upper (lower) panel shows the $\YE-\JE$ ($\JE-\HE$) colour. The dashed lines represent colours measured in six different sectors. The colour uncertainties, shown by the shaded areas, are propagated from the background uncertainties (see Appendix\,\ref{sec:background_uncertainties}). 
    The horizontal lines are \citet{Bruzual2003} models at a fixed age of 10 Gyr and different [Fe/H] metallicities (as labelled). The hatched area indicates the region with uncertain colours due to persistence residuals.}
    \label{fig:colour-profiles}
\end{figure}

Figure \ref{fig:colour-profiles} shows the $\YE-\JE$ (top panel) and $\JE-\HE$ (bottom panel) radial colour profiles for the BCG+ICL of Perseus. The error in the colour profiles, represented by the shaded regions, is the quadratic sum of the errors in the individual radial surface brightness profiles. 
The greyed-out area corresponds to radii where there was strong persistence in the individual NISP exposures. Residuals from the imperfectly modelled persistence contaminate this region and therefore the measured colours are not indicative of the true BCG colours. 

Both colour profiles exhibit negative gradients within $1\sigma$ confidence out to $a=50$\,kpc and $a=90$\,kpc for the $\YE-\JE$ and $\JE-\HE$ colours, respectively. Beyond these distances, uncertainties due to background inhomogeneities and residual ICL in the outermost annuli increase (see Appendix~\ref{sec:background_uncertainties} for details). 
We also measured colour profiles using the same elliptical annuli as before but subdivided into six different sectors (dashed lines in Fig.\,\ref{fig:colour-profiles}), each with an opening angle of $60\degr$. The scatter among these sectors is consistent with the background uncertainties.

To interpret the colour profiles, we assume a constant age for the BCG and ICL. This assumption is justified by studies of nearby clusters which show that the age of the ICL is old \citep[$\gtrsim 10$ Gyr, ][]{Williams2007, Coccato2010, Mihos2017, Gu2020}, and that the expected light-weighted ages are typically between 9 and 13\,Gyr \citep{Coccato2010, Greene2015, Gu2020}. The colours of such an old stellar population change little within this age range \citep{Bruzual2003,Vazdekis2016}, especially in the NIR colours (as we show in Appendix \ref{app:models}).

Under this assumption, the negative colour gradient is likely due to negative metallicity gradients: the stellar population of the BCG+ICL becomes more metal-poor towards the outskirts. A strong gradient in metallicity has also been observed more directly using resolved red-giant-branch stars \citep{Lee2016,Hartke2022} in the Brightest Group Galaxy M\,105. \cite{Lee2016} found that the gradient results from a second, distinct metal-poor component whose relative amplitude increases outwards.
We compare the measured colours to predictions from single stellar population models \citep{Bruzual2003} with varying metallicities [Fe/H] (horizontal lines in Fig.\,\ref{fig:colour-profiles}). The measured colours are consistent with slightly subsolar metallicities ${\rm [Fe/H]}\sim-0.3$ at $a=20$\,kpc and strongly subsolar metallicities ranging from ${\rm [Fe/H]}\sim-0.6$ to $-1.0$ at $a=80$\,kpc. While being consistent near the centre, we find a $3\sigma$ discrepancy at $a=50$\,kpc between the two metallicities inferred from the different colours, with the $\YE-\JE$ colour indicating a significantly lower metallicity of ${\rm [Fe/H]}\sim-1.03\pm0.17$ compared to ${\rm [Fe/H]}\sim-0.47\pm0.08$ for $\JE-\HE$. This suggests a radial variation in the stellar populations that is not captured by our simple model. It is unlikely that photometric zero point offsets can explain this behaviour because we measure a consistent metallicity near the galaxy centre. Furthermore, it is unlikely that background offsets could produce this behaviour because we find consistency in the scatter of the sector profiles with the total uncertainties. Further investigations using more recent stellar population models, composite stellar populations to test the dual-halo scenario \citep{Lee2016}, and larger BCG samples to assess the robustness are needed to understand these trends.

\section{Discussion}
\subsection{A new window opens to study the ICL}
The results presented here show the extraordinary potential of \Euclid\ to observe ICL and ICGCs in nearby galaxy clusters. For the first time, we have mapped the distribution of the ICL and ICGCs in Perseus out to a radius of 200--600\,kpc (up to $\sim\frac{1}{3}r_{200}$, depending on the parameter) from the BCG, and measured their properties. Note that the analysis presented here was possible despite the significant Galactic cirri contamination in the field of view. 

\Euclid\ has enabled us to improve these measurements from previous studies in Perseus \citep{Harris2017, Harris2020, Kluge2020} for three main reasons: it has a large field of view relative to other optical and NIR space telescopes due to the Korsch three-mirror anastigmat optical design, it is diffraction limited with tight stray light control which yields a compact and highly stable PSF, and it has high spatial resolution in the \IE\ image which results in a large point source depth.  

The large field of view enables us to model both bright central galaxies iteratively to identify each galaxy's contribution to the light or GCs in the intracluster region. We have shown that NGC\,1272 is relatively compact and only minimally contributes to the diffuse light or GCs in the intracluster region. The large field of view also allowed us to map the shape of the GC and ICL distributions out to 350\,kpc, which is 150\,kpc further than typically achieved using ground-based observations \citep{Kluge2020, Montes2021}. Previous works on massive clusters at similar distances, \citep[e.g.][]{Peng2011, Harris2020} were only able to study the ICGCs in small pockets, constrained by the need to use space-based images to identify GCs, but limited by HST's and JWST's small field of view. 

While the 0.7\,deg$^2$ field of view of the \Euclid\ ERO enables us to identify intracluster stellar components out to 600\,kpc from the BCG, it also limits our analysis as we need to estimate the background from this same field of view and the intracluster stellar populations may extend further (see Appendix \ref{sec:background_constant}). The analysis of galaxy clusters in the EWS \citep{Scaramella-EP1} will not be limited by the field of view since that survey will provide nearly continuous coverage of a third of the sky. It will therefore be possible to trace the ICL and ICGCs out to the splashback radius -- if they indeed exist out to this distance (\citealt{Gonzalez2021}; Bellhouse et al. in prep.).

The compact and stable PSF allows us to reliably remove the outer profile of bright stars using only a hundred stars located within the field of view. This is most dramatically demonstrated by the bright star that lies between NGC\,1275 and NGC\,1272 and severely contaminates the ICL in this region.  

The detection of the ICGCs is possible only thanks to the large point source depth of \Euclid, which results from the combination of the compact PSF and high spatial resolution of the VIS instrument that produced the \IE\ image (see Fig.\,\ref{fig:cfht_euclid_zoom}, middle panel). Unfortunately, the undersampling of the PSF for the NIR data prevented us from identifying the NIR counterparts to the \IE\ sources (see Fig. \ref{fig:cfht_euclid_zoom}, right panel) and measuring their colours. While ICGCs have previously been detected in Perseus \citep{Harris2017,Harris2020}, the \Euclid\ ERO have enabled the first assessment of their global distribution. These data allow a direct comparison to the ICL, and quantification of the GCLF as a function of distance from the BCG, which enables the first accurate measure of the total number of ICGCs in this cluster.  Euclid Collaboration: Voggel et al.~(in prep.) 
has shown that GCs will be detectable in the EWS in galaxies up to a distance of 100\,Mpc ($z\sim0.023$). Extrapolating using the \citet{yang07} group catalogue from the SDSS spectroscopic data release 7, we estimated that there are 90 groups and clusters in the 14\,000\,deg$^{2}$ EWS footprint with a total mass greater than $10^{13}$\,M$_{\odot}$ for which we can perform similarly detailed ICGCs studies.

\subsection{Evidence for a distinct intracluster stellar population}  \label{sec:tracers}

The surface brightness profiles of the BCG+ICL (Fig. \ref{fig:1D_dist}) are adequately fit by a combination of two S\'ersic profiles: one compact S\'ersic (S1) and one extended (S2). The outer component, S2, dominates the light profile beyond $a\sim60$ kpc, where the diffuse light follows the GCs. It is tempting to associate these components with the physical systems of the BCG (S1) and the ICL (S2).  This division implies that $48\pm12$\% (\IE) to $67\pm7$\% (\HE) of the light of the BCG+ICL system is from the ICL. \citet{Vazdekis2016} show that the stellar mass-to-light ratio, $M/L$, in \HE\ is approximately 1.0 for a wide range of metallicities. Under these assumptions, 67\% of the mass of the BCG+ICL system lies in the ICL.

On the other hand, there has long been a debate as to whether a physically meaningful division can be made between the BCG and the ICL from radial surface brightness profiles alone\footnote{A distinction between the BCG and the ICL stellar populations can be more robustly made in both simulations and observations when considering their kinematics \citep[e.g.][]{Dolag2010, Longobardi2013, Longobardi2015, Hartke2022}, but these kinetically distinct populations overlap in space.}, and if so, where this division is physically located \citep[e.g.][]{gonzalez05,Dolag2010,Bender2015,Remus2017,kluge21,Contini2022,Brough2024}. Simulations show that BCGs are made up of as much as 70\% of accreted, ex-situ material \citep[e.g.][]{Pillepich2018}, so the BCG and ICL appear to have a similar origin. Therefore separating the components using radial profiles might not be physically meaningful.

While we are reluctant to define a hard boundary between the BCG and ICL components from the radial profiles alone, we have presented further evidence that the BCG and ICL of Perseus should be considered distinct systems of stars, albeit sometimes overlapping.
\begin{itemize}
    \item We showed that the ICL component is not centred on the BCG, but rather at a point $\Delta\,{\rm RA}\approx60$\,kpc west of the BCG (see Fig. \ref{fig:offsets}). This implies the systems of stars that make up the BCG and ICL are not centred at the same location.
    \item The ellipticity of the BCG+ICL changes with radius. It rises rapidly from the BCG core up to $a=150$\,kpc, and flattens thereafter (Fig. \ref{fig:ellp_PA}). This suggests that the orbital paths of the stars in the BCG and ICL components differ.
    \item The colour of the BCG+ICL, shown in Fig.\,\ref{fig:colour-profiles}, gradually becomes bluer at least until $a=50$--$90$\,kpc.
    This suggests the BCG and ICL comprise stellar populations with different metallicities.
\end{itemize}  

\begin{figure}
    \centering
    \includegraphics[width=\linewidth]{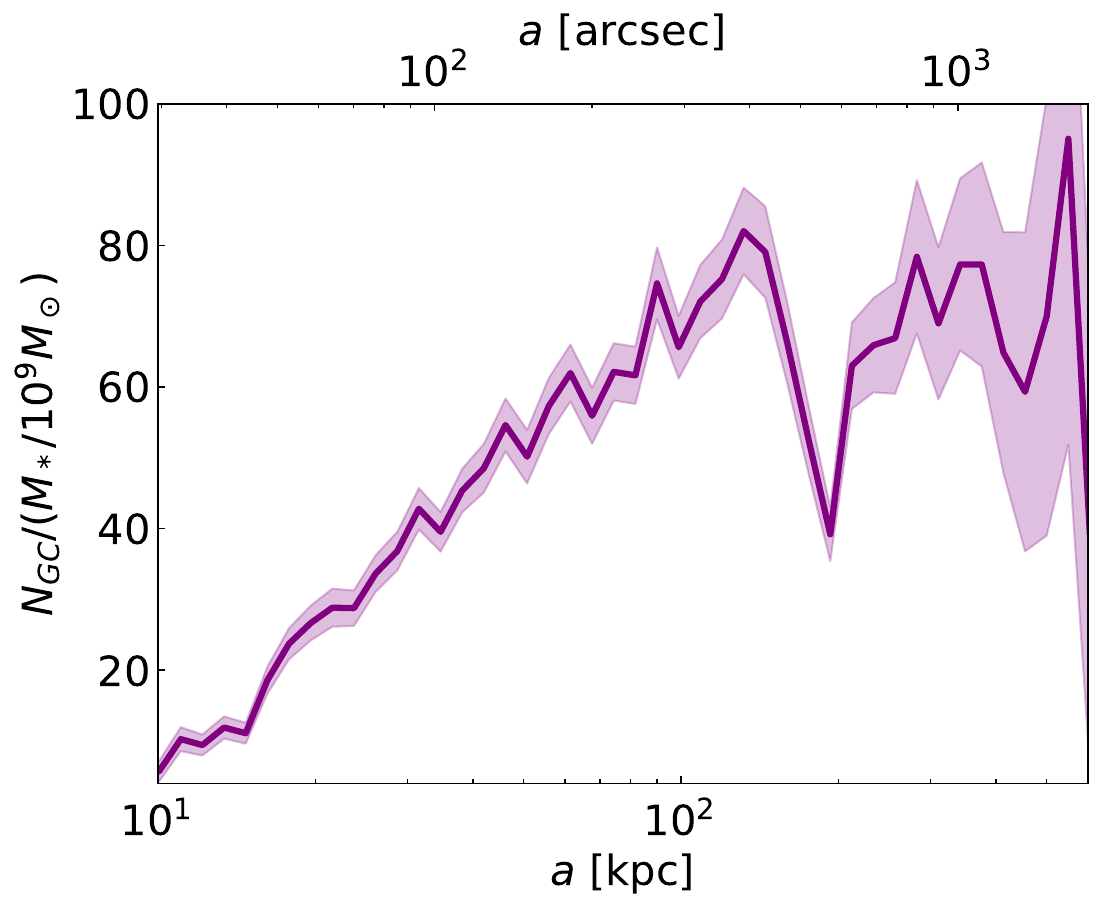}
    \caption{Radial profile of the specific frequency of GCs in NGC\,1275+ICGCs, defined as the ratio of GCs to stellar mass in units of $10^{9}\,{\rm M}_\odot$. The specific frequency within the central 10\,kpc is consistent with other BCGs, but then increases to a value typically found only in dwarf galaxies or the outskirts of massive galaxies.}
    \label{fig:T}
\end{figure}

There is also evidence that the GCs in the BCG are distinct from the ICGCs. We illustrate this through the radial profile of the specific frequency of GCs \citep[defined as the parameter $T$ by][]{Zepf1993}. The specific frequency was calculated as the ratio of the GC radial surface density (corrected to account for incompleteness using the GCLFs measured in Sect.\,\ref{sec:GCLF_res}) to the BCG+ICL surface mass density (calculated from the \HE\ surface brightness profile assuming $M/L=1.02\,{\rm M_\odot/L_\odot}$; \citealt{Vazdekis2016}). 

In Fig.\,\ref{fig:T}, we show that the specific frequency increases from 10 GCs per $10^{9}$\,M$_{\odot}$ at 10\,kpc to $\sim80$ GCs per $10^{9}$\,M$_{\odot}$ at $a\sim130$\,kpc. 
An increasing GC specific frequency within annuli was also discovered recently in the BCG of the Hydra I cluster, starting from $S_N=2$ in the centre and reaching $S_N=9$ at 200\,kpc radius \citep{Spavone2024}. Beyond $a=60$\,kpc, the intracluster specific frequency in the Perseus cluster stabilises in the range $60\lesssim T\lesssim 80$ until $a=600$\,kpc.
The different specific frequencies implies that the GCs in the inner ($a\lesssim 60$\,kpc) and outer ($a\gtrsim 60$\,kpc) parts of the BCG and intracluster space trace different stellar populations, and perhaps had different origins. 

Catalogues of the specific frequency of GCs are provided by \citet{Peng2008}, \citet{Liu2019}, and \citet{Carlsen2022}, who show that $T=10$ is consistent with the GC content of galaxies with masses $M>10^{11}$\,M$_{\odot}$. Larger specific frequencies are associated with dwarf galaxies with $M<10^{9}$\,M$_{\odot}$, and are inconsistent with galaxies in the mass range $10^{9.7}<M/{\rm M}_{\odot}<10^{11}$, which have the lowest values of specific frequency.   

The specific frequency of GCs commonly rises with radius in elliptical and S0 galaxies, similar to what we observe at $a<60$\,kpc, as a consequence of the radial profile of light being steeper than the profile of GC number densities \citep{Hargis2014}. This is interpreted as evidence that the outskirts of giant ellipticals and S0s were built from the gradual accretion of dwarf galaxies -- a process that also leaves radial gradients in the age and metallicity of the stellar population. On the other hand, a constant $T\sim60$--$80$ GCs per $10^{9}$\,M$_{\odot}$ at $a>60$\,kpc suggests that the ICGCs and ICL at these large radii are well-mixed.

\subsection {Intracluster stars as tracers of the dark matter halo of Perseus}
\label{subsec:ICLDM}
Recent work by \citet{MT19} showed that ICL traces the global distribution of dark matter extremely well for six clusters observed with HST. This sparked a strong interest in using the ICL to study massive dark matter halos. This work has been followed by a plethora of simulations that have confirmed that the global ICL distribution is expected to trace the underlying dark matter \citep{Alonso-Asensio2020,Yoo2022,ContrerasSantos2024}.
Confirming this, the ICL has been shown to be a reliable tracer of the position angle of the hosting dark matter halo as well as its spatial offset with respect to the BCG \citep{kluge21}.

The shape of the ICL and ICGC distribution in Perseus is consistent with an ellipse of $\epsilon=0.4$, which agrees with the predictions of triaxial halo ellipticity from $\Lambda$ cold dark matter \citep{Oguri2010}. This ellipticity matches the galaxy distribution but deviates from the large-scale distribution of the intracluster medium traced by X-rays (see Fig.\,\ref{fig:xrays}) and the Sunyaev-Zeldovich signal in the {\it Planck} data, which displays a rounder shape. \citet{Ettori1998} finds the X-ray emission has an ellipticity of $\epsilon=0.06\pm0.04$ at $a=80$\,kpc and only $\epsilon=0.22\pm0.07$ at $a=480$\,kpc from the BCG. As a collisionless tracer, the ICL is a more faithful tracer of the dark matter distribution than the collisional X-ray emitting gas \citep{MT19} and is therefore likely to be a better direct tracer of the dark matter halo's ellipticity. \citet{Pulsoni2021b} used simulations to show that, in situations such as BCGs that are surrounded with a large fraction of accreted material (i.e. the ICL), there is a strong coupling between the outer stellar halo and the dark matter halo. They conclude that the outer profile of stellar halos, such as the ICL, can be used to infer the intrinsic shape and principal axes of the dark matter halo.
The ICL measured in the EWS will provide ellipticity measurements of the ICL for many thousands of clusters which will enable a statistical exploration of the ellipticity distribution of dark matter halos to test dark matter models. 

Since light at \HE\ wavelengths closely traces the stellar mass, we derived the stellar mass density profile for Perseus using Eq.\,(1) in \citet{MT14} assuming a $M/L = 1.02$ \citep{Vazdekis2016}. We measured the slope $\alpha_{\rm 2D}$ of the BCG+ICL from 20\,kpc to 470\,kpc, calculated as $\delta{\mathrm{log}}\,\Sigma_* / \delta{\mathrm{log}}\,a$, obtaining $-1.47\pm 0.02$. This value translates into a 3D slope of $\alpha_{\rm 3D}\sim-2.47$, using Eq.\,(5) in \citet{Stark1977}, which agrees with measurements of the radial slope of the BCG+ICL stellar mass density in other clusters \citep[][]{MT18, MT22}. As shown in Fig.\,\ref{fig:GC_1d}, the ICGCs of Perseus also follow this slope \citep[see also][]{Reina-Campos2023, Kluge2023a}. 

\citet{Pillepich2014, Pillepich2018} found that within high-mass simulated halos, the 3D slope of the stellar halo becomes as shallow as that of the dark matter with  $-2.6<\alpha_{\rm 3D}<-2$. This implies that the slopes of both the ICL and the ICGCs measured here are similar to that of the dark matter and they are good luminous tracers of the dark matter in clusters of galaxies.
However, we note that we do not know the exact slope of the dark matter halo of Perseus, and studies of other clusters have shown that the slope of the stellar light of clusters can be shallower than the slope of the dark matter \citep[e.g.][]{Diego2023}.

\subsection{Interpretation of the spatial offset between the BCG and the ICL or ICGCs of the Perseus cluster}

\label{subsec:cluster centre}
\begin{figure}
    \centering
    \includegraphics[width=\linewidth]{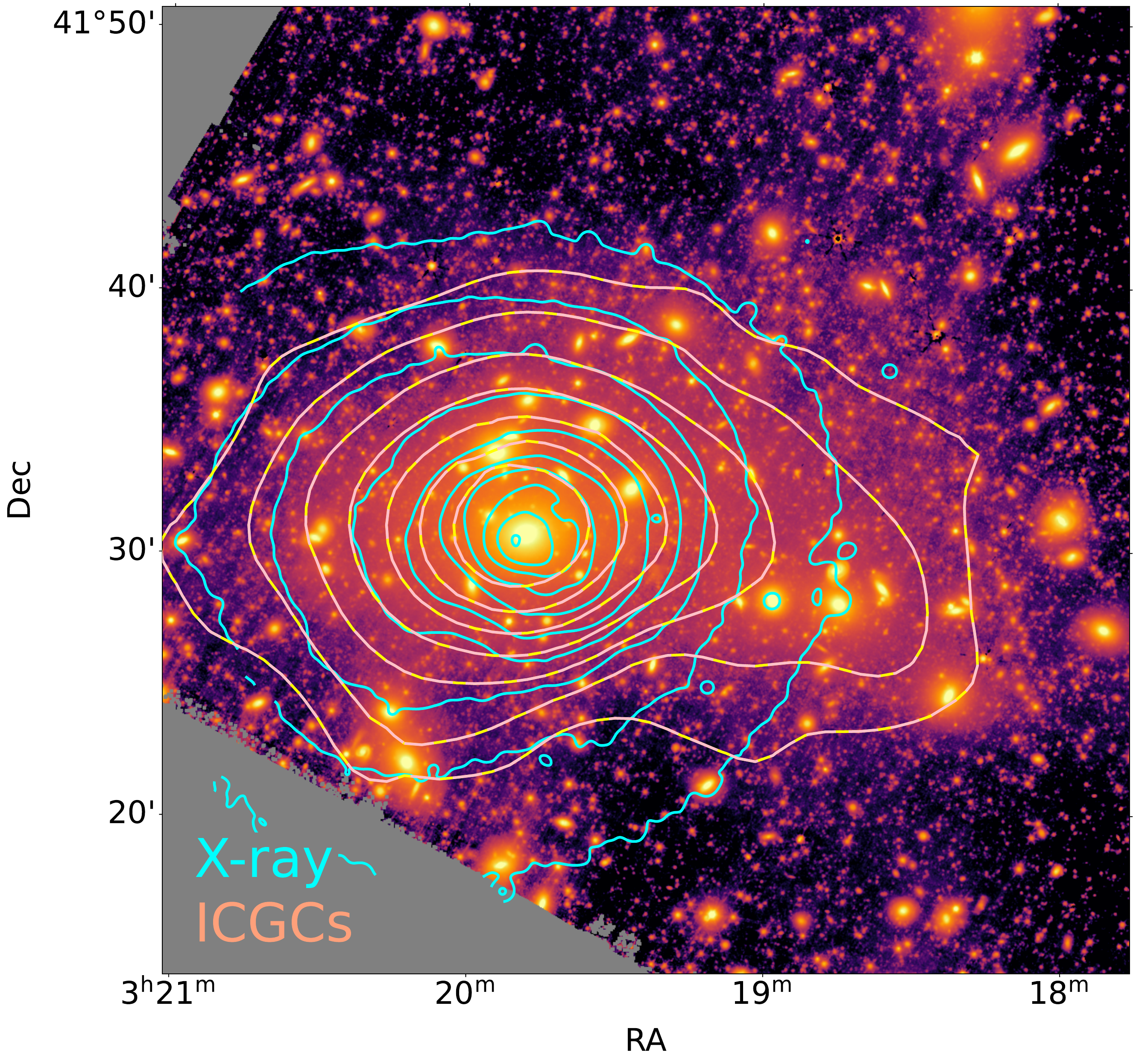}
    \caption{Comparison between the X-ray, the ICL and the ICGCs of Perseus. The background image is the \HE\ image with the light from NGC\,1272 modelled and subtracted. Cyan contours show the X-ray emission of the intracluster medium observed by {\it Chandra} \citep{Sanders2016}, adaptively smoothed to reach a $S/N=15$, and further smoothed with a Gaussian kernel of $\sigma=12$\arcsec. Orange contours show the distribution of GCs, not including those from NGC\,1272, binned and smoothed with a Gaussian kernel on scales of 40\arcsec.}
    \label{fig:xrays}
    
\end{figure}

The centre of a cluster can be estimated using various proxies: the X-ray centroid, the location of the BCG, the centroid of the large-scale dark matter distribution determined using weak gravitational lensing, or the centroids of the ICL and ICGC distributions.  
In an idealised, fully relaxed cluster all of these proxies for the cluster centroid should agree, and any disagreement between these proxies can imply that the cluster is not relaxed.

Perseus is known to be a fairly relaxed cluster \citep{Simionescu2012}, but several studies suggest that Perseus is not completely virialised \citep{andreon94, churazov03} and there are signs that the core is merging with a smaller system \citep{Fabian2011, HyeongHan24}. Indeed, there is disagreement between many of the proxies for the Perseus cluster's centroid. Recent weak gravitational lensing analysis of the Perseus cluster \citep{HyeongHan24} report a westward offset of the total mass distribution from the BCG. They also report a westward mass bridge connecting the BCG to NGC\,1264, which lies 430\,kpc west of the BCG. On the other hand, the gaseous intracluster medium, as traced by the isophotes of the X-ray emission, is offset to the east of the BCG, as shown in Fig.\,\ref{fig:xrays} and originally discussed in \citet{churazov03}. Here we report on a misalignment of $\Delta\,{\rm RA}\sim60$\,kpc (3\% of $r_{200,{\rm c}}$) between the BCG and the centroid of the large-scale ICL and ICGCs distributions (especially the isophotes and iso-denisty contours at $a>200$\,kpc), such that the ICL and ICGCs are centred to the west of the BCG as shown in Fig.\,\ref{fig:xrays}.

Spatial offsets between the BCG and ICL centroids are commonly observed \citep{Kluge2020}, and many BCGs display a velocity offset from the mean velocity of the satellite galaxy distribution of a cluster \citep[e.g.][]{Gerhard2007,Longobardi2015, Bender2015, Ye2017,Barbosa2018}. This velocity offset can result in a spatial offset between the BCG and dark matter halo centroid that may be as large as 3\%\,$r_{200,{\rm c}}$ for massive clusters like Perseus \citep{vandenbosch2005}.

Simulations by \citet{vandenbosch2005} and  \citet{DePropris2021} postulate that there are two explanations for the offset: the BCG is offset from the centre of the halo (in velocity and/or position) after being disturbed by a merger (the non-relaxed galaxy scenario), or the dark matter halo core is offset by the merger and oscillates until it relaxes (labelled as the non-relaxed halo scenario). Standard cold dark matter models predicted cuspy halos, which would cause the BCG to rapidly return to equilibrium , but \citet{Harvey2017} shows that residual wobbling of the BCG may be a long-term phenomenon after the merger if the dark matter halo is cored.

In the case of NGC\,1275, we find no evidence for a velocity offset along the line of sight: \citet{Aguerri2020} find that the peak of the velocity distribution of all cluster members is 5258\,km\,s$^{-1}$, which is in good agreement with the velocity of NGC\,1275  \citep[5264\,km\,s$^{-1}$,][]{Huchra1999}. Therefore it is unlikely that NGC\,1275 is oscillating around or travelling towards the halo centre unless it is at the turnaround point of the oscillation or moving in the plane of the sky.

The velocity distribution of the dwarf galaxies, however, provides some evidence that the core of the dark matter halo, traced by the ICL and ICGCs, is not yet relaxed and may possibly be oscillating. \citet{EROPerseusDGs} find that the dwarf galaxies in Perseus trace the distribution of the ICL and share a similar centroid as the large-scale ICL and ICGCs. It is therefore possible that the dwarf galaxies, ICGCs, and ICL share the same kinematic distribution as well. \citet{Aguerri2020} find the peak velocity of the Perseus dwarfs is 5049\,km\,s$^{-1}$, that is, offset by $-215$\,km\,s$^{-1}$ from the BCG velocity. Therefore, it is plausible that the spatial offset between the BCG and ICL is because the core of Perseus oscillating after a recent or ongoing merger; the dark matter halo core, traced by the ICL, ICGCs, and the dwarf galaxies, may be oscillating around the BCG.

\subsection{The origin of the stellar populations in the intracluster region}\label{sec:origin}
The colours of the ICL and the luminosity function of the ICGCs allow us to infer the mass range of the progenitor galaxies of these stars and, therefore, improve our understanding of how the intracluster stellar component assembled over time. In Sect.\,\ref{sec:colours}, we presented the radial $\YE-\JE$ and $\JE-\HE$ colour profiles of the BCG+ICL up to $a=200$\,kpc from the centre of the BCG.

From $a=20$\,kpc to $a\sim50-90$\,kpc, we measure a significant negative gradient in colour which we interpret as due to a metallicity gradient in the stellar population because the NIR colours are mostly sensitive to metallicity, and relatively insensitive to stellar age (see Appendix \ref{app:models}). This interpretation is supported by \citet{Penny2012}, whose finding of a blue radial gradient in the colours of the GCs out to 40\,kpc also implies a metallicity gradient. This finding is consistent with several stellar population metallicity gradients found in the BCG and ICL within local clusters \citep[e.g.][]{Spavone2020,Montes2021, Hartke2022}. Furthermore, such gradients in the stellar population metallicities are predicted in the halos of massive galaxies due to the cumulative results of violent relaxation, dynamical friction, and tidal interactions over time \citep{Amorisco2017}.

In Fig.\,\ref{fig:colour-profiles} we showed that the ICL has subsolar metallicities ([Fe/H] $\sim-0.6$ to $-1.0$) at $a\gtrsim60$\,kpc. This is consistent with ICL metallicities derived for both nearby and intermediate redshift clusters \citep[e.g.][]{Williams2007, Coccato2010, MT14, MT18, Gu2020}, and suggests that the progenitors of the ICL are galaxies of a few times $10^{9}$\,M$_\odot$ (using the mass-metallicity relation of \citealt{Gallazzi2005}).
The formation scenario of disrupted dwarf galaxies is also supported by the narrow width of stellar streams in the ICL of the Virgo cluster \citep{Rudick2010}.

On the other hand, the total stellar mass of dwarf galaxies that fall into clusters is too small to account for the large stellar mass of the ICL. Dwarfs can provide less than $10$\% of the total ICL mass \citep[][]{Martel2012,DeMaio2015, Kluge2023b}. This suggests there must be another source of metal-poor stars to populate the ICL. A good candidate is the outskirts of more massive satellites ($\sim10^{10}$\,M$_\odot$) which are easily stripped and exhibit similar metallicities \citep{Pastorello2014,Greene2015,Marian2018}.

The properties of the ICGCs provide additional clues to the masses of the progenitor galaxies that contributed to the ICL. Dwarf galaxies ($M<10^9$\,M$_\odot$) have a narrower GCLF \citep[$\sigma\sim0.6$--1.0\,mag,][]{Villegas2010} than giant elliptical galaxies ($\sigma=1.4$\,mag) and the peak of the GCLF is $\sim$0.3\,mag fainter for the dwarfs \citep{Jordan07,Villegas2010,Harris14}.

We find that the GCLF of the BCG, measured in an annulus of $20<a\leq40$\,kpc matches the GCLF typically found in giant elliptical galaxies \citep{Jordan07,Villegas2010,Harris14}: the peak of the Gaussian distribution lies at $M_I=-8.1$\,mag, with a standard deviation of $\sigma=1.4$\,mag. We conclude that the GCs in this annulus formed in situ within the BCG or in the progenitors of massive elliptical galaxies.

The ICGCs at $a>100$\,kpc are neither consistent with the GCLF of the BCG nor can their GCLF be adequately fit by a single Gaussian. By fitting the data with a more complex model, we find that the GCs at $a>40$\,kpc are better fit by two Gaussian distributions: one of which matches the width of that found in the BCG and is fixed to have the same turn-over magnitude, the other, which is fixed to have a turn-over magnitude that is 0.3\,mag fainter, is significantly narrower $\sigma\sim0.78$\,mag. 
This narrower component is consistent with the GCLF of dwarf galaxies (see above).
Therefore the ICGC luminosity function suggests that the progenitor galaxies for these GCs are likely a mix of both giant and dwarf satellite galaxies. 

Inferring the progenitors of the ICL from the progenitors of the ICGCs is not straightforward because the specific frequency of GCs is higher for dwarfs with stellar masses $M<10^9$\,M$_\odot$ than for galaxies with intermediate masses $10^{9.7}<M/{\rm M_\odot}<10^{11}$ (see Sect.\,\ref{sec:tracers}). This disproportionately larger GC contribution from dwarfs motivates the simplicity of the two-component GCLF model but we expect a more continuous mass range of ICL progenitors. Nevertheless, we can readily conclude at least some contribution from dwarf galaxies to the ICL at $a>100$\,kpc.

\subsection{Direct versus indirect accretion of dwarf galaxies}

We have shown in Sect.\,\ref{sec:origin} that GCs which are consistent with having formed in dwarf satellite galaxies contribute significantly to the ICGCs at $a>100$\,kpc. Given this finding, it is puzzling how dwarfs can be destroyed at large clustercentric distances since the tidal forces of the cluster are too weak to directly disrupt them. The typical effective radius of a dwarf galaxy in Perseus is only 1\,kpc \citep{EROPerseusDGs}. Whereas the tidal radius, $r_{\rm T}\approx D\,(M_{\rm sat}/3M_{\rm cluster})^{1/3}$ \citep{Binney2008}, is large with $r_{\rm T}=3.2$\,kpc for a typical dwarf galaxy with total mass $M_{\rm sat}=10^{11}$\,M$_\odot$ \citep[corresponding to a stellar mass $M=10^9$\,M$_\odot$;][]{Wang2015} in a cluster with a total mass $M_{\rm cluster}=10^{15}$\,M$_\odot$, at a clustercentric distance of $D=100$\,kpc, and it increases further with distance. 

A possible reconciliation takes into account the high orbit eccentricity of cluster member galaxies \citep{Stark2019,Aguirre2021}. On such orbits, the pericentres can reach the central cluster regions where the large tidal forces are sufficient enough to disrupt dwarfs.
The liberated stars continue their trajectory and form tidal streams with large apocentres \citep[e.g.][]{MartinezDelgado2023}. Dynamical friction is inefficient in circularising the orbits of dwarf galaxies, whereas the orbits of massive galaxies are more efficiently circularised by this process \citep[deceleration ${
\rm d}v/{\rm d}t\propto M_{\rm sat}$,][]{Chandrasekhar1943}. Hence the tidally stripped stars from massive galaxies will reach lower apocentres than those from dwarf galaxies.
This procedure can build up gradients in colour, metallicity, age, and GC properties.

Another possibility is an indirect accretion channel. 
Dwarf galaxies are first accreted onto the outskirts of intermediate-mass galaxies via the procedure described above. These outskirts get subsequently tidally stripped onto the ICL (Bahé et al., in prep.). This process is hierarchical and closely related to the pre-processing formation channel of ICL \citep[][]{Rudick2006}.

Further support for the indirect accretion channel is found in the colours of the GCs in Perseus. While the GCLFs are compatible with a large fraction of the ICGCs originating from massive ellipticals, there is evidence to suggest that these galaxies were not entirely destroyed during ICL and ICGCs assembly. 
\citet{Harris2020} have shown that 90\% of the ICGCs in Perseus are blue and therefore metal-poor. Blue, metal-poor GCs generally dominate dwarf galaxies and the outskirts of massive galaxies, whereas the central regions of massive galaxies are dominated by red (metal-rich) GCs \citep[e.g.][]{Pota2013}.  
This finding is consistent with GCs from dwarf galaxies being temporarily accreted onto the outskirts of intermediate-mass \lq feeder\rq\ galaxies that are then tidally stripped and deposited into the intracluster region.

\section{Conclusions}
We present an in-depth analysis of the intracluster stars of the Perseus cluster with the Early Release Observations taken by \Euclid. To enable analysis of the LSB features we removed contamination by Galactic cirri and large-scale gradients from the \IE\ image, but no such additional processing was needed for the NIR images. GC candidates were selected as compact point-like sources and we subtracted a background population to derive the distribution and luminosity function of the GCs associated with the cluster. 

We detected ICL and ICGCs out to 600\,kpc from the BCG, NGC\,1275. The BCG+ICL contains a total stellar mass of $1.7\times10^{12}$\,M$_{\odot}$, making up 38\% of the total cluster light within 500\,kpc, and there are $\sim$70\,000 GCs in this same volume. 

The ICL and the ICGCs have a quantitatively similar spatial distribution (Figs.\,\ref{fig:IE_map} and \ref{fig:GC_map}) and follow the same radial slope (beyond 60\,kpc; Fig.\,\ref{fig:GC_1d}), where we find $\sim60$--80 GCs per $10^9$\,M$_{\odot}$ of diffuse light (Fig.\,\ref{fig:T}). The centroid of the isophotes of the BCG+ICL and ICGCs show an increasing offset with radius to the west (decreasing RA), reaching $\sim60$ kpc from the centre of the BCG (Fig.\,\ref{fig:offsets}). 

The large-scale distribution of the ICL and ICGCs is qualitatively similar to that of the cluster galaxies, but is inconsistent with the gaseous intracluster medium.  The intracluster gas has a lower ellipticity (Fig.\,\ref{fig:xrays}), and on the largest scales its contours are asymmetric towards the east of the BCG.  

This mismatch in the distribution of the cluster components suggests that Perseus is undergoing a merger. The existence of a line-of-sight-velocity offset of the dwarf galaxy population relative to the BCG prefers a scenario where the spatial offset results from the dark matter halo core oscillating around the BCG. The dark matter halo, ICL, ICGCs, and dwarf galaxies potentially trace this oscillatory motion.

The isophotes and iso-density contours of the ICL and ICGC distribution have an ellipticity $\epsilon\sim0.4$ (beyond $a=150$\,kpc), and a radial 3D slope of $\delta{\mathrm{log}}\,\Sigma_* / \delta{\mathrm{log}}\,a_{\rm 3D}~=-2.47$ (fitted in the range $20<a<470$\,kpc), which is consistent with the expectations of the shape and radial profile of typical cluster-sized dark matter halos. This suggests that these luminous intracluster stellar populations are good tracers of the underlying dark matter.

We present several pieces of evidence that suggest the ICL and ICGCs were predominantly stripped from the low-metallicity outskirts of massive cluster galaxies, with a possible contribution from disrupted dwarf galaxies. The first piece of evidence comes from the NIR colours of the ICL beyond 60\,kpc (Fig.\,\ref{fig:colour-profiles}). We show that the ICL is relatively blue suggesting the stars consist of a low-metallicity population consistent with that of the dwarf galaxies in Perseus or the outskirts of more massive satellite galaxies ($\sim10^{10}$\,M$_{\odot}$). 

Next, we show that the ICL and ICGCs were not assembled from the full destruction of massive cluster galaxies. Most stars and GCs in massive cluster elliptical galaxies are metal-rich and have correspondingly red colours, so the complete destruction of the galaxy would produce a redder ICL than measured. Furthermore, the specific frequency of the GCs in the ICL region is extremely large (60 GCs per $10^9$\,M$_{\odot}$ of diffuse light) compared to the typical values found in most galaxies \citep{Liu2019}.  If the ICL and ICGCs came from the complete destruction of massive galaxies the specific frequency of the ICGCs would be similar to that of their progenitor galaxies, whereas such high specific frequencies as we observe can be built up by stripping the GC-rich outskirts of massive galaxies. 

The ICGCs provide further clues to determine the progenitors of the intracluster stars. First, the two components of the ICGC luminosity function (Fig.\,\ref{fig:GCLF1} and Table\,\ref{tab:GCLF2}) imply that $\sim$50\% of the ICGCs beyond $a=100$\,kpc belonged to massive cluster satellites at one time. The remaining $\sim$50\% have a smaller dispersion in luminosity, consistent with that of dwarf galaxies with stellar masses $\leq10^9$\,M$_\odot$.
The blue colour of the ICGCs (measured by \citealt{Harris2020}) supports their origin from low-metallicity progenitors. This can be either dwarf galaxies or the outskirts of intermediate-mass ellipticals. A direct accretion of dwarf galaxies is unlikely because their tidal radius is large at large clustercentric distances. A possible reconciliation is a hierarchical accretion scenario: where dwarf galaxies are accreted onto the outskirts of intermediate-mass \lq feeder\rq\ galaxies, which are subsequently stripped and deposited into the ICL.

The results presented in this work highlight the power of constraining the stellar assembly history in the intracluster region by combining a large field of view and the resolution required to explore the GC systems of nearby clusters of galaxies. The inclusion of unprecedentedly deep NIR colours in our analysis provides a consistent inference of the progenitor galaxies that make up the ICL and ICGCs.

\begin{acknowledgements}
The authors thank the referee for helpful and constructive
comments on the draft.
NAH and JBGM gratefully acknowledge support from the Leverhulme Trust through a Research Leadership Award, and acknowledge support from the UK Science and Technology Facilities Council (STFC) under grant ST/X000982/1.
MM acknowledges support from the Project PCI2021-122072-2B, financed by MICIN/AEI/10.13039/501100011033, and the European Union “NextGenerationEU”/RTRP and IAC project P/302302.
TS, AL acknowledge support from Agence Nationale de la Recherche, France, under project ANR-19-CE31-0022.  

\AckEC  

\AckERO

This research made use of Photutils, an Astropy package for detection and photometry of astronomical sources (Bradley et al. 2023).

\end{acknowledgements}

%
%

\bibliography{ref}

\appendix
\section{Residual background constant} \label{sec:background_constant}
\begin{strip}
    \setlength{\baselineskip}{0.9\baselineskip}
    \includegraphics[width=\linewidth]{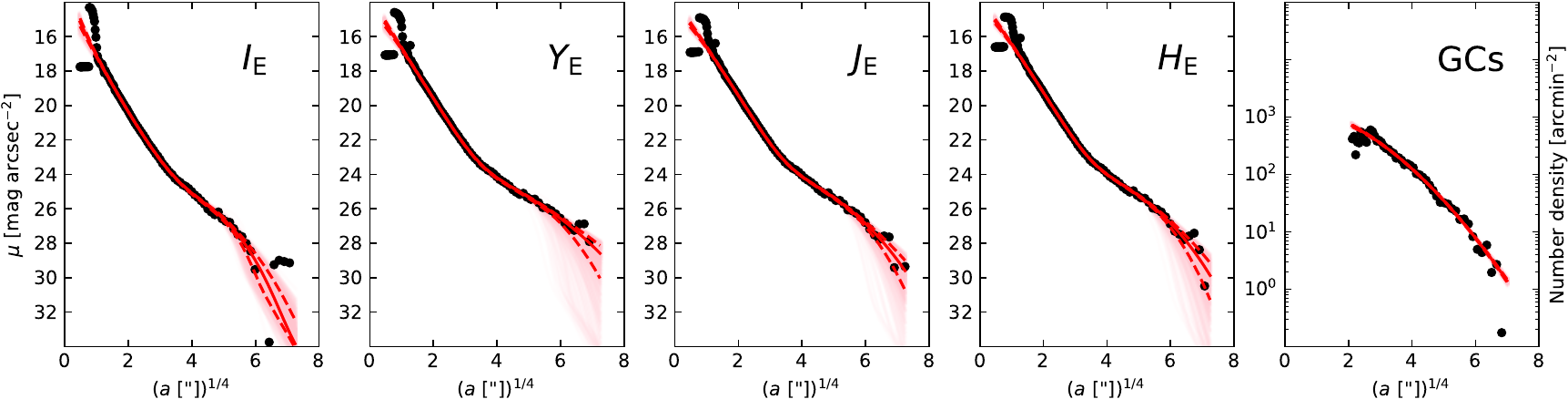}
    \refstepcounter{figure}
    \vspace{-4.5pt}
    {\normalfont\small\\ \textbf{Fig. \protect\NoHyper\ref{fig:box_scatter3}\endNoHyper.} Surface brightness profiles of NGC\,1275 are shown by black points. The solid red line is the best-fit double-S\'ersic function. Semi-transparent red lines show Monte Carlo realisations of the fitted profiles, accounting for the covariance matrix. The 16th and 84th percentiles of these profiles are given by the red dashed lines.}
    \label{fig:box_scatter3}
    
    \vspace{12pt}
    \includegraphics[width=\linewidth]{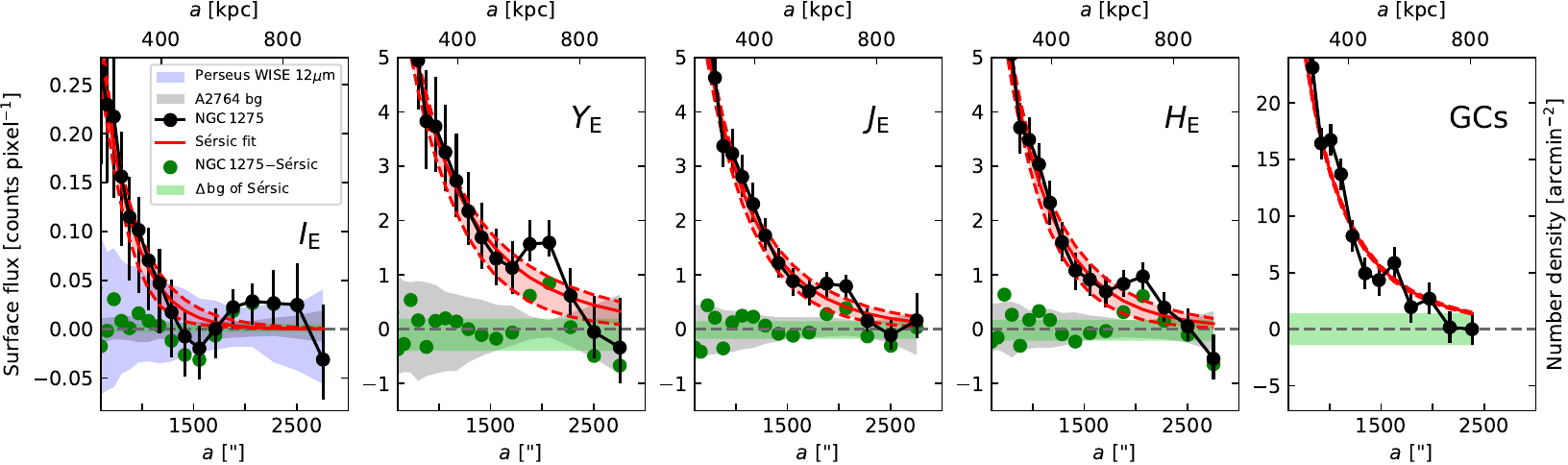}
    \refstepcounter{figure}
    \vspace{-4.5pt}
    {\normalfont\small\\ \textbf{Fig. \protect\NoHyper\ref{fig:box_scatter4}\endNoHyper.} Surface flux profiles of NGC\,1275 are shown by black lines. The solid and dashed red lines correspond to the best-fit double-S\'ersic profiles and the uncertainties as in Fig. \ref{fig:box_scatter3}. These profiles are subtracted from the surface flux profiles of NGC\,1275, yielding the green data points. For the ICL, the residual background constant is estimated by averaging the outermost eight of these points, with that constant set to zero in this figure. The green shading represents the uncertainties on the residuals. For the ICGCs, the background is set to the number density of the outermost data point. The grey (blue) shading represents the uncertainties from the background surface flux profiles of Abell\,2764 (the Perseus WISE 12$\mu$m map). Both uncertainties combined and flipped result in the black error bars of the surface flux profiles of NGC\,1275. One count in \IE\ (\YE, \JE, \HE) corresponds to 27.4 (25.1)\,mag\,arcsec$^{-2}$.}
    \label{fig:box_scatter4}
\end{strip}

To calculate the surface brightness and number density profiles of the ICL and ICGCs, we first need to subtract a background value from each of the annuli. Residual background signals affecting the ICL arise from time-varying detector backgrounds, Zodiacal light, Galactic cirri, or scattered light. For the ICGCs, the background consists of foreground stars and background objects that passed the compactness criteria set out in Sect.\,\ref{sec:GC_detection}, which we assume are distributed homogeneously on large scales.

The residual background constant is typically determined as the value to which the surface flux or number density profiles converge at large radii. In our case, the spatial extent of the BCG+ICL is comparable to the field of view of the \Euclid observations. Residual ICL at the outermost radii biases the background estimate high. To account for this possible flux we fitted double-S\'ersic functions to the surface brightness profiles (shown as red opaque continuous lines in Fig.\,\ref{fig:box_scatter3}). We subtracted these models from the measured surface flux profiles (which are shown as the black solid lines with vertical error bars in Fig.\,\ref{fig:box_scatter4}) to leave behind the residual background (which we display as the green points). Finally, we estimated the residual background constant as the average value of the eight outermost data points in the residual profile. This process was iterated eight times until convergence.

The uncertainty in the residual background constant (green shading) was estimated by subtracting from the observed surface flux profile 1000 double-S\'ersic profiles (semi-transparent lines in Fig.\,\ref{fig:box_scatter3}) generated based on the best-fit parameters and their covariance matrices, and computing the background constant for every case, following the method described above. The 16th and 84th percentiles of the distribution of all background constants were taken as the lower and upper uncertainty bounds, respectively.

The best-fit double-Sérsic profiles in Fig.\,\ref{fig:box_scatter3} are the converged results after correction for the residual background constant and correspond to the fits listed in Table \ref{tab:sersic} for the \IE and \HE profiles.
This method assumes that the best-fit double-S\'ersic profile remains (within the statistical uncertainties of the best-fit parameters) a valid description of the real surface brightness profile beyond the largest fitted radius.

Unfortunately, this method is not applicable to the ICGCs. Their number density profile is well-described by a single S\'ersic function with limited flexibility at large radii. Hence, the iteration did not converge.
Fortunately, the ICGC number density profile suffers much less from background inhomogeneity.
Therefore, we set the outermost number density to zero and examine the uncertainty given that choice in Sect.\,\ref{sec:icgc_background_uncertainties}. Note that the background of the IGCGs is determined differently when we needed to examine the luminosity function of the background sources; this is described in detail in  Sect.\,\ref{sec:GCLF_res}.

\section{Background uncertainties} \label{sec:background_uncertainties}

Uncertainties in surface flux profiles due to the background can be estimated using several methods. These include the standard deviation or the standard error of the mean of the outermost data points or the variation in the background at different spatial scales estimated using randomly placed boxes.
These techniques are ideal in the Poissonian regime where the uncertainties are purely statistical and uncorrelated. However, in deep images of extended galaxies, the uncertainties in the faint outskirts are dominated by systematic background inhomogeneities. Correlations between surface fluxes measured in adjacent annuli arise from background patterns on preferred spatial scales, such as those introduced by a time-variable bias level in each detector.

\subsection{Quantifying the background homogeneity using randomly placed apertures}

\begin{figure}
    \centering
    \includegraphics[width=\linewidth]{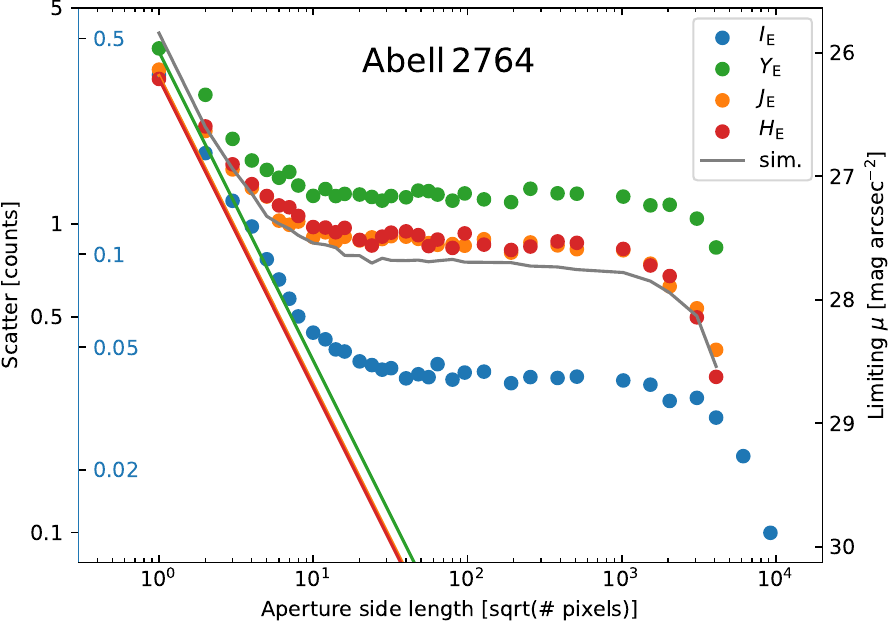}
    \caption{Standard deviation of the mean pixel values in randomly placed square apertures with increasing size. Diagonal lines correspond to pure Poissonian noise while upward deviations indicate systematic background inhomogeneities. The grey line is derived from simulations of white noise with random constants added to each simulated NISP-detector image.}
    \label{fig:box_scatter}
\end{figure}

To investigate this effect, we have utilised the ERO of Abell\,2764. This field captures part of the galaxy group Abell\,2764, but most of the field of view is relatively empty of large, nearby galaxies. This ERO data is processed using the same method as the Perseus ERO, which is described in \citet{EROData}. To be onsistent with the procedure described in Sect. \ref{sec:dataprocessing}, we also fitted and subtracted a 2D fourth-order polynomial from the masked \IE image of Abell\,2764 to account for the large-scale gradient.
Following \cite{Matthews1997} and \cite{Iodice2002}, we then estimated the background inhomogeneity on different spatial scales. We began by placing 1000 squared apertures of a given size at random locations within the masked images. We calculated the mean pixel value within each aperture and then computed the standard deviation of these mean values for all valid apertures.
This process was repeated for various aperture sizes that are shown on the $x$-axis in Fig.\,\ref{fig:box_scatter}.
To ensure that the reduction in effective size of an aperture due to masked pixels does not bias the scatter high, we considered an aperture valid only when at least 100 pixels remained unmasked. We have tested and confirmed with our simulations (see below) that the effect of masking on the noise estimate is negligible because systematic background inhomogeneities dominate over the Poissonian scatter on these scales.
If an aperture had fewer than 100\,pixels in total, then it must have been completely unmasked to be considered valid.

The scatter in the background determined by this method is shown in Fig.\,\ref{fig:box_scatter} for the Abell\,2764 field in the \IE, \YE, \JE, and \HE bands.
In the ideal case of purely Poissonian noise, the scatter should decrease linearly with the spatial scale (we demonstrate this expected behaviour by the diagonal lines in Fig.\,\ref{fig:box_scatter}). The Poissonian noise is normalised using the scatter of the next-neighboring pixels to avoid correlated noise by the weighted combination of neighboring pixels during image resampling \citep{ERONearbyGals}.

We find that the scatter between random single pixels is dominated by Poissonian noise: the lines converge to the leftmost data point in Fig. \ref{fig:box_scatter}. Around $10\,{\mathrm {pixels}}\times 10\,{\mathrm {pixels}}$ aperture sizes, the scatter enters a saddle point around $27\lesssim SB_{\rm lim} \lesssim 27.5$\,mag\,arcsec$^{-2}$. On these scales, systematic effects like flat-fielding errors, scattered light, cirri, or detector instabilities dominate over the Poissonian noise. Beyond aperture sizes of $2000\,{\mathrm {pixels}}\times 2000\,{\mathrm {pixels}}$, which is consistent with the size of one NISP detector, the scatter in the background decreases again for the NISP filters. The turnover point for \IE is around $5000\,{\mathrm {pixels}}\times 5000\,{\mathrm {pixels}}$, which is consistent with the typical size of cirrus patches (see Sect. \ref{sec:dataprocessing}).

We tested whether a time-variable background constant in each detector is likely the dominant source of background inhomogeneity in both the VIS and NISP cameras by exploring whether the shape of the curve in Fig.\,\ref{fig:box_scatter} can be reproduced by mock observations. We generated a four-by-four grid of squares with $2250\,{\mathrm {pixels}}\times 2250\,{\mathrm {pixels}}$\footnote{Each NISP detector has only $2040\,{\mathrm {pixels}}\times 2040\,{\mathrm {pixels}}$ but we used a larger size to account for the gaps of $\sim$40--80\arcsec\ between the detectors.}, which were populated with random values drawn from a normal distribution with a standard deviation of 4 counts. The pixels within each square (our mock detector) was then offset systematically by a random number drawn from a normal distribution with a standard deviation of 1 count to mimic a time-variable bias level. The resulting scatter profile is shown in Fig.\,\ref{fig:box_scatter} by the grey line and follows the same shape as the scatter curves measured in the background for the NISP filters.
This finding implies that a time-variable background constant in each detector is likely the dominant source of background inhomogeneity in both the VIS and NISP cameras (although we revise this statement for the VIS camera in Sect. \ref{sec:correlated_bgscatter}). We confirmed this variation by visually inspecting single NISP exposures.
Moreover, increasing our simulated grid size to $16\times16$\,squares did not impact the shape of the curve, which means that the decrease in scatter at large box sizes cannot be explained by overlapping random boxes.

\subsection{Correlated scatter in the background: estimation using Abell\,2764} \label{sec:correlated_bgscatter}

\begin{figure*}
    \centering
    \includegraphics[width=\linewidth]{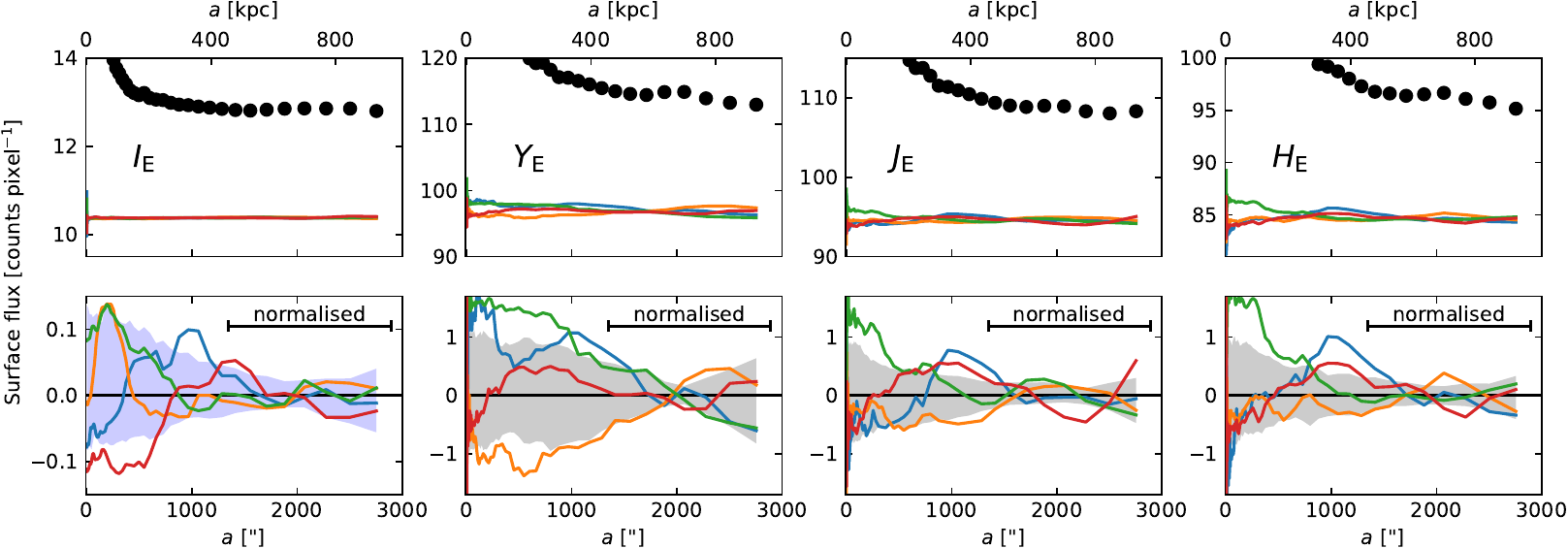}
    \caption{Residual background constant and the background uncertainties of the surface flux profiles. The black dots in the upper panels are the surface flux profiles of NGC\,1275. The coloured lines are four randomly selected background surface flux profiles in Abell\,2764, except for the bottom left panel, where we show profiles measured on the WISE 12$\mu$m map for Perseus. Shades in the lower panels mark the 16th and 84th percentiles. One count in \IE (\YE, \JE, \HE) corresponds to 27.4 (25.1)\,mag\,arcsec$^{-2}$.}
    \label{fig:box_scatter2}
\end{figure*}

The background inhomogeneities measured in the previous section cannot be translated directly to uncertainties of the surface flux profiles of NGC\,1275. In contrast to randomly placed apertures, the elliptical annuli placed around NGC\,1275 are adjacent and, hence, surface fluxes measured from neighbouring annuli are correlated \citep{zhang19}. To demonstrate this, we placed the elliptical annuli (whose construction is described in Sect.\,\ref{sec:isomodelling}) on the masked images of the Abell\,2764 ERO, but added a random global spatial offset to the elliptical centroids. The surface flux profiles were measured in these annuli and the process was repeated 130 times using different global spatial offsets. 
We show four randomly selected surface flux profiles in Fig.\,\ref{fig:box_scatter2} by the coloured lines. 
The top panel of Fig.\,\ref{fig:box_scatter2} shows that the global background in the Perseus images is considerably higher than in the Abell\,2764 ERO in every band. This is likely due to increased zodiacal light because the Perseus cluster is closer to the ecliptic plane than Abell\,2764. The bottom panel of Fig.\,\ref{fig:box_scatter2} shows the same example profiles in \YE, \JE, and \HE\ (while for \IE, we show the profiles measured on the WISE 12$\mu$m map; see below) but after subtracting the mean of the outermost eight surface flux data points ($1400<a[\arcsec]<2800$). Here, the correlation between neighboring surface fluxes becomes apparent by the trends in the individual profiles. We expect that such correlations also exist in the background of the Perseus ERO images as they were processed in the same way. The grey shaded regions in the bottom panel of Fig.\,\ref{fig:box_scatter2} mark the region between the 16th and 84th percentiles of all 130 normalised background surface flux profiles and provide uncertainties in the measured background flux from each elliptical annulus in the NISP filters.

The uncertainties derived from the Abell\,2764 ERO (grey shades) are only consistent for the NISP filter bands with the scatter of the surface flux data points around NGC\,1275 in Fig.\,\ref{fig:box_scatter4}. For \IE, the uncertainties estimated in this way from the Abell\,2764 ERO significantly underestimate the observed scatter in the \IE\ surface flux profile around NGC\,1275. We therefore conclude that the dominant contribution to the background inhomogeneities in the \IE\ band is not a time-variable bias level. We therefore examined whether the sky flux variation due to Galactic cirri could be the dominant source of uncertainty in the background measurement. To do this we assume the cirri in \IE\ has the same large-scale distribution as seen in the dust emission map constructed from WISE 12$\mu$m data of the Perseus ERO region (shown in Fig.\,\ref{fig:vis_steps}f). This map was normalised to the average background properties of the \IE\ image (as described in Sect.\,\ref{sec:dataprocessing}) and then we used this image to measure the surface flux profile in the same elliptical annuli as described above, shifted by 130 different global spatial offsets. The bottom left panel of Fig.\,\ref{fig:box_scatter2} shows four example profiles after subtracting the mean of the points at $1400<a[\arcsec]<2800$. The blue shaded area marks the region between the 16th and 84th percentiles of all 130 normalised background surface flux profiles, and these uncertainties are much more in agreement with the scatter of the outermost surface flux profile around NGC\,1275 in \IE. We conclude that the dominant contribution to the background inhomogeneities in the \IE\ band is not a time-variable bias level but, instead, the remaining cirrus. The discrepancy is most likely due to the higher amount of cirrus in Perseus than in the Abell\,2764 ERO region, as evidenced by the tenfold increase in extinction caused by Galactic dust. Therefore, we adopt the uncertainties derived from the normalised dust emission WISE 12$\mu$m map for the \IE\ surface brightness profiles.

\subsection{Background uncertainties for the ICGCs} \label{sec:icgc_background_uncertainties}

To estimate the background uncertainty for the ICGCs, we fitted a single-S\'ersic function to the measured number density profile (red line in Figs. \ref{fig:box_scatter3} and \ref{fig:box_scatter4}, right panel). The difference between the value of the S\'ersic profile at the outermost radius and the value of the outermost data point (set to zero) gives an estimate of the background uncertainty (green shades and, equivalently, black error bars in Fig. \ref{fig:box_scatter4}, right panel).

\subsection{Combining ICL uncertainties and background uncertainties}

To estimate the total uncertainty in the surface flux profiles of NGC\,1275, we combined the uncertainty in the residual background constant (Appendix \ref{sec:background_constant}) with the correlated background uncertainties measured from the Abell\,2764 ERO data (for the NIR profiles) or normalised dust emission WISE 12$\mu$m map for the \IE\ profile. We used an empirical Monte Carlo approach. In each realisation, we first subtracted a randomly selected background constant obtained using the double-S\'ersic fits from the surface flux profile of NGC\,1275. Next, we subtracted a randomly selected background surface flux profile from the Abell\,2764 ERO data or the normalised dust emission WISE 12$\mu$m map. This process is repeated 130 times, with each background constant and profile used only once.

The lower and upper uncertainties on the surface flux profiles were then adopted from the 16th and 84th percentiles of the resulting profiles. These uncertainties are represented by black error bars in Fig.\,\ref{fig:box_scatter3} and correspond to the grey or blue and green shaded regions, combined in quadrature. The grey (blue) shading reflects the uncertainties in the background profile as estimated from the Abell\,2764 ERO data (normalised WISE 12$\mu$m map), while the green shading captures the uncertainties in the residual background constant based on the extrapolated double-Sérsic profiles of NGC\,1275. Note that the black error bars are flipped because the two components are subtracted.
A summarising number for the 1$\sigma$ limiting depth $\mu_{\rm lim}$ of the 1D profiles can be calculated by the mean of 8 outermost error bars. We obtain $\mu_{\rm lim}=28.9$ ($28.1, 28.8, 28.7$) mag\,arcsec$^{-2}$ in \IE (\YE, \JE, \HE).

To propagate these uncertainties on the structural parameters, we integrated the 130 surface flux profiles and obtained 130 results. The 16th and 84th percentiles of these distributions were adopted as the lower and upper uncertainties for the structural parameters which are listed in Table \ref{tab:struc_params} and given in the text.

\newpage

\section{Masks, models \& residual images} \label{sec:masks_resids}

The derived ICL distributions are sensitive to the masking of high surface-brightness sources, so we show the masked regions in both Figs.\,\ref{fig:masked_zoomin} and \ref{fig:isopy_4panels}.

The BCG is a complex system with diffuse light emitted from an emission line nebula that extends more than 50\,kpc from the nucleus \citep{Conselice2001}. We therefore applied a manual re-masking of NGC\,1275 to ensure that we removed the signal from these regions. The result of this manual re-masking is shown in Fig.\,\ref{fig:masked_zoomin}.

In Fig.\,\ref{fig:isopy_4panels}, we show: the original images (with the model for NGC\,1272 subtracted; first column), the masked images (second column), the residual images (third column), and the isophote models (fourth column). The residuals were obtained by subtracting the model images from the original images. The first four rows correspond to the filter bands \IE, \YE, \JE, and \HE, while the last row shows the map of the GCs.

The isophotes were fitted to the \IE\ band image and then also used for the other filter bands to measure the surface fluxes. This was done to obtain colour profiles that are unaffected by the slightly different apparent shapes of the ICL in the different filter bands.
The only exception is the ICGCs, for which we used the iso-density contours determined on the ICGC image instead of \IE.

\vfill

\vspace{-40em}

\begin{strip}
    \setlength{\baselineskip}{0.9\baselineskip}
    \includegraphics[width=\linewidth]{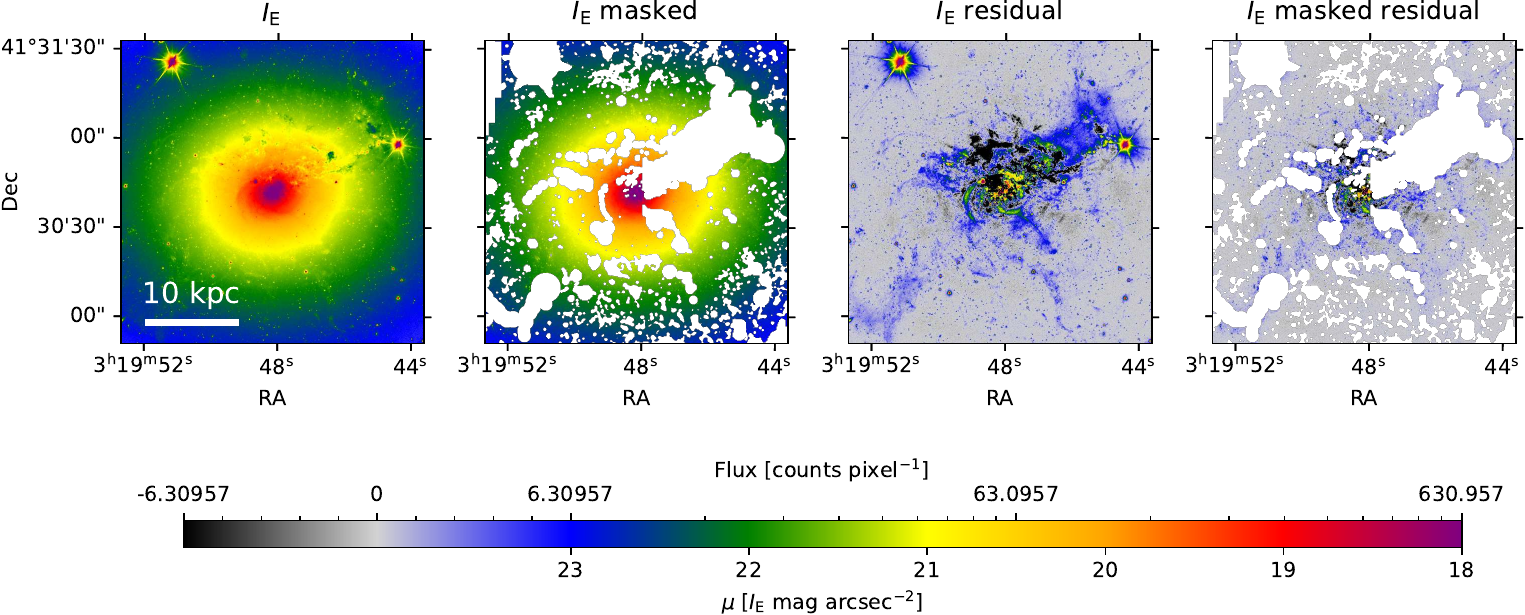}
    \refstepcounter{figure}
    \vspace{-4.5pt}
    {\normalfont\small\\ \textbf{Fig. \protect\NoHyper\ref{fig:masked_zoomin}\endNoHyper.} Zoomed-in view of NGC\,1275 highlighting the inner masks. North is up and east is left. The original image in the left panel shows star formation and dust towards the north and north-west. These regions were masked in the second panel. The third panel shows the residuals after subtracting the isophote model from NGC\,1275. Here, the regions mentioned before plus more filamentary structures in the sound-east and south become readily visible. The masked residual image in the fourth panel shows the accurate masking of these regions.}
    \label{fig:masked_zoomin}
\end{strip}

\clearpage
~

\begin{figure*}[h]
    \centering
    \includegraphics[width=0.8\linewidth]{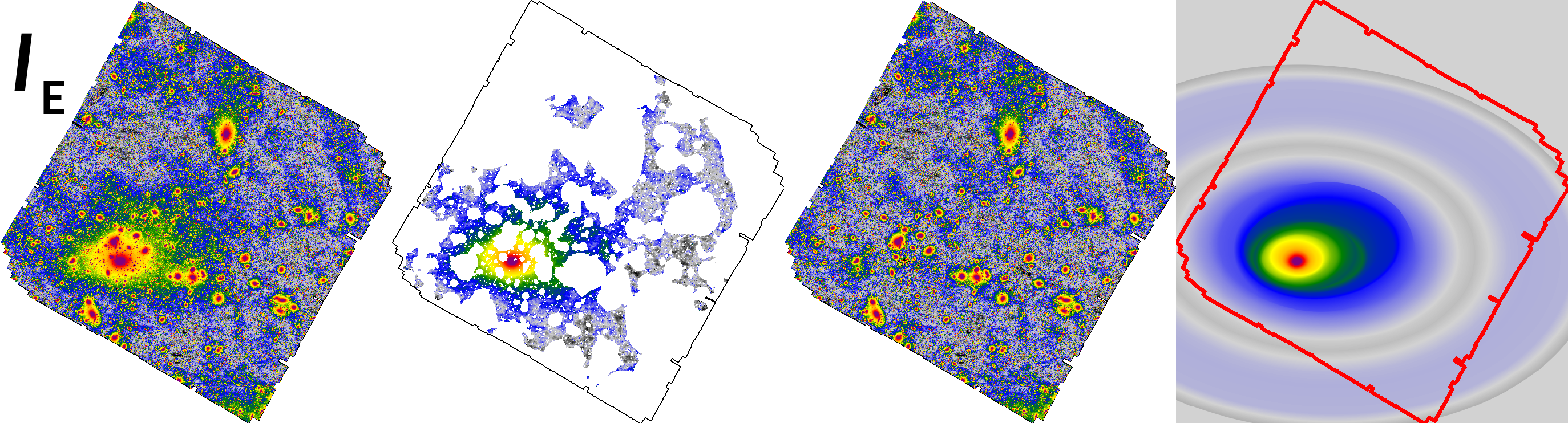}\\
    \includegraphics[width=0.8\linewidth]{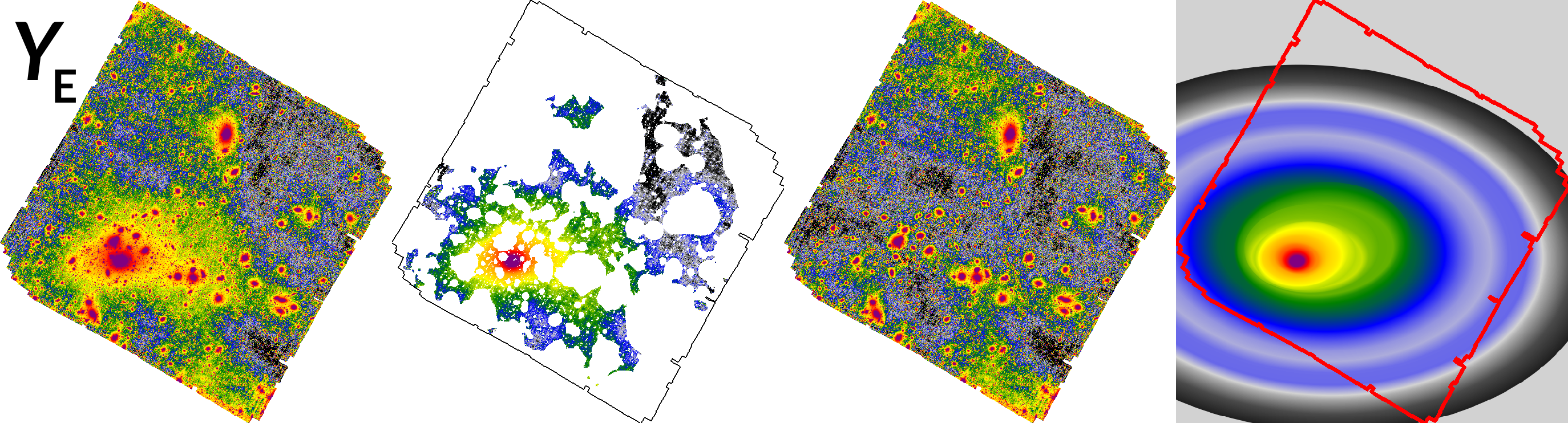}\\
    \includegraphics[width=0.8\linewidth]{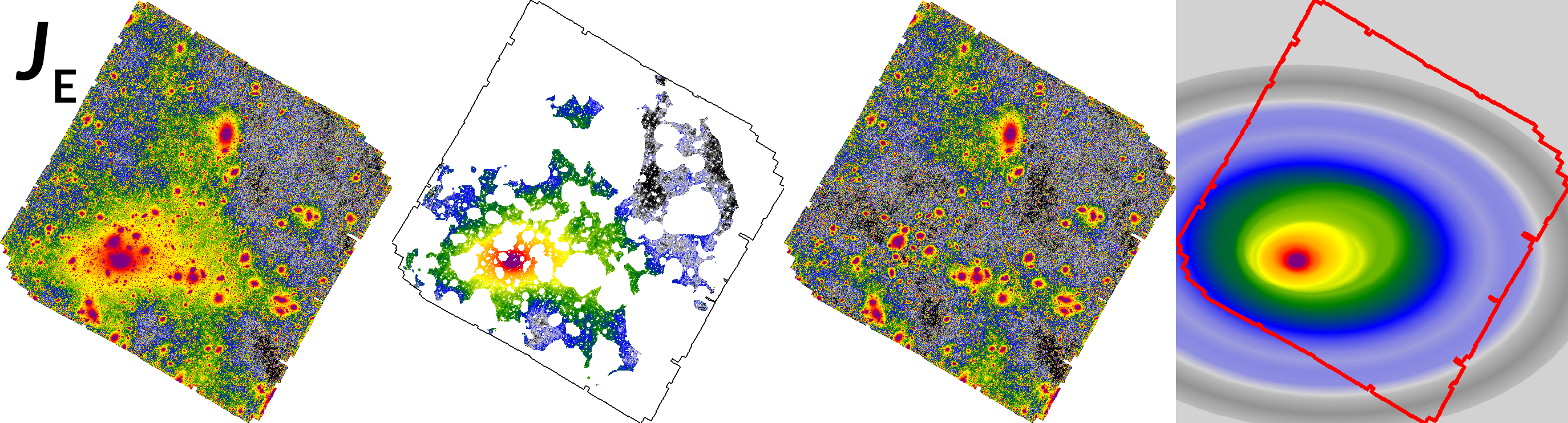}\\
    \includegraphics[width=0.8\linewidth]{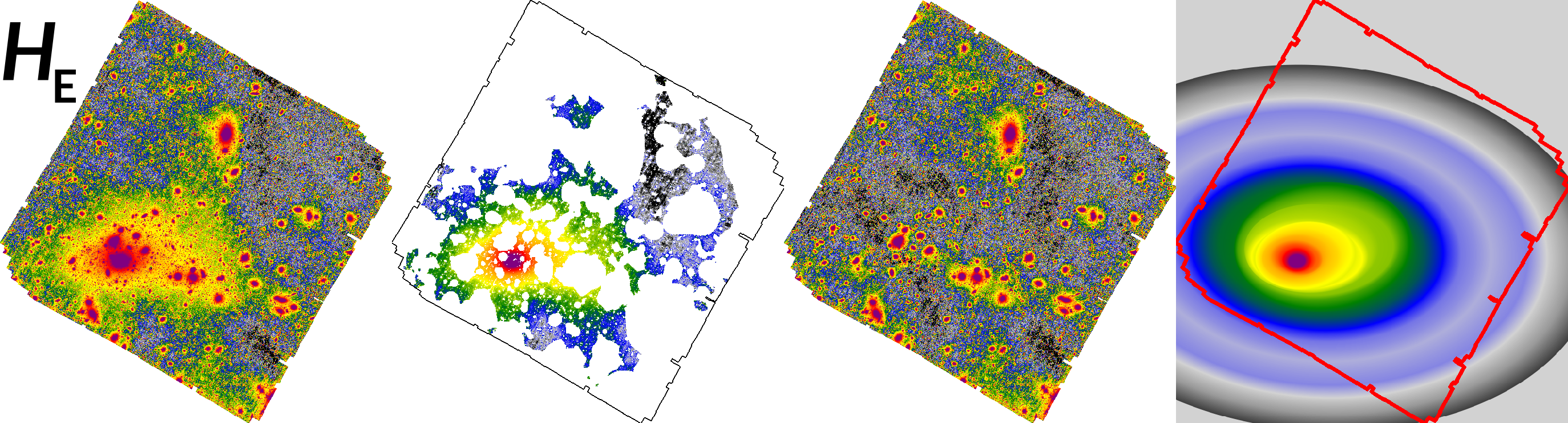}
    \includegraphics[width=0.8\linewidth]{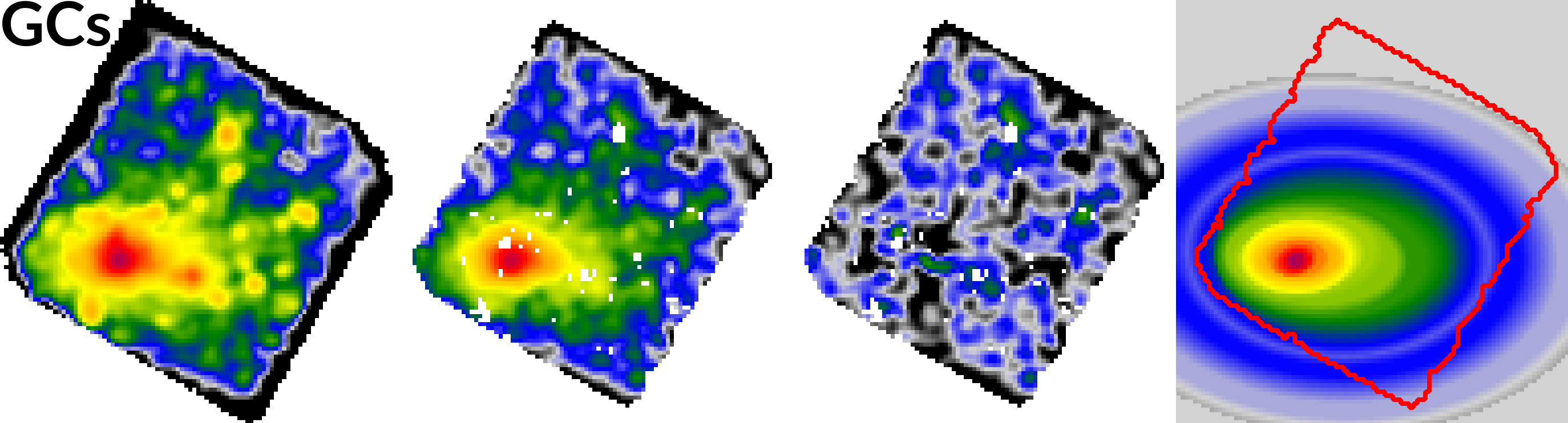}
    \includegraphics[width=0.8\linewidth]{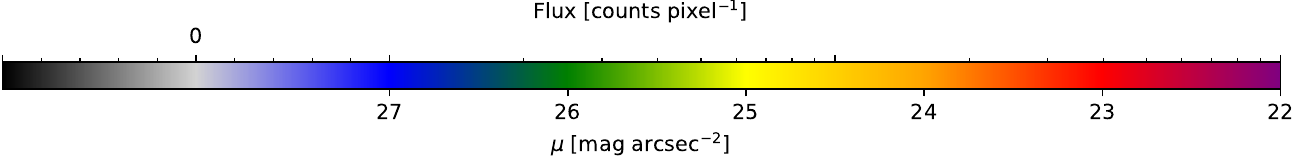}
    \caption{Original images (first column), masked images (second column), residual images (third column), and isophote models (fourth column). The first four rows correspond to the filter bands \IE, \YE, \JE, \HE, while the last row shows the map of the GCs.
    The original images have an isophote model for NGC\,1272 already subtracted.
    For the GCs, the mask has been applied to the original image before smoothing. The outlines in columns two and four show the footprint of the images in column one for the integrated light and column two for the GCs.}
    \label{fig:isopy_4panels}
\end{figure*}

\clearpage

\section{Absolute magnitudes of the Sun and absorption in \Euclid\ filters for Perseus}\label{app:magabs}
\begin{table}[h]
 \caption{Absolute AB magnitudes of the Sun and median Galactic absorption for the Perseus ERO field of view.}
    \centering
    \begin{tabular}{l|l|l}
    \hline\hline
     Filter & AB mag & $A_{\rm Filter}$ \\
    \hline
     \IE  & 4.60 & 0.33\\
     \YE  & 4.54 & 0.17\\
     \JE  & 4.58 & 0.11\\
     \HE  & 4.78 & 0.07\\
    \hline\hline
    \end{tabular}
   
    \label{tab:mags}
\end{table}

To obtain the absolute magnitudes of the Sun in the \Euclid\ filters, we convolved the combined spectrum of the Sun as provided in \citet{Willmer2018}, with the filter response curves of the \Euclid\ photometric system. The convolution was done following Eq.\,(1) in \citet{Montes2014}. 

We also list the median Galactic absorption within each of the \Euclid\ bands for the field of view of the Perseus ERO. These were derived using the {\it Planck} thermal dustmap \citep{Planck_dust2014}, \citet{Gordon2023} extinction law, and assuming an SED of a 5700\,K blackbody. The ICL measurements have been corrected using these median extinctions.

\section{Stellar population models in the \Euclid\ NIR filters}\label{app:models}

\begin{figure}[t]
    \centering
    \includegraphics[width=1\linewidth]{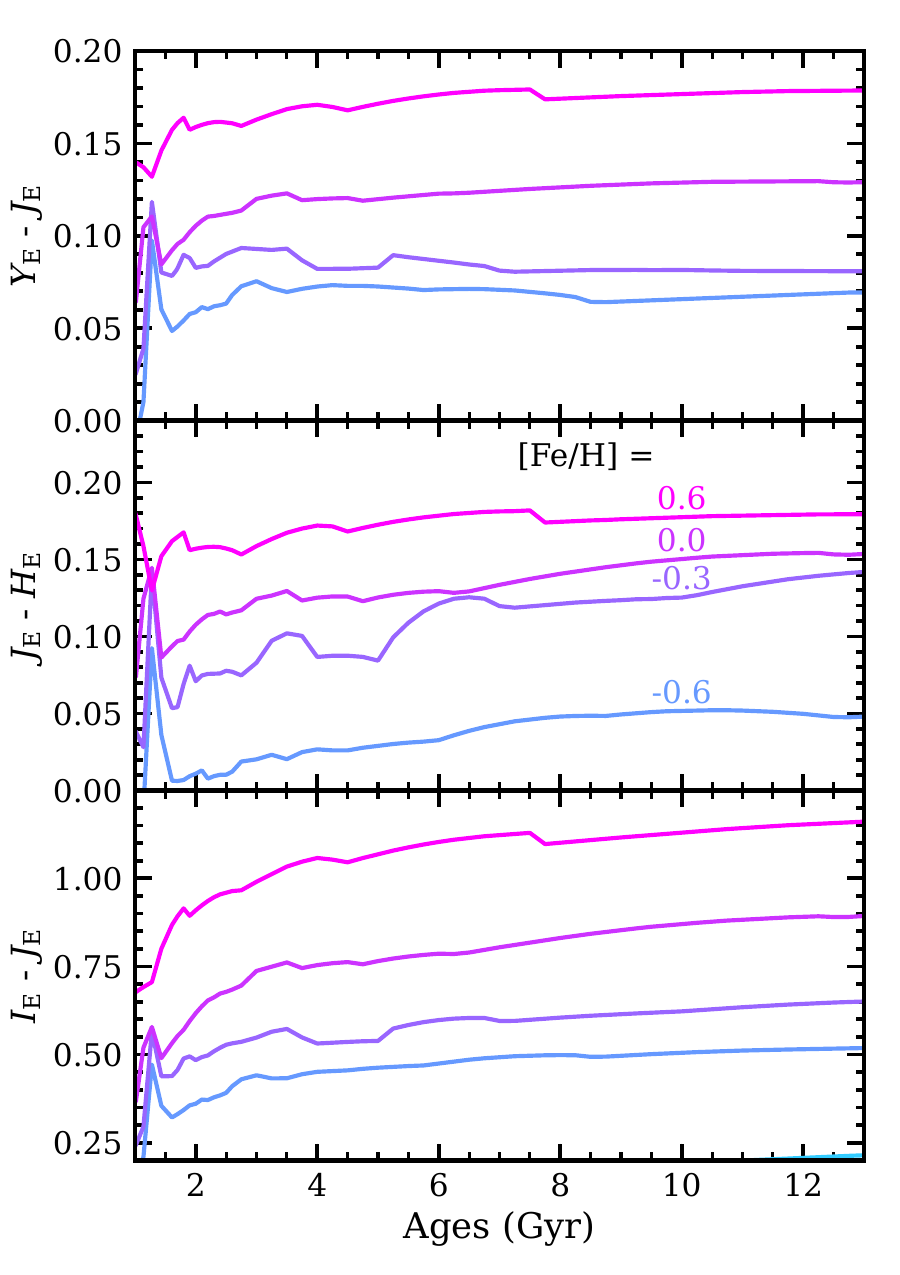}
    \caption{\Euclid\ NIR colours of the \citet{Bruzual2003} single stellar population models shown as a function of stellar age. When the stellar population is older than 2\,Gyr, its NIR colours, $\YE-\JE$ and $\JE-\HE$, are fairly constant with age but vary with metallicity.} 
    \label{fig:models}
\end{figure}

In this section, we explore how the \Euclid\ NIR colours of stellar populations vary with stellar age.
In Fig.\,\ref{fig:models} we show the NIR colours of the single stellar population models from \citet{Bruzual2003} assuming a \citet{Chabrier2003} initial mass function, convolved with the \Euclid\ filters following the prescriptions in \citet{Montes2014}. The models are colour-coded according to the metallicity, as labelled in the middle panel, and are shown as a function of stellar age. The [$\alpha$/Fe] abundances of the models is solar.
The NIR colours $\YE-\JE$ and $\JE-\HE$ of the stellar populations show almost no variation with age once the stellar ages are $>2$\,Gyr.

Several studies have shown that the ICL is $\gtrsim 10$\,Gyr old \citep{Williams2007, Coccato2010, Gu2020}, and therefore, the variation in the NIR colours of the ICL that we observe in Fig.\,\ref{fig:colour-profiles} mainly trace a metallicity variation in the stellar population of the ICL. 

\end{document}